\documentclass[12pt]{article}
\let\ssection=\section
\renewcommand{\section}{\setcounter{equation}{0}\ssection}

\newcommand\mathC{\mkern1mu\raise2.2pt\hbox{$\scriptscriptstyle|$}
        {\mkern-7mu\rm C}} 
\newcommand{\mathR}{{\rm I\! R}}         

\newcommand\bi{\begin{itemize}}
\newcommand\ei{\end{itemize}}
\newcommand\be{\begin{equation}}
\newcommand\ee{\end{equation}}
\newcommand{\F}{\ensuremath{{\bf F}}}
\newcommand{\rr}{\ensuremath{{\bf r}}}
\newcommand{\dd}{\ensuremath{{\delta}}}
\newcommand{\al}{\ensuremath{{\alpha}}}

\newcommand{\pl}{\ensuremath{{\partial}}}

\begin{document}
\begin{titlepage}

\begin{center}
{\large\bf Between Laws and Models:\\
 Some Philosophical Morals of Lagrangian  Mechanics}
\end{center}

\vspace{0.3 truecm}

\begin{center}
        J.~Butterfield\footnote{email: jb56@cus.cam.ac.uk;
            jeremy.butterfield@all-souls.oxford.ac.uk}\\
			[10pt] All Souls College\\ Oxford OX1 4AL
\end{center}

\begin{center}
  Friday 3 September 2004
\end{center}

\begin{abstract}
I extract some philosophical morals from some aspects of Lagrangian  
mechanics. (A companion paper will present similar morals from  
Hamiltonian mechanics and Hamilton-Jacobi theory.) One main moral 
concerns  methodology: Lagrangian mechanics provides a level of 
description of phenomena  which has been largely ignored by 
philosophers, since it falls between their accustomed levels---``laws 
of nature'' and ``models''. Another main moral concerns ontology: the 
ontology of Lagrangian mechanics is both more subtle and more 
problematic than philosophers often realize.

The treatment of Lagrangian  mechanics provides an introduction to the 
subject for philosophers, and is technically elementary. In 
particular, it is confined to systems with a finite number of degrees 
of freedom, and for the most part eschews modern geometry.
\end{abstract}

\vspace{0.5 truecm}
\begin{center}
\end{center}

\noindent Newton's fundamental discovery, the one which he considered 
necessary to keep secret and published only in the form of an anagram, 
consists of the following: {\em Data aequatione quotcunque fluentes 
quantitae involvente fluxiones invenire et vice versa}. In 
contemporary mathematical language, this means: ``It is useful to 
solve differential equations''.\\
\indent\indent V. Arnold, {\em Geometrical Methods in the Theory  of 
Ordinary Differential Equations}, Preface

\end{titlepage}

\tableofcontents

\newpage

\section{Introduction}\label{Intr}
\subsection{Against the matter-in-motion picture}\label{AgstMiM}
Lagrangian mechanics is one of the three great schemes of analytical 
mechanics, which forms a major part of classical mechanics. The other 
two schemes are Hamiltonian mechanics and Hamilton-Jacobi theory; 
(sometimes, Hamilton-Jacobi theory is considered a part of Hamiltonian 
mechanics). This paper is only about Lagrangian mechanics. But its 
companion paper will give a similar discussion---with similar 
morals---of Hamiltonian mechanics and Hamilton-Jacobi theory. So I 
shall begin by celebrating analytical mechanics as a whole, and 
arguing that all of  it, i.e. all three schemes, deserves much more 
philosophical attention than it currently gets.  

Analytical mechanics is one main part, and one of the glories, of 
classical mechanics.  Its development is one of the triumphs, of both 
mathematics and physics, in the last three hundred years. It runs from 
the early discoveries of Maupertuis and d'Alembert, through those of 
Euler, Lagrange, Hamilton and Jacobi, and the application of their 
theories to continuous systems (as in classical field theory), to the 
work of Poincar\'{e} and his twentieth-century successors in chaos 
theory, and catastrophe theory. This development reveals the enormous 
depth and power  of just a handful of ideas, such as configuration 
space, ``least action'', phase space and the Legendre transformation. 
Besides, these ideas, and the theories of analytical mechanics that 
use them, underpin in various ways the twentieth-century theories, 
quantum theory and relativity, that overthrew classical mechanics. In 
short: a triumph---and a rich legacy.

But in recent decades, analytical mechanics, indeed all of classical 
mechanics, has been largely ignored by philosophers of science. They 
have focussed instead on the interpretative problems of quantum theory 
and relativity. There are of course good reasons for this: among them, 
the enormous influence of the new theories on analytic philosophy of 
science, their undeniably radical innovations (e.g. indeterminacy, 
dynamical spacetime) and philosophers' understandable desire to 
address the issues raised by currently accepted physical  theories.

But I fear there is also a worse reason: worse because it is false. 
Namely, philosophers think of classical mechanics as unproblematic. 
Indeed, there are two errors here, corresponding to what I will call 
the `matter-in-motion picture' and the `particles-in-motion picture'. 
Here I will discuss the first, leaving the second to Section 
\ref{Acce}.   According to the matter-in-motion picture, classical 
mechanics pictures the world as made out of bodies (conceived either 
as swarms of a vast number of tiny particles separated by void, or as 
made of continuous space-filling `stuff'), that move through a vacuum 
in Euclidean space and interact by forces such as gravity, with their 
motions determined by a single deterministic law, viz. Newton's second 
law.

 Nowadays this picture is sufficiently part of ``common sense'' to 
seem unproblematic. Or at least, it is part of the common sense of the 
``educated layperson'', with memories of high school treatments of 
falling balls and inclined planes! Certainly a great deal of work in 
contemporary analytic metaphysics of science uses this picture of 
classical mechanics.  Writings on such topics as laws of nature and 
physical properties often appeal to the matter-in-motion picture, as a 
source of examples or counterexamples for various metaphysical theses. 
For example, in debates about such theses as whether laws could  be 
`oaken', or Lewis' doctrine of Humean  supervenience, the speculative 
examples often concern particles, conceived in an essentially 
classical way, and how they might interact with one another when they 
collide.\footnote{Indeed, I think much contemporary philosophy is 
unduly wedded to the idea that the ontology (``world-picture'') of any 
physical theory must be close to this  matter-in-motion picture: 
consider for example discussions of ``naturalism'', or of the contrast 
between causes and reasons. But I shall duck out of trying to 
substantiate these accusations, since they are irrelevant to my aims.}

So I agree that classical mechanics suggests the matter-in-motion 
picture; or to be more specific, the elementary approach to classical 
mechanics, familiar from high school, suggests it. But a moment's 
thought shows that the matter-in-motion picture is problematic. For 
whether we conceive bodies as swarms of  tiny particles, or as made of 
continuous `stuff', there are troublesome questions about how bodies 
interact. 

For example, if they are swarms of particles, how can two bodies 
interact? And how can we explain the impenetrability of solids, or the 
difference between liquids and solids? Indeed, how should we take 
individual particles to interact? A force between them, across the 
intervening void (``action-at-a-distance''), seems  mysterious: indeed 
it seemed so, not only to critics of Newton's theory of gravity, but 
to Newton himself.\footnote{For discussion and references, cf. e.g. 
Torretti 1999, p. 78.} 

On the other hand, if individual particles interact only on contact 
(as the seventeenth century corpuscularians proposed), we face the 
same sorts of question as confront the alternative conception of 
bodies as made of continuous ``stuff''. For example, how should we 
conceive the boundary of a continuous body (including as a special 
case, a tiny particle), in such a way that there can be contact, and 
so interaction, between two such bodies? And even if we have a 
satisfactory account of boundaries and contact, there are questions 
about how to understand bodies' interaction. Considering for 
simplicity two continuous spheres that collide and touch, either at a 
point, or if deformed, over a finite region: how exactly does each 
sphere exert a force on the other so as to impede the other's motion 
(and limit its own further deformation)? Does each particle (i.e. 
point-sized bit of matter) somehow exert a force on ``nearby'' 
particles, including perhaps those in the other sphere?

These are hard, deep questions about the foundations of classical 
mechanics. They were pursued and debated, not only by the giants of 
seventeenth century natural philosophy; but also by their successors,  
the giants of classical mechanics from 1700 to 1900, including the 
heroes of analytical mechanics: Euler, Lagrange, Hamilton and Jacobi. 
But despite the supreme empirical success that classical mechanics had 
achieved by 1900, these and similar questions remained 
controversial---and were recognized as scientifically significant. So 
much so that in 1900 Hilbert proposed the rigorous axiomatization of 
mechanics (and probability) as the sixth of his famous list of open 
problems. But soon afterwards research in the foundations of classical 
mechanics was overshadowed by the quantum and relativity revolutions. 
It was only after 1950 that it flourished again, pursued as much by 
engineers and mathematicians, as by physicists. (It formed part of a 
multi-faceted  renaissance in classical mechanics;  other aspects 
included celestial mechanics, much stimulated by spaceflight, and 
``chaos theory'', stimulated by numerical analysis on computers.) And 
it remains a very active research area. 

But not for philosophers! That is: most philosophers will recognize my 
list of questions as ones with which the  seventeenth century natural 
philosophers struggled. But with the one exception of the mystery of 
action-at-a-distance, our philosophical culture tends to ignore the 
questions, and (even more so) the fact that they are still a focus of 
scientific research. There are at least two reasons, both of them 
obvious and understandable, for this.  

First (as I said above), the quantum and relativity revolutions have 
led philosophers of science  to concentrate on those theories' 
interpretative problems. 

Secondly, there is a humdrum pedagogical reason, relating to the 
educational curriculum's inevitable limitations.  In the elementary 
mechanics that most of us learn in high school, these questions are in 
effect suppressed. The problems discussed are selected so that they 
can be solved successfully, while ignoring the  microscopic 
constitution of bodies, and the details of contact between them. For 
example,  the bodies are often assumed to be small and rigid enough to 
be treated as point-particles, as in elementary  treatments of 
planetary motion; (where, as I conceded,   the main mystery one faces 
is the one acknowledged by philosophers---action-at-a-distance). And 
for some simple problems about extended bodies, like a block sliding 
down a plane, one can manage by adopting a broadly ``instrumentalist'' 
approach, in which for example, one just assumes that each body is 
rigid (whatever its microscopic constitution), contact is 
unproblematic, and  all the forces on a body  (including friction 
exerted at a boundary) act on the body's centre of mass. One then 
determines the motion of each body by determining the vector sum of 
all forces on it. (Pedagogy apart, I shall return to the limitations 
of this approach to mechanics at the start of Section \ref{Anal}.) In 
short, one never faces the questions above.\\
\indent Of course, philosophers often augment high school mechanics 
with some  seventeenth-century mechanics, through studying natural 
philosophers such as Descartes, Hobbes and Leibniz. But again 
curricular limitations impinge. Most philosophers' acquaintance with 
mechanics ends there. For around 1700, natural philosophy divided into 
physics and philosophy, so that few philosophers know about how 
mechanics developed after 1700, and how it addressed these 
foundational  questions. In particular, as regards the eighteenth 
century: philosophers read Berkeley, Hume and Kant, but not such 
figures as Euler and Lagrange, whose monumental achievements in 
developing analytical   mechanics, and in addressing such questions, 
changed the subject out of all recognition: a transformation continued 
in the nineteenth century by figures like Hamilton and Jacobi.

In addition to these two reasons, there may well be others. For 
example, we now know that quantum theory underpins the empirical 
success of classical mechanics in the macroscopic world. (Somehow! 
There are many open questions, both philosophical and technical, about 
the details: the quantum measurement problem, and the physics of 
decoherence, are active research areas.) From this, some philosophers, 
especially those with more instrumentalist inclinations, will conclude 
that we need not worry too much about foundational questions about 
classical mechanics. 

These reasons are obvious and understandable. But their effect---the 
belief that the matter-in-motion picture is unproblematic (apart 
perhaps from action-at-a-distance)---is unfortunate. It is not just 
that good foundational questions about classical mechanics get 
ignored. There is also a loss to our understanding of modern 
philosophy of science's origins. For it was not only the quantum and 
relativity theories that had an influence on philosophy of science; 
(as I mentioned). Before they arose, the lively debate over these 
foundational questions also strongly influenced philosophy of science. 
For example, Duhem's instrumentalism was largely a response to the 
intractability of these questions. 

To be sure, today's specialist philosophers of physics know perfectly 
well that the matter-in-motion picture is problematic. Even in what 
seem the most unproblematic cases, there can be both a wealth of 
intricate mathematical structure, and plenty of philosophical issues 
to pursue. 
The familiar case of Newtonian point-particles interacting only by 
gravity affords good examples. Most philosophers would say that surely 
the mathematics and physics of this case is  completely understood, 
and the only problematic aspect is gravity's acting at a distance. But 
the mathematics and physics {\em is} subtle.\\
\indent For example, Painlev\'{e} conjectured in 1898 that there are 
collision-free singularities: that is, roughly speaking, that a system 
of Newtonian point-particles interacting only by gravity could all 
accelerate to spatial infinity within a finite time-interval (the 
energy being supplied by their infinite gravitational potential 
wells), so that a solution to the equations of motion did not exist 
thereafter. Eventually (in 1992) Xia proved that five particles could 
indeed do this.\footnote{For more details, cf. e.g. Diacu and Holmes' 
splendid history (1996; Chapter 3).}\\
\indent As to philosophy,   relationist criticisms of the absolute 
Euclidean space postulated by Newton, and endorsed by the 
matter-in-motion picture, continue to be a live issue (Belot 2000). 
And instead of the usual definition  of determinism in terms of 
instantaneous states across all of space determining the future, 
alternative definitions in terms of the states on open spacetime 
regions have been proposed and investigated (Schmidt 1997, 1998).

So much by way of exposing the errors of the matter-in-motion picture; 
(for more discussion, cf. my (2004: Section 2, 2004a)). I now turn to 
the specific theme of this paper and its companion: the philosophical 
morals of analytical   mechanics.

\subsection{Prospectus}\label{Prosp}
I will draw four such philosophical morals: (the morals will be the 
same in this paper and its companion). Two concern methodology, and 
two concern ontology; and in each pair, there will be a main moral, 
and a minor one. No doubt there are other morals: after all, 
analytical mechanics is a vast subject, inviting philosophical 
exploration. I emphasize these four partly because they crop up 
throughout analytical mechanics; and partly because they all 
(especially the two main morals) arise from a common idea.\\
\indent Namely: by considering the set of all possible states of the 
systems one is concerned with, and making appropriate mathematical 
constructions on it, one can formulate a general scheme for 
representing these systems by a characteristic family of differential 
equations---hence the quotation from Arnold which forms this paper's 
motto. As we shall see, such a scheme has several merits which greatly 
extend the class of problems one can solve; and even when one cannot 
solve the problem, the scheme often secures significant information 
about it. Though the details of the schemes of course vary between 
Lagrangian, Hamiltonian and Hamilton-Jacobi mechanics, they have 
several merits, and morals, in common.\footnote{Of course, the ideas 
of a space of states (state-space), and general schemes for 
representing and solving problems, also occur in other physical 
theories, quantum as well as classical. I believe similar morals can 
often be drawn there. One salient example is  catastrophe theory, a 
framework that grew out of analytical mechanics, but is in various 
ways much more general; I discuss its morals in Butterfield (2004b).} 

These  schemes  lie at a level of generality between two others on 
which philosophers have concentrated: the very general level of ``laws 
of nature'' or ``the laws'' of classical mechanics; and the level of a 
``model'' (which in some philosophers' usage is so specific as to be 
tied to a single physical problem or phenomenon). Hence my title.

Lagrangian mechanics  is  a large subject, and I cannot fully expound 
even its more elementary parts. On the other hand, it is largely 
unfamiliar to philosophers, to whom I want to advertise the importance 
of its details. So I must compromise: I shall describe some central 
ideas---with a minimum of formalism, but with  enough detail to bring 
out my morals. But in a few places (including some whole 
Subsections!), I indulge in  expounding details that are not used 
later on, and so can be skipped: I will announce those indulgences 
with a {\em Warning} in italics.

But fortunately, my morals will be in many ways straightforward, and I 
will only need elementary pieces of formalism to illustrate them. 
There will be nothing arcane or recherch\'{e} here.\footnote{Many fine 
books contain the pieces of formalism I cite (and vastly more!). I 
will mainly follow and refer to just two readable sources: Goldstein 
et al. (2002), a new edition of a well-known text, with the merit of 
containing significant additions and corrections to previous editions; 
and Lanczos (1986), an attractively meditative text emphasising 
conceptual aspects. Among more complete and authoritative books, I 
recommend: Arnold's magisterial (1989), the beautifully careful and 
complete textbooks of Desloge  (1982) and Johns (2005), and 
Papastavridis' monumental and passionate (2002). All these books also 
cover Hamiltonian mechanics and Hamilton-Jacobi theory.} (I suppose 
the reason these straightforward morals have  apparently been 
neglected is, again, the matter-in-motion picture: i.e. the widespread 
belief that the ontology of classical  mechanics is unproblematic, and 
that every problem is in principle solved by the deterministic 
laws---so why bother to look at the details of analytical mechanics?) 

To indicate the road ahead, I shall also  state the morals first, in 
Section \ref{Mora}. Then I illustrate them in two further Sections. In 
Section \ref{Anal}, I describe the basic ideas of analytical 
mechanics, using the principle of virtual work and d'Alembert's 
principle to lead up to the central equations of Lagrangian mechanics: 
Lagrange's equations. Then in Section \ref{Lagr}, I discuss 
variational principles, the reduction of a problem to a simpler 
problem, and the role of symmetries in making such reductions: I end 
by proving a simple version of Noether's theorem. Most of the morals 
will be illustrated, usually more than once, in each of Sections 
\ref{Anal} and \ref{Lagr}. 

Two more preliminary remarks: both of them about the limited scope of 
this paper, and its companion covering Hamiltonian mechanics and 
Hamilton-Jacobi theory.

\indent (1): {\em Eschewing geometry}:---\\
 One main way in which my treatment will be elementary is that I will 
mostly eschew the use of modern geometry to formulate mechanics. 
Agreed, the use of geometry has in the last century formed a large 
part of the glorious development of analytical mechanics that I began 
by celebrating. For example, the {\em maestro}  is no doubt right when 
he says `Hamiltonian mechanics cannot be understood without 
differential forms' (Arnold, 1989: 163). But apart from a few 
passages, I will eschew geometry, albeit with regret. One cannot 
understand everything in one go---sufficient unto the day is the 
formalism thereof!

(2): {\em Finite systems: (Ideal)}:---\\
 Another way in which my treatment will be elementary is that I will 
only consider analytical mechanics' treatment of finite-dimensional 
systems (often called, for short: finite systems). These are systems 
with a finite number of degrees of freedom. That is: they are systems 
whose configuration (i.e. the positions of all component parts) can be 
specified by a finite number of (real-valued) variables. So any finite 
number of point-particles is an example of a finite system.

If one takes bodies (i.e. bulk matter) to be continua, i.e. as having 
matter in every region, no matter how small, of their apparent volume, 
then one is taking them as infinite-dimensional, i.e. as having 
infinitely many degrees of freedom. For there will be at least one 
degree of freedom  per spatial point. In that case, to treat a body as 
finite-dimensional represents a major idealization. But as we shall 
see, analytical mechanics does  this.  Indeed, it idealizes yet more. 
For even if bodies were ultimately discrete, the total number of their 
degrees of freedom would be enormous, though finite. Yet analytical 
mechanics typically describes bodies using a small finite number of 
variables; for example, it might describe the configuration of a bead 
on a ring by just the position of the centre of mass of the bead. So 
whether or not bodies are in fact continua (infinite systems), 
analytical mechanics typically makes the idealization of treating them 
with a finite, indeed small finite, number of coordinates. These are 
in effect collective coordinates that aggregate information about many 
underlying microscopic degrees of freedom.

This  idealization---of treating infinite or ``large-finite'' systems 
as ``small-finite''----will occur often in what follows; so it will be 
convenient to have a label for it. I will call it (Ideal). 

But (Ideal) will not affect my morals; (and not just because, being an 
idealization adopted by analytical mechanics itself, my morals must 
anyway ``follow suit''). There are two reasons. First, the very same 
morals could in fact be drawn from the analytical mechanics of 
infinite systems: but to do so in this paper would make it unduly long 
and complicated.\footnote{Anyway, it is  pedagogically indispensable 
to first treat the case of finitely many degrees of freedom. It was 
also no doubt historically indispensable: even a genius of Lagrange's 
or Hamilton's stature could not have first analysed the infinite 
system case, treating the finite system case as a limit.} 

Second, (Ideal) can of course be justified in various ways. It can be 
justified empirically. For like any idealization, it amounts to 
limiting the theory  to those physical situations where the error 
terms arising from the idealization  are believed to be negligible; (I 
shall be more precise about idealization in Section \ref{schemes}). 
And the countless empirical successes of analytical mechanics using 
(small!) finitely many degrees of freedom shows that indeed, the 
errors often are negligible. Besides, (Ideal) can in various ways be 
justified theoretically; for example, by theorems stating that for 
systems with many degrees of freedom, a collective coordinate like the 
system's centre of mass will evolve according to a formula involving 
only a small number of degrees of freedom.\footnote{Such theorems are 
not special to analytical mechanics: in particular, what is often 
called `the  vectorial approach' to mechanics has such theorems. For 
more discussion of (Ideal), cf. Section \ref{Acce}'s statement of  my 
fourth moral.} 

\section{Morals}\label{Mora}
Recall the idea which I announced as central to all of analytical 
mechanics: that by considering the set of all possible states of 
systems, and making appropriate mathematical constructions on it, one 
can formulate a general scheme with various merits for solving 
problems---or if not solving them, at least getting significant 
information about them. 

With this idea in hand, I can give a general statement of my four 
morals. For all four arise from this idea. They fall into two pairs, 
the first pair concerning scientific method, and the second pair 
ontology. In each pair, one of the morals is the main one since it is 
more novel (and perhaps controversial!) than the other one. For the 
minor morals rebut the idea that classical mechanics is unproblematic, 
just matter-in-motion. As I said in Section \ref{Intr}, that idea is 
wrong. But as I admitted, for specialist philosophers of physics its 
being wrong is old news: so for them, these two rebuttals will be less 
important. 

To help keep track of the morals, I shall give each moral a label, to 
be used in later Sections. The main moral about method will be 
labelled (Scheme); the minor one will have two labels, (Reformulate)  
and (Restrict). The main moral about ontology will be  (Modality); the 
minor one (Accept). (So main morals get nouns, and minor ones verbs.)

The  morals are presented in the following Sections. First, the morals 
about method:\\
\indent\indent (Scheme): Section \ref{schemes}. (Sections \ref{Solv} 
to \ref{ODESTori} give necessary preliminaries.) This moral is about 
analytical  mechanics' schemes for representing problems. Section 
\ref{schemes} will also introduce labels for four merits the 
Lagrangian  scheme enjoys; (merits also shared by the Hamiltonian and 
Hamilton-Jacobi schemes). Those labels are: (Fewer), (Wider), 
(Reduce), (Separate).\\
\indent\indent (Reformulate)  and (Restrict): Section \ref{Refo}. This 
moral is about the methodological value of reformulating and 
restricting theories.\\
\indent Then the morals about ontology:---\\
\indent\indent (Modality) : Section \ref{Moda}. This moral is about 
analytical mechanics' involvements in modality (necessity and 
possibility). I will distinguish three grades of modal involvement, 
labelled (Modality;1st) through to (Modality;3rd). Lagrangian 
mechanics will exhibit all three grades; (as will Hamiltonian and 
Hamilton-Jacobi mechanics).\\
\indent\indent (Accept): Section \ref{Acce}. This moral is about the 
subtle and various ontology of analytical mechanics, in particular 
Lagrangian mechanics.

\subsection{Method}\label{Meth}
 My main moral about method, (Scheme), is that the provision of 
general schemes for representing and solving problems is a significant 
topic in the analysis of scientific theories: not least because it 
falls between two topics often emphasised by philosophers, ``laws of 
nature'' and ``models''.  I will urge this moral by showing that 
Lagrangian  mechanics is devoted to providing such a scheme for 
mechanical problems. But I will lead up to Section \ref{schemes}'s 
description of Lagrangian  mechanics' scheme, and its merits, by :\\
\indent (i): discussing some of physics' various senses of `solve a 
problem': an ambiguous phrase, which repays philosophical analysis! 
(Section \ref{Solv} and \ref{Generalizing});\\
\indent (ii): reporting some results about differential equations that 
I will need (Section \ref{ODESTori}).\\
Finally, Section \ref{Refo} states my minor moral about scientific 
method.\footnote{I shall duck out of trying to prove my accusation 
that the philosophical literature has overlooked a significant topic, 
between ``laws of nature'' and ``models'. But here is one example. 
Giere gives two extended discussions of classical mechanics to 
illustrate his views about laws, models and related notions like 
scientific theories and hypotheses (1988: 62-91; 1999: 106-117, 
165-169, 175-180). Very roughly, his view (partly based on ideas from 
cognitive science) is that science hardly needs laws, and that 
theories  are appropriately structured clusters of models. Maybe: but 
it is noteworthy that he does not articulate the level of description 
provided by analytical mechanics. For him, Newton's laws are typical 
examples of laws, and the harmonic oscillator, or inverse-square 
orbital motion, are typical models; (in fact, the closest Giere comes 
to articulating analytical mechanics is in a brief discussion of 
axiomatic approaches: 1988: 87-89). Similarly, I submit, for the rest 
of the literature.} 

\subsubsection{(Scheme): what is it to be ``given a 
function''?}\label{Solv}
Throughout physics we talk about `solving a problem'. The broad 
meaning is clear. To solve a problem is to state the right answer to a 
question.\footnote{For present purposes, `right' here can mean just 
`what the theory dictates' not `empirically correct'. That is, I can 
set aside the threat of empirical inadequacy, whether from neglecting 
non-negligible variables, or from more fundamental flaws, such as 
those that quantum theory lays at the door of classical mechanics. Of 
course, nothing I say in this paper about analytical mechanics' 
schemes for solving all problems is meant to deny the quantum 
revolution!} And similarly for related phrases like `reducing one 
problem to another': to reduce one problem to a second one is to show 
that the right answer to the second immediately yields the right 
answer to the first; and so on. 

But there is a spectrum of meanings of `stating a right answer', 
ranging from the ``in-principle'' to the ``useful''. The spectrum 
arises from a point familiar in philosophy: the ambiguity of what it 
is to be ``given'' an object (in our case, an answer to a question). 
One is always given an object under a mode of presentation, as Frege 
put it; and the mode of presentation may be useless, or useful, for 
one's purposes.

A standard example in the philosophy of mathematics (specifically, 
discussion of Church's thesis) makes the point clear. Suppose I give 
you a function $f$ on positive integers by defining: $f(n) := 1 \;\; 
\forall n$, if Goldbach's conjecture is true; and $f(n) := 0 \;\; 
\forall n$, otherwise. Have I given you an effectively computable 
function? Intuition pulls both ways: Yes, since both the constant-1 
and constant-0 functions are effectively computable; No, since the 
mode of presentation that the definition used makes it useless for 
your purpose of calculating a value of the function.

Similarly in physics, in particular  mechanics. The solution to a 
problem, the right answer to a question, is typically a function, 
especially a function  describing how the position of a body changes 
with time. And a number or function can be ``given'' uselessly (one 
might say: perversely) as $f$ was in the example.
 So the extreme in-principle meaning for `stating the right answer' is 
logically weak: it tolerates any way of giving the answer. In this 
sense, one may take a specification of appropriate initial conditions 
(and perhaps boundary conditions) for any deterministic set of 
equations to ``solve'' any problem about the variables' values in the 
future. Though this in-principle sense is cavalier about whether we 
could ever ``get our hands on'' the answer to the problem (i.e. state 
it in useful form), it is important. It is connected with significant 
mathematics, about the existence and uniqueness of solutions to 
differential equations, since for the initial conditions to ``solve'', 
even in this in-principle sense, any such problem, requires that they 
dictate a unique solution. (Hence, the jargon: `a well-posed problem', 
`the initial-value problem' etc.)

On the other hand, there are logically stronger meanings of 
`solution'. There is obviously a great variety of meanings, depending 
on just which ways of stating the right answer (what modes of 
presentation) are accepted as ``useful''. Broadly speaking, there is a 
spectrum, from the tolerance of ``in-principle'' to very stringent 
conceptions of what is acceptable: 
 for example that the function that solves a problem must be one of an 
elite family of functions, e.g. an analytic function. 

Here we meet the long, rich history of the notion of a mathematical 
function. For, broadly speaking, physics has repeatedly (since at 
least 1600) come across problems in which the function representing 
the solution (or more generally, representing a physical quantity of 
interest) does {\em not} belong to some select family of functions; 
and in   attempting to handle such ``rogue'' functions, physics has 
often  prompted developments in mathematics. In particular, these 
problems in physics have been among the  main causes of the successive 
generalizations of the notion of function: a process which has 
continued to our own time with, for example, distributions and 
fractals. 

The  general schemes provided by analytical mechanics (whether 
Lagrangian, Hamiltonian or Hamilton-Jacobi) do not  secure solutions 
in any of these stringent senses, except in a few cases. As we shall 
see, these schemes lie in the middle of the  spectrum that ranges from 
the tolerance of ``in-principle'' to the stringent senses. So neither 
these stringent senses, nor the history of how physics had repeatedly 
to go beyond them, is my main topic; (and each is of course a large 
topic on which many a book has been written!). But I need to say a 
little about these senses and this history, just because analytical 
mechanics' schemes are themselves examples---and historically very 
significant ones---of how physics' conception of solving problems has 
gone beyond the stringent senses.\\
\indent In other words: it will help locate these schemes, and  the 
middle of the spectrum, if I report some of the pressure towards the 
middle from the stringent extreme. So Section \ref{Generalizing} gives 
a (brutal!) summary of the history of the notion of function. Then 
Section \ref{ODESTori} reports some basic results about differential 
equations, as preparation for Section \ref{schemes}'s return to the 
schemes.

\subsubsection{Generalizing the notion of 
function}\label{Generalizing}

\paragraph{2.1.2: Preamble}\label{212B} Since at least the seventeenth 
century, the notion of a mathematical function (and allied notions 
like that of a curve) has been successively generalized---and often it 
has been problems in physics that prompted the generalization. I shall 
sketch this development, giving in  Paragraphs 2.1.2.A-C three 
examples of how solving problems, in particular solving differential 
equations, prompted going beyond stringent conceptions of function. 
(For more details of the history, cf. Bottazini (1986), Lutzen (2003), 
Kline (1972) and  Youschkevitch (1976).)\\
\indent {\em Warning}:--- A reassurance at the outset:  none of the 
details in these three Paragraphs will be needed later in the paper, 
though the main idea of the second Paragraph, viz. quadrature, will be 
important later. 

Nowadays, our  conception of  function is utterly general: any 
many-one mapping between arbitrary sets. But this represents  the 
terminus of a long development. In the seventeenth century, a much 
narrower concept had emerged, mainly from the study of curves and its 
principal offspring, the calculus. Authors differed; but speaking 
broadly (and in modern terms), a function was at first taken to be a 
real function (i.e. with $\mathR$ as both domain and codomain) that 
could be expressed by a (broadly algebraic) formula.  One main  
episode in the formation of this concept was Descartes' critique in 
his {\em La Geometrie} (1637: especially Book II) of the ancient 
Greeks' classification of curves; and in particular, their 
disparagement of mechanical curves, i.e. curves constructed by 
suitable machines. More positively, Descartes accepted (and classified 
as `geometric') all curves having an algebraic equation (of finite 
degree) in $x$ and $y$;  thereby including some curves the Greeks 
disparaged as `mechanical', such as the conchoid of Nicomedes and the 
cissoid of Diocles (But Descartes himself, and his contemporaries, of 
course  also studied non-algebraic curves. For details, cf. Kline 
(1972: 117-119, 173-175, 311-312, 335-340) and Youschkevitch (1976: 
52-53).)

But the exploration of new problems, in (what we would call!) pure 
mathematics and physics, gradually generalized the concept: in 
particular, so as to include trigonometric and exponential functions, 
and functions represented by infinite series---though there was of 
course  never an agreed usage of the word `function'. (The word 
`function' seems to be due to Leibniz; as are `constant', `variable' 
and `parameter'.) Kline (1972:  339) reports that the most  explicit 
seventeenth century  definition of `function' was Gregory's (1667): 
that a function is a quantity obtained from other quantities by a 
succession of operations that are  either\\
\indent (i) one of the five familiar algebraic  operations (addition, 
subtraction, multiplication, division and extraction of integral 
roots), or\\
\indent (ii) the operation  we would call taking a limit.\\
But Gregory's (ii) was lost sight of. The contemporary emphasis was on 
(i), augmented with the trigonometric and exponential functions and 
functions represented by series.

Let us put this more precisely, in some modern jargon which is 
customary, though not universal (e.g. Borowski and Borwein 2002). The 
{\em elementary operations} are: addition, subtraction, 
multiplication, division and extraction of integral roots, in a given  
field. (Of course,  the general notion of a field dates from the late 
nineteenth century, together with group, ring etc.; so here we just 
take the field to be the reals.) An {\em algebraic function} is any 
function  that can be constructed in a finite number of steps from the 
elementary operations, and the inverses of any function already 
constructed; e.g. $\surd(x^2 - 2)$.  An {\em elementary function} is 
any function built up from the exponential and trigonometric functions 
and their inverses by the elementary operations, e.g. 
$\log[\tan^{-1}\surd(\exp x^2)+ 1]$. A {\em transcendental function} 
is one that is not elementary.\footnote{A related usage: an algebraic 
number is any number that is a root of a polynomial equation  with 
coefficients drawn from the given field, and a transcendental number 
is any number that is not algebraic. But in this usage, the field is 
almost always taken to be the rationals, so that $\surd 2$ is 
algebraic (and there are only denumerably many algebraic numbers), 
while $\pi$ and $e$ are transcendental.}

So by the early eighteenth century, it had emerged that the function 
that answers a physical  problem, in particular an indefinite integral 
expressing the solution of a differential equation, is often not 
elementary. (Though this is now a commonplace of the pedagogy of 
elementary calculus, it is of course   hard to {\em prove} such 
functions are not elementary.) 
Accordingly, what was regarded as a solution, and as a general method 
of solution, was generalized beyond the elementary functions and 
techniques associated with them. 

  To illustrate, I will discuss (in succeeding Paragraphs) three main 
ways in which such a generalization was made, namely:\\
\indent (A): to include infinite series; a topic which leads to the 
foundations of the calculus.\\
\indent (B): to include integrals, even if these integrals could 
themselves only be evaluated numerically. (Jargon: {\em quadrature} 
means the integration, perhaps only numerical, of a given function).\\
\indent (C): to include weak solutions; which originated in a famous 
dispute between d'Alembert and Euler. \\
\indent {\em Warning}: The details of (A)-(C) are not needed later on, 
and can be skipped. But the main idea of (B), the idea of reducing a 
problem to a quadrature,  will be  prominent later on.

\paragraph{2.1.2.A Infinite series, and the rigorization of the 
calculus}\label{212A} I have already mentioned the admission of 
infinite series. That is to say: such series were not required to be 
the series for an elementary function, especially since one sought 
series solutions of differential equations.

For simplicity and brevity, I will only consider first-order ordinary 
differential equations, i.e. equations of the form $\frac{dx}{dt} = 
f(x,t)$. To find an infinite series solution of such an equation, one 
assumes a solution of the form 
\be
x = a_0 + a_1t + a_2t^2 + \dots \;\; ; \;\; {\dot x} = a_1 + 2a_2t + 
3a_3t^2 + \dots \; ,
\label{seriessoln}
\ee
substitutes this into the question, and equates coefficients of like 
powers of $t$. Though Newton, Leibniz and their contemporaries had 
provided such solutions for various equations (including higher order 
equations), the method came to the fore from about 1750, especially in 
the hands of Euler. Such series of course raise many questions about 
convergence, and about what information might be gained by proper 
handling of divergent series.\\
\indent  Here we meet another large and complicated story: the 
development of the calculus, especially its rigorization in the 
nineteenth century by such figures as Cauchy and Weierstrass 
(Bottazini 1986: Chapters 3, 7; Boyer: 1959 Chapter 7; Kline 1972: 
Chapter 40). But for present purposes, I need only note two aspects of 
this story.\\
\indent (i): With the rigorization of analysis, a function became an 
arbitrary  ``rule'' or ``mapping''; and with the late nineteenth 
century's set-theoretic foundation for mathematics, this was made 
precise as a kind of set, viz. a set of ordered pairs. Note the 
contrast with the eighteenth century conception: for Euler and his 
contemporaries, a function was first and foremost {\em a formula}, 
expressing how one ``quantity'' (cf. Gregory's definition above) 
depended on another.\\
\indent (ii): Again, physics' need for functions that were not 
``smooth'' (in various stringent senses) was a major motivation for 
this rigorization. It was also a significant reason why, after the 
rigorization was accomplished, one could not go back to founding 
analysis solely on such smooth functions. (But even in the late 
nineteenth century, many mathematicians (including great ones like 
Weierstrass and Poincar\'{e}!) hankered after such a return. Cf. 
Lutzen 2003: 477-484, and Paragraph 2.1.2.C below.)\\
\indent Several episodes illustrate both these aspects, (i) and (ii). 
For example, Dirichlet's famous example  of a function that cannot be 
integrated occurs in a paper (1829) on  the convergence of Fourier 
series (a physically motivated topic). He suggests
\be
\phi(x) := c \;\; \mbox{ for $x$ rational } ; \;\; \phi(x) := d \;\; 
\mbox{ for $x$ irrational } ;
\ee
the first explicit statement of a function other than through one or 
several analytic expressions (Bottazini (1986: 196-201), Lutzen (2003: 
472), Kline (1970: 950, 966-967). Another example is the realization 
(in the 1870s, but building on Liouville's work in the 1830s) how 
``rarely'' is the solution of a differential equation algebraic; (for 
details, cf. Gray (2000: 49, 70f.).)

\paragraph{2.1.2.B Examples of quadrature}\label{212B}
Again I will, for simplicity and brevity,  only consider first-order 
ordinary differential equations $\frac{dx}{dt} = f(x,t)$. Here are 
three standard examples of methods which reduce the integration of 
such an equation $\frac{dx}{dt} = f(x,t)$ to a quadrature. \\
\indent (i): The most obvious example is any autonomous equation, i.e. 
an equation
\be
\frac{dx}{dt} = f(x)
\label{autonmsfoode}
\ee
whose right hand side is independent of $t$. Eq. \ref{autonmsfoode} 
gives immediately $t = \int \frac{1}{f(x)} \; dx$; inverting this, we 
obtain the solution $x=x(t)$.\\
\indent (ii): Another obvious example is the separation of variables. 
That is: If $f(x,t)$ is a product, the problem immediately reduces to 
a quadrature:
\be
\frac{dx}{dt} = g(x)h(t) \;\; \Rightarrow \;\;
\int \frac{1}{g(x)} \; dx = \int h(t) \; dt \; .
\label{sepblefoode}
\ee
\indent (iii): Less obvious is the method of {\em integrating 
factors}, which applies to any linear first-order ordinary 
differential equation, i.e. an equation of the form 
\be
\frac{dx}{dt} + p(t) x = q(t) \; .
\label{linfoode}
\ee
This can be integrated by multiplying each side by the integrating 
factor $\exp [\int^t p(t') dt']$. For the left hand side then becomes 
$\frac{d}{dt}(x(t). \exp(\int^t p))$, so that multiplying the equation 
by $dt$, integrating and rewriting the upper limit of integration as 
$t$, the solution is given by
\be
x(t). \exp(\int^t p) = \int^t dt' \; \{ \; \exp (\int^{t'} p(t'')dt'') 
\; . \;\; q(t') \; \} \; .
\label{solnlinfoode}
\ee
These three examples, indeed all the elementary methods of solving 
first-order equations, were known by 1740. In particular, the third 
example (and  generalizations of it to non-linear first-order  
equations) was given by Euler in 1734/35; (and independently by 
Clairaut in 1739/40: Kline 1972: 476). 

So much by way of examples of quadrature. The overall effect of the 
developments sketched in this Paragraph and 2.1.2.A was that by the 
middle of the eighteenth century, the prevalent conception of a 
function had become: an analytic expression formed by the processes of 
algebra and calculus.\\
\indent Here `processes of algebra and calculus' is deliberately 
vague, so as to cover the developments I have sketched. And `analytic 
expression' emphasises the point above (in (i) of Paragraph 2.1.2.A)  
that a  function was defined as a formula: not (as after the 
rigorization of analysis) as an arbitrary mapping, indeed a 
set-theoretic object.\\
\indent  That a function was a formula meant that it was {\em ipso 
facto} defined for all values of its variable(s), and that identities 
between functions were to be valid for all such values. These features 
were at the centre of a famous dispute, that led to our third 
generalization of the notion of function ...

\paragraph{2.1.2.C Vibrating strings and weak solutions}\label{212C} 
Namely, the dispute between Euler and d'Alembert over the nature of 
the solutions of d'Alembert's wave equation (1747) describing the 
displacement $f(x,t)$ of a vibrating string:
\be
\frac{\partial^2 f}{\pl t^2} = a^2 \; \frac{\partial^2 f}{\pl x^2} 
\;\; .
\label{waveeqn} 
\ee
Though this paper will be almost entirely concerned with ordinary 
differential equations, whose theory is enormously simpler than that 
of partial differential equations, it is worth reporting this dispute. 
Not only did it represent the first significant study of partial 
differential equations. More important: Euler's viewpoint foreshadowed 
important nineteenth century  developments in the notions both of 
function, and of solution of an equation---as we shall see. (For 
details of this dispute, cf. Bottazini (1986: 21-43), Kline (1972: 
503-507),  Lutzen (2003: 469-474), and Youschkevitch (1976: 57-72); 
Wilson (1997) is a philosophical discussion).

More precisely, the dispute was about whether eq. \ref{waveeqn} could 
describe a plucked string, i.e. a string whose configuration has a 
`corner'. In the simplest case, this would be a matter of an initial 
condition in which, with the string extending from $x=0$ to $x=d$:\\
\indent (i) $f(x,0)$ consists of two straight lines, with a corner at 
$x = c < d$. That is: $f(x,0) = kx$, $k$ a constant, for $0 \leq x 
\leq c $ and $f(x,0) = (\frac{-kc}{(d-c)})x + \frac{kcd}{(d-c)}$, for 
$c \leq x \leq d $.\\
\indent (ii) the string has zero initial velocity: i.e. $\frac{\pl 
f}{\pl t}\mid_{t=0} \;\; \equiv \;\; 0$.\\
More generally, the question  was whether  the wave equation can 
describe a waveform, in particular an initial condition, that is (as 
we would now say) continuous but not (even once) differentiable.

\indent  D'Alembert argued that it could not. His reasons lay in the 
prevalent contemporary conception of a function just outlined. But his 
arguments also come close to formulating what later became the 
standard requirement on a solution of a second-order equation such as 
eq. \ref{waveeqn}: viz. that it be twice differentiable in both 
variables.

\indent Euler took the view that analysis should be generalised so 
that it could indeed describe a plucked string: as he puts it (1748) 
`so that the initial shape of the string can be set arbitrarily ... 
either regular and contained in a certain equation, or irregular and 
mechanical'. More generally, Euler advocates allowing functions given 
by various analytic expressions in various intervals; or even by 
arbitrary hand-drawn curves for which, he says, the analytic 
expression changes from point to point. (He calls such functions 
`discontinuous'.)\\
\indent To be precise: according to Truesdell (1956: p. xliii, 1960: 
p. 247-248), Euler proposes to mean by `function' what we now call a 
continuous function with piecewise continuous slope and curvature. 
Accordingly, he disregards the differentiability conditions implicit 
in eq. \ref{waveeqn} and focusses on the general solution he finds for 
it, viz. $f(x - at) + f(x + at)$ with $f$ an arbitrary function in his 
sense.\\
\indent Truesdell stresses the scientific and philosophical importance 
of Euler's innovation. Indeed, he goes so far as to say that `Euler's 
refutation of Leibniz's law [i.e.: Leibniz's doctrine that natural 
phenomena can be described by what we now call analytic functions] was 
the greatest advance in scientific  methodology in the entire century' 
(1960: p.248).\footnote{Truesdell's reading is endorsed by Bottazini 
(1986: 26-27) and Youschkevitch (1976: 64, fn 18, 67).   But I should 
add that---as all these scholars of course recognize---some of Euler's 
work (even after 1748) used, and sometimes even explicitly endorsed, 
more traditional and restrictive notions of function.}

That may well be so:  I for one will not question either Truesdell's 
scholarship or Euler's genius! In any case: several major developments 
thereafter---some of them well into the nineteenth 
century---vindicated Euler's viewpoint that analysis, and in 
particular the conception of solutions of differential equations, 
should be generalised.

 To illustrate, I will sketch one such development: {\em weak 
solutions} of partial differential equations. Though the idea is 
anticipated by Euler and contemporaries (e.g. Lagrange in 1761: cf. 
Bottazzini 1986: p. 31-33), it was properly established only in the 
nineteenth and twentieth centuries; partly through the investigation 
of shock waves---another example of physics  stimulating 
mathematics.\footnote{For a philosophical discussion of different 
kinds of ``optimism'' about mathematics' ability to describe natural 
phenomena (including praise of Euler's optimism), cf. Wilson (2000). 
St\"{o}ltzner (2004) is a sequel to this paper, arguing that optimism 
can and should be combined with what Wilson calls ``opportunism''.}

The idea is to multiply the given partial differential equation 
$L[f]=0$ by a test function $g$ (roughly: a function that is 
sufficiently smooth and has compact support), and then to integrate by 
parts (formally) so as make the derivatives fall on $g$. A function 
$f$ satisfying the resulting equation for all test functions $g$ is 
called a `weak solution' of the original equation. Thanks to the 
integration  by parts, such an $f$ will in general not obey the 
standard differentiability conditions required of a solution of 
$L[f]=0$. (But I should add that one common strategy for finding 
solutions in the usual sense is to first construct a weak solution and 
then prove that it must be a solution in the usual sense.)

To give more details in  modern but heuristic terms, I will consider 
only a linear first-order partial differential equation for the 
unknown function $f$ of independent variables $x$ and $t$; (I follow 
Courant and Hilbert 1962: Chap.V.9, p. 486-490). So the equation is, 
with partial derivatives indicated by subscripts: 
\be
L[f] := A(x,t)f_x + B(x,t)f_t + C(x,t) = 0 \; \; .
\label{linfopde}
\ee 
We define the operator $L^*$ adjoint to $L$ by the condition that 
$gL[f] - fL^*[g]$ is a divergence expression. That is, we define 
\be
L^*[g] := -(Ag)_x - (Bg)_t + Cg {\rm{\;\;\;\; so \;\; that\;\;\;\;}}
gL[f] - fL^*[g] = (gAf)_x + (gBf)_t \;\; .
\label{adjointgetdivexpression}
\ee
For the domain $R$ in which $f$ is considered, we now consider 
functions  $g$ that have compact support in a subregion $S$ of $R$ 
(called {\em test functions}); so that integrating eq. 
\ref{adjointgetdivexpression} over $S$, we obtain by Gauss' theorem 
\be
\int \int_S (gL[f] - fL^*[g]) \;\; dx \; dt = 0 \; .
\label{divintegral0}
\ee
If $f$ is a solution of the partial differential equation, i.e.  $L[f] 
=0$, then
\be
\int \int_S \;\; fL^*[g] \;\; dx \; dt = 0 \; .
\label{divintegral0}
\ee
(There is a converse, roughly as follows. If eq. \ref{divintegral0} 
holds for $f$ with continuous derivatives, for all suitably smooth 
test functions $g$ with compact support in any suitable subregion $S$, 
then eq. \ref{divintegral0} yields $\int \int_S \; gL[f] \; dx dt = 
0$: which implies that $L[f] = 0$.)

This motivates the following (admittedly, non-rigorous!) definition. 
Suppose a function $f(x,t)$ and its partial derivatives are piecewise 
continuous; (i.e. at worst, each possesses  jump discontinuities along 
piecewise smooth curves). Such a function $f$ is called a {\em weak 
solution} of $L[f]=0$ 
in $R$ if for all suitable subregions $S$ of $R$, and suitably smooth 
test functions $g$ with compact support in $S$
\be
\int \int_S \;\; fL^*[g] \;\; dx \; dt = 0 \; .
\label{defineweaksoln}
\ee 

I shall not further develop the idea of a weak solution, since it will 
not be used in the rest of the paper. But I note that it gives yet 
more examples of the theme at the end of Section \ref{AgstMiM}: the 
subtleties of determinism in classical mechanics.\\
\indent (i): It was discovered (with a shock!) in the mid-nineteenth 
century that for a non-linear equation, a solution that begins with 
smooth, even analytic, initial data can develop discontinuities in a 
finite time. And: \\
\indent (ii): For weak solutions of the Euler equations for fluids, 
determinism is strikingly false.  Scheffer (1993) and Shnirelman 
(1997) exhibit weak solutions on $\mathR^3 \times \mathR$ with compact 
support on spacetime. This means that a fluid is initially at rest 
($t=0$) but later on ($t=1$) starts to move with no outside stimulus, 
and later still ($t=2$) returns to rest: the motion being always 
confined to a ball $B \subset \mathR^3$!\footnote{These weak solutions 
are discontinuous unbounded $L^2$ functions.  Underlying
Shnirelman's example is the physical idea (which had already been 
recognized) of an inverse energy cascade   in two-dimensional 
turbulence. Given a force $f(x,t)$ with a small spatial scale, energy 
is transported via the non-linearity of the Euler equations to the 
lower frequencies and longer spatial scales. In particular, if $f$'s 
spatial scale is infinitely small, simple dimension considerations 
show that it nevertheless takes only a finite time for the energy to 
reach the low-frequency range. I am very grateful to Tim Palmer for 
explaining these examples to me.}

To conclude this Subsection: in this history of generalizations of the 
notion of function, I have emphasised the influence of differential 
equations. I of course admit that other influences were equally 
important, though often related to differential equations: e.g. the 
rigorization of analysis.\\
\indent But my emphasis suits this paper's purposes; and is surely 
``not too false'' to the history. Recall the paper's motto; and these 
two famous remarks by late nineteenth-century masters. Lie said `The 
theory of differential equations is the most important branch of 
modern mathematics'; and---more evidence of mathematics being 
stimulated by physics---Poincar\'{e} said `Without physics, we would 
not know differential equations'. 

\subsubsection{Solutions of ordinary differential equations;  
constants of the motion}\label{ODESTori}
I now report some basic results  of the theory of ordinary 
differential equations: results which are crucial for Lagrangian 
mechanics (and for Hamiltonian and Hamilton-Jacobi mechanics). 
Incidentally, this Subsection will also give a glimpse of a fourth way 
that functions have been generalized: viz. to include discontinuous 
functions, so as to describe dynamical chaos. But this paper (and is 
companion) will not discuss chaos. And here my emphasis will be on 
very elementary aspects of differential equations. I shall report 
that:\\
\indent (A) this theory  guarantees the local existence and uniqueness 
of solutions, and of constants of the motion; but\\
\indent (B) for most problems, these constants only exist locally.\\
Both these points, (A) and (B), will be fundamental for later 
Sections.

\paragraph{2.1.3.A The local existence and uniqueness of 
solutions}\label{213A}
I shall follow an exposition of the basic theorems about  solutions to 
systems of ordinary differential equations, by a {\em maestro}: Arnold 
(1973).

 For our purposes, these theorems can be summed up in the following 
four propositions. Arnold calls the first, about first-order ordinary 
differential equations in Euclidean space $\mathR^n$, the `basic 
theorem'. It not only secures the local existence and uniqueness of 
solutions; it also characterizes the local constants of the motion; 
and it underpins the corresponding propositions about differential 
equations that are of higher order than the first, or that are defined 
on differential manifolds rather than $\mathR^n$.

\indent\indent (i): {\em The Basic Theorem} (Arnold 1973: 48-49).\\
Consider a system of $n$ first-order ordinary differential equations
\be
{\dot {\bf x}} ={\bf v}({\bf x}) \;\; , \;\; x \in U
\label{arnold10de}
\ee 
on an open set $U \subset \mathR^n$; equivalently, a vector field $\bf 
v$ on $U$. Let ${\bf x}_0$ be a non-singular point of the vector  
field, i.e. ${\bf v}({\bf x}_0) \neq 0$. Then in a sufficiently small 
neighbourhood $V$ of ${\bf x}_0$, there is a coordinate system 
(formally, a diffeomorphism $f:V \rightarrow W \subset \mathR^n$) such 
that, writing $y_i: \mathR^n \rightarrow \mathR$ for the standard 
coordinates on $W$ and ${\bf e}_1$ for the first standard basic vector 
of $\mathR^n$, eq. \ref{arnold10de} goes into the very simple form
\be
{\dot {\bf y}} ={\bf e}_1 ; {\rm{\;\; i.e. \;\;}} 
{\dot y_1} = 1, \;\; {\dot y}_2 = \dots = {\dot y}_n = 0 {\rm{\;\; in 
\;\;}} W \;\; .
\label{arnold10deparallelized}
\ee 
(In geometric terms: $f_*({\bf v}) = {\bf e}_1$ in $W$.) On account of 
eq. \ref{arnold10deparallelized}'s simple form, Arnold suggests the 
theorem might be called the `rectification theorem'.  NB: For 
simplicity, I have here set aside the non-autonomous case where ${\bf 
v} = {\bf v}(t,{\bf x})$; for details of that case, cf. ibid., p. 56.

\indent\indent (ii): {\em Solutions}; (ibid.: 12, 50).  \\
Let us define a {\em solution} (aka: {\em integral curve}) of eq. 
\ref{arnold10de} to be a differentiable mapping $\phi: I \rightarrow 
U$ of the real open interval $I \subset \mathR$ to $U$ such that for 
all $\tau \in I$
\be
\frac{d}{dt}\mid_{t=\tau} \phi(t) = {\bf v}(\phi(\tau)) \; .
\label{arnoldsoln}
\ee 
The image $\phi(I)$ is called the {\em phase curve} (also sometimes, 
the integral  curve).\\
\indent Then Proposition (i) implies that there is a solution of eq. 
\ref{arnold10de} satisfying the initial condition $\phi(t_0) = {\bf 
x}_0$. But NB: this solution need only exist locally in time: as I 
mentioned in Section \ref{Intr}, a solution need not globally exist 
even for familiar ``deterministic'' theories, such as point-particles 
interacting by Newtonian gravity.\\
\indent Proposition (i) also  implies  that the local solution  is 
unique in the obvious sense that any two solutions with the same 
initial condition are equal on a common sub-interval.

\indent\indent (iii): {\em Constants of the motion}; (ibid.: 75-78).\\
 A differentiable function $f:U \rightarrow \mathR$ is called a {\em 
constant of the motion} (aka: {\em first integral}) of eq. 
\ref{arnold10de} iff its  derivative in the direction of the vector 
field $\bf v$ vanishes. Equivalently: iff $f$ is constant along every 
solution $\phi:I \rightarrow U$; iff every phase curve $\phi(I)$ 
belongs to a unique {\em level set} $f^{-1}(\{c \}), c \in \mathR$, of 
$f$.\\
\indent Typically, eq. \ref{arnold10de} has no first integrals other 
than the trivial constant functions $f(U) = c \in \mathR$.

\indent But it follows from the Basic Theorem that {\em locally} there 
are non-constant first integrals. That is, in the notation of (i):\\
\indent \indent  (a): There is a neighbourhood $V$ of ${\bf x}_0$ such 
that eq. \ref{arnold10de} has $n-1$ functionally independent first 
integrals $f_1,\dots,f_{n-1}$ in $V$. (We say $f_1,\dots,f_m:U 
\rightarrow \mathR$ are {\em functionally independent} in a 
neighbourhood of $x \in U$ if their gradients are linearly 
independent. More precisely: if the rank of the derivative $f_*\mid_x$ 
of the map $f:U \rightarrow \mathR^m$ determined by the functions 
$f_1,\dots,f_m$ equals $m$.)\\
\indent \indent  (b): Moreover, any first integral of eq. 
\ref{arnold10de} in $V$ is a function of $f_1,\dots,f_{n-1}$.\\
\indent \indent  (c): Of course, in the coordinate system in which  
eq. \ref{arnold10de} take the very simple form eq. 
\ref{arnold10deparallelized}, the coordinates $y_2,\dots,y_n$ give us 
$n-1$ functionally independent first integrals; and the other first 
integrals are all the arbitrary differentiable functions of these 
coordinates. 

Assertions (a)-(c) give a sense in which the Basic Theorem secures the 
existence of a coordinate system in which the problem of integrating 
eq. \ref{arnold10de} is {\em  completely solved, locally}. But I 
stress that this  does {\em not} mean it is easy to {\em write down} 
this coordinate system: to write down the diffeomorphism $f$. In 
general, it is very hard to do so!\\
\indent This point can hardly be over-emphasised. It will be a 
recurrent theme in this paper---and I will discuss it in philosophical 
terms already in Section \ref{schemes}.

\indent\indent (iv): {\em Other cases}:\\
 Corresponding definitions and propositions hold for:\\
\indent  (a) ordinary differential equations of higher order, 
principally by the standard device of writing down an equivalent 
system of first-order equations in which new variables represent the 
higher derivatives of the original variable or variables  (ibid.: 
59-61); and \\
\indent  (b) collections of such equations of varying orders  (p. 
62-63); and\\
\indent  (c)  ordinary differential equations defined, not on a patch 
of $\mathR^n$ but on a differential manifold (p. 249-250).

\indent There are countless details about (iv) which I will not 
report; (some more details are in Sections \ref{LagEqAccScheme} and  
\ref{VecfieldsSymmies}). Here I note only the following:---\\
\indent \indent In Lagrangian mechanics, the dynamics  of a system 
with $n$ configurational degrees of freedom is essentially described 
by $n$ second-order ordinary differential equations; or equivalently, 
by $2n$ first-order equations. (In most cases, this system of 
equations is  defined on a manifold, not a patch of $\mathR^n$.) This 
system  has a locally unique solution, specified by $2n$ arbitrary 
constants; (roughly, the initial positions and velocities of the 
system's constituent particles). Besides, we define a first integral 
of a differential equation of arbitrary order (or of a system of them) 
as a first integral of the equivalent system of first-order equations. 
This means that for this system, as in (iii) above: global constants 
of the motion are rare; but  locally, we are guaranteed that there are 
$2n-1$ of them.

\paragraph{2.1.3.B The rarity of global constants of the motion: the 
circle and the torus}\label{213B} I said in (iii) of Paragraph 2.1.3.A 
that typically, eq. \ref{arnold10de} has first integrals other than 
the trivial constant functions $f(U) = c \in \mathR$, only {\em 
locally}. We will later be much concerned with the few global 
constants such as energy  that arise in mechanics. But it is worth 
giving at the outset two (related) examples of a system with no global 
constants of the motion. For they are a simple and vivid illustration 
of the rarity of such constants. (They are also a prototype for: (i)  
topics that will loom large in the companion paper, viz. 
Poincar\'{e}'s theorem, and the theory of completely integrable 
systems; (ii) structural stability, a central topic  in catastrophe 
and bifurcation theory.)

One or both examples  occur in many  books. But again I recommend 
Arnold. For proofs of the results below, cf. his (1973: 160-167); or 
more briefly, his (1989: 72-74). For further results (including 
details about topics (i) and (ii)), cf. his (1983: 90-112; 1989: 
285f.).  

{\em First example: the circle}:---\\
The first example is a ``toy-model'' in discrete time, rather than a 
classical mechanical system. Consider a circle $S$, and let $\tau: S 
\rightarrow S$ be a rotation through an angle $\al$. (Think of $S$ as 
a space of states, and $\tau$ as time-evolution in discrete time.) If 
$\al$ is {\em commensurable} (aka: commensurate) with $2\pi$, i.e. 
$\al = 2\pi(m/n)$ with integers $m,n$ , then $\tau^n$ is the identity. 
So for any point $x \in S$, the orbit of $x$  under the repeated 
action of $\tau$ (the set of images of $x$) is closed: it eventually 
rejoins itself. But if $\al$ is incommensurable with $2\pi$ (i.e. $\al 
\neq 2\pi(m/n)$ for any integers $m,n$), then the set of images 
$\{\tau^i(x) \mid i \in {\bf Z} \}$ of any point $x$ is everywhere 
dense in $S$.

\indent Now suppose we define a constant of the motion for this 
discrete-time dynamical system on analogy with (iii) of Paragraph 
2.1.3.A. We need only require continuity, not differentiability: so we 
say a continuous function $f: S \rightarrow \mathR$ is a constant of 
the motion iff throughout each orbit $f$ is constant. (Here 
`throughout' emphasises that the definition is ``global in time''.)\\
\indent (i): If $\al = 2\pi(m/n)$ with integers $m,n$ in their lowest 
terms, there are many constants of the motion: any continuous 
real-valued function $f$ defined on an arc of $\frac{2\pi}{n}$ radians 
defines one. But:---\\
\indent (ii): If $\al$ is incommensurable with $2\pi$, the only  
constants of the motion are the trivial constant functions $f(S) = \{c 
\}, c \in \mathR$.

\indent It is worth expressing (ii) in terms of discontinuous 
functions. For that will show how even simple systems prompt a general 
notion of function. (This point gets greatly developed in the study of 
chaos---though as I said at the start  of this Subsection, I will not 
discuss chaos.) Suppose we partition $S$ under the equivalence 
relation: $x \equiv y$ iff $x$ is an image of $y$, or vice versa, 
under repeated application of $\tau$. Then any function $f: S 
\rightarrow \mathR$ that is constant on the cells of this partition 
(i.e. whose level sets $f^{-1}(c)$ are cells, or unions of cells) is 
either discontinuous at every point of $S$, or a trivial constant 
function $f(S) = \{c \}, c \in \mathR$.

{\em Second example: the torus}:---\\
The second example is a harmonic oscillator in two spatial dimensions, 
but with different frequencies in the two dimensions. So the  
equations of motion are
\be
{\ddot x}_i + \omega^2_i x_i = 0 \;\; , \;\; i = 1,2 \;\; ;
\label{2dofHO}
\ee
which have the energy in each dimension as two constants of the motion 
\be
E_i = \frac{1}{2}{\dot x}^2_i + \frac{1}{2}\omega^2_i x^2_i \;\; ;
\label{Efor2dofHO}
\ee
and solutions 
\be
x_i = A_i \cos(\omega_i t + \phi_i ) \;\; , \;\; A_i = \frac{\surd 
(2E_i)}{\omega_i} \;\; , \;\; i = 1,2 \;\; .
\label{soln2dofHO}
\ee
Each $E_i$ defines an ellipse in the $(x_i,{\dot x}_i)$ plane, so that
a pair of values $(E_1,E_2)$ defines a two-dimensional torus $T$ (i.e. 
a product of two ellipses).\\
\indent As we shall  discuss in more detail for Lagrangian 
mechanics:--- Here, the original system has four degrees of freedom, 
two for configuration and two for velocity, i.e. $x_1,x_2,{\dot 
x}_1,{\dot x}_2$. But the two constants of the motion have reduced the 
problem to only two variables. That is: given a pair $(E_1,E_2)$, the 
system's state---both positions and velocities---is defined by a pair 
of variables, say $x_1,x_2$. Besides, the variables are separated: the 
equations for them are independent of each other.\\
\indent Furthermore, we can introduce on the surface of the torus $T$, 
angular coordinates $\theta_1, \theta_2$ mod $2\pi$, each winding 
around one of the two ellipses, in terms of which the equations of 
motion become even simpler:
\be
{\dot \theta}_i = \omega_i  \;\; , \;\; i = 1,2 \;\; . 
\label{thetafor2dofHO}
\ee
So
\be
\theta_i = \omega_i t + \theta_i(0) \;\; , \;\;  i = 1,2 \;\; .
\label{solnthetafor2dofHO}
\ee
In terms of the $\theta_i$ (which we naturally call {\em longitude} 
and {\em latitude}), the motion winds around the torus uniformly. 

\indent The qualitative nature of the motion on $T$ depends on whether 
the frequencies $\omega_1, \omega_2$ are {\em commensurable} (aka: 
commensurate, or rationally dependent). That is: on whether the ratio 
$\frac{\omega_1}{\omega_2}$ is rational. More precisely: it follows 
readily from the results for the discrete-time system on the circle 
$S$ that:---

\indent (i): If $\omega_1, \omega_2$ are commensurable, then every 
phase curve of eq. \ref{thetafor2dofHO} on $T$ is closed: it 
eventually rejoins itself. We say the motion is {\em periodic}. And 
similarly to the discrete-time system: any differentiable real-valued 
function defined on a curve $\gamma$ in $T$ that intersects every 
phase curve once (so that the orbit of $\gamma$ is the whole of $T$) 
defines a constant of the motion which is independent of the two 
energies. 

\indent (ii): But if $\omega_1, \omega_2$ are incommensurable, then 
every phase curve of eq. \ref{thetafor2dofHO} on $T$ is everywhere 
dense on $T$: for any neighbourhood of any point on it, the curve 
eventually re-enters the neighbourhood. We say the motion is {\em 
quasiperiodic}.  Now recall that we defined constants of the motion 
(in (iii) of Paragraph 2.1.3.A) to be differentiable functions. So as 
in the discrete-time system on $S$: phase curves being everywhere 
dense implies that the only  constants of the motion independent of 
the energies are the trivial constant functions $f(T) = \{c \}, c \in 
\mathR$. That is: these are the only constants for all time.  Locally, 
i.e. in a sufficiently small neighbourhood $V$ of any point 
$(\theta_1,\theta_2)$, there {\em are} constants of the motion 
independent of the energies. They are given as in case (i) by any 
differentiable real-valued function defined on a curve $\gamma$ that 
intersects every phase curve in $V$ just once; and the time-scale on 
which they ``hold good'' as constants is set by how long it takes for 
some such phase curve to wind around the torus and re-enter $V$.\\
\indent And again, one could express these points  by saying that any  
function that is constant on every phase curve is either discontinuous 
at every point of $T$, or a trivial constant function $f(T) = \{c \}$. 
(For more details, cf.  Arnold (ibid.) who also discusses the 
analogues using higher-dimensional tori.)

\subsubsection{Schemes for solving problems---and their 
merits}\label{schemes}
As I announced at the end of Section \ref{Solv}, analytical mechanics' 
schemes for representing and solving problems operate in  the middle 
of the spectrum of meanings of `solve a problem': neither very 
tolerant (even of the useless), nor very intolerant (e.g. accepting 
only algebraic functions). To further explain this, I must first 
distinguish two main topics. The first has received philosophical 
attention and I will set it aside. The second has not---and is my 
topic. 

The first topic is that of approximation techniques. Mathematics and 
physics have of course developed an armoury of such techniques, 
precisely in order to overcome the predicament of how few problems are 
soluble in more stringent senses (e.g. analytically). That armoury is 
impressively large and powerful: and in recent years---not before 
time!---philosophicers have   studied it, often as part of studying 
scientific models.\footnote{The literature is of course vast, but 
Morgan and Morrison (1999) is a useful recent anthology. One reason it 
is vast is that `model' is used in so many senses: here, Emch's 
distinction (2002) between $L$-models ($L$ for `logic' or `language') 
and $H$-models ($H$ for `heuristic') is helpful. Emch and Liu (2002) 
is a gold-mine of information about approximations and models (and 
much else) in thermodynamics and statistical physics.}
 
But I shall not enter into details about this topic. For my point is 
that analytical mechanics' general schemes bring out another topic. 
Though these schemes do not (of course!) ``solve all problems'' in 
some stringent sense, they are {\em not} simply examples of the 
armoury of approximation techniques, for two (related) reasons.

 First,  there is a sense in which these schemes do not involve 
approximations---though they do involve idealizations. Agreed, there 
is no established philosophical usage distinguishing approximation and 
idealization.  But I distinguish between them as 
follows.\footnote{Some authors propose similar distinctions. For 
example, in McMullin's excellent discussion of Galileo (1985), `causal 
idealization' (264f.) is like my `approximation', and `construct 
idealization' (254f.) is like my `idealization'; and more briefly, 
Teller (1979: 348-349) makes almost exactly my distinction. But 
agreed: other authors vary, sometimes using other words, such as 
`abstraction', e.g. Suppe (1989: 82-83, 94-96) and Cartwright (1989: 
183-198); (thanks to Anjan Chakravartty for these two references). In 
any case, discussion suggests my distinction  is  an acceptable 
stipulation; though  I admit that for some colleagues, the terms have 
other connotations, e.g. the semantic  one that any idealization is 
strictly speaking false, while some approximations are true. Earman 
and Roberts (1999) is a fine discussion of the related topic of {\em 
ceteris paribus} clauses in laws of nature.} Both approximation and 
idealization  involve neglecting some quantities believed (or hoped!) 
to make little difference to the answer to a problem. But 
approximation  does not involve simplifying, or in any way revising, 
the mathematical form in which the problem is {\em posed}. Rather, one 
applies the approximation in the course of solving the problem as 
posed. On the other hand, idealization involves neglecting such 
quantities, precisely by simplifying or otherwise revising the 
mathematical formulation of the problem; (maybe the simplification 
occurs implicitly, when formulating a mathematically well-defined  
problem, once given a verbal physical description). So as I use the 
terms, idealization is neglecting what you believe negligible when you 
first pose a problem, while approximation is doing so while solving 
it. As we shall see, the schemes of analytical mechanics involve, in 
this sense, idealizations---a major example being  (Ideal), discussed 
in Section \ref{Intr}---but not approximations. 

Second and more important, these schemes give information---indeed, an 
amazing amount of information---not just about solutions to individual 
problems, but about the  structure of the {\em set} of solutions to 
all problems of a large class. Of course, the nature of this 
information, and of the large class, can only become clear when I 
expound the scheme in question, be it Lagrangian (in this paper) or 
Hamiltonian or Hamilton-Jacobi (in the companion paper). For they vary 
from one scheme to another. But for all three schemes, this 
information  is independent of approximations.  

For these reasons, I think the schemes give a sense of `solve a 
problem', that lies in the middle of our spectrum---and is  
distinctive, because not a matter of approximations. Or rather: they 
give a group of senses, since the information the scheme provides   
varies both with the problem and with the scheme:---\\
\indent (i):  Variation from one problem  to another is already clear. 
After all, despite the ``pessimism'' of Section \ref{Generalizing}'s 
review of ``rogue'' functions: Happily, {\em some} problems can be 
solved in one of the stringent senses, e.g. the position of a particle 
being given at all times by  an analytic formula. And as we shall see, 
the other problems are a mixed group: the information a scheme can 
provide varies. For example, some problems  can be reduced to 
quadratures (cf. Paragraph 2.1.2.B); others cannot be.\\
\indent (ii): Variation from one scheme to another. This variation 
will of course  only  be clear once the schemes are expounded. But 
broadly speaking, it is fair to say that for a given problem:\\
\indent \indent (a): The schemes agree about what kind of solution 
(stringent, or middle-of-the-spectrum) it has. After all, the schemes 
can hardly disagree about whether the position of a particle in a 
well-defined problem is given at all times by  analytic formula---{\em 
nobody} can disagree about it!\\
\indent \indent (b): But the schemes can and do differ in the 
information they provide about the given problem. And this information 
is not always ``just theoretical'': it can bear very directly on how 
to solve the problem (to the extent that it can be solved). The 
obvious example is the way that Jacobi's invention of transformation 
theory (for Hamiltonian mechanics) enabled him to solve problems that 
had thitherto been intractable. 

This variation across problems and schemes, points  (i) and (ii), 
reinforces the important  point that these schemes are by no means 
``algorithms'' for solving problems. Indeed, they are not such 
algorithms, even in some single middle-of-the-spectrum sense of `solve 
a problem', such as reduction to quadratures. If only such 
``middling'' solutions  were always possible---what a neat world it 
would be!\\
\indent In particular, be warned:---  In Section \ref{Anal} onwards, 
one  recurrent theme will be the schemes' allowance of arbitrary 
variables, so that we can adopt those variables best suited to the 
problem we face---i.e. yielding as stringent a solution to it as is 
possible. So that allowance will be a major advantage of the schemes. 
But it will {\em not} mean that the schemes tell us what are the best 
variables for our problem. If only! 

It would be a good project to define more precisely various 
middle-of-the-spectrum senses of `solution', and classify the various 
problems and schemes with respect to these senses. But I shall duck 
out of this  project, and restrict myself to laying out the kinds of 
information the schemes provide. In effect this sort of information   
would be the raw material to use for making  such definitions and 
ensuing classification. 

We will see in Sections \ref{Anal} and \ref{Lagr} that the Lagrangian  
scheme has the following four merits---as do the other two  schemes, 
Hamiltonian and Hamilton-Jacobi. It will be convenient to have 
mnemonic labels for these merits, just as it was for the morals.

\indent (Fewer): The use of fewer functions  to describe the motion of 
a complicated system that the number of degrees of freedom suggests. 
Indeed, in each of the three schemes  an arbitrarily complicated 
system is described by {\em just one} main function. (Their symbols 
are respectively $L, H$ and $S$: so this paper will be concerned with 
$L$, where `L' stands for `Lagrangian'.)

\indent (Wider): A treatment of a wider class of problems. One main 
way this happens (in all three schemes) is that the scheme allows a 
choice of variables, to suit the problem at hand. (If there is a 
symmetry, or another way to separate (de-couple) variables, the best 
choice will almost always exploit it; see (Reduce) and (Separate) 
below.) But all choices of a certain wide class are equally 
legitimate: this will give a sense in which the general scheme's 
equations are covariant.\footnote{The morals arising from allowance of 
arbitrary coordinates turn out to be very different from those arising 
from general covariance in spacetime theories. This is not surprising: 
after all, since the manifold we are concerned with is a state-space, 
not spacetime, at most one point of the manifold is ``occupied'' or 
``realized'' at any one time.}

\indent   (Reduce): The ability to reduce the number of variables in a 
problem. By this I do not mean the general idea of holding some 
variables negligible: whether as a matter of what I labelled 
approximation or of idealization. Nor do I mean the specific 
idealization labelled (Ideal) at the end of Section \ref{Prosp}: i.e. 
treating infinite or ``large-finite'' systems as ``small-finite''.\\
\indent Rather, I mean a specific kind of elimination of variables 
from the description of a problem, where the elimination is rigorously 
justified by the scheme in question. Two main sorts of example of such 
elimination will recur in this paper. Namely, elimination of:\\
\indent (i)  Variables that describe constraints, or the forces that 
maintain constraints. This will be prominent in Section \ref{Anal}.\\
\indent (ii) Variables that describe different values of a constant of 
the motion. Recall from Section \ref{ODESTori} that knowing the value 
$c \in \mathR$ of a constant of the motion $f$ means we can analyse 
the motion wholly within the level set $f^{-1}(\{c\})$. As we shall 
see, constants of the motion typically arise from a {\em symmetry} of 
the system. This will be prominent in Section \ref{ssecGeneMom} 
onwards.
 
A simple ``toy'' example of all of (Fewer), (Wider) and (Reduce) is 
provided by one-dimensional conservative systems: by which is meant 
any system described by a differential equation, for a single real 
variable as a function of time $x(t)$, of the form
\be
{\ddot x} = F(x) \; ; \;\; F {\rm \; a \; differentiable \; function 
\; defined \; on \; a \; real \; interval}.
\label{onedimcons}
\ee
In mechanical terms, these are systems with one configurational degree 
of freedom, such as a point-particle moving frictionlessly in one 
spatial dimension.\\
\indent It is easy to show that for any such system, the total energy 
$E$, the sum  of the potential and kinetic energies $T$ and $V$
\be
E \; := \; T + V \; := \frac{1}{2}\;{\dot x}^2 \; - \; \int^x_{x_0} \; 
F(\xi) \; d\xi
\label{EnyConsFirst}
\ee  
is a constant of the motion. And this implies that the motion of the 
system can be explicitly solved in the sense of being reduced to a 
single integration; (though we may  be able to do the integration only 
numerically: `quadrature'). For by solving eq. \ref{EnyConsFirst} for 
${\dot x}$, the problem of integrating the second-order eq. 
\ref{onedimcons} is  reduced to integrating
\be
{\dot x} = \pm  \surd\{ 2E - V(x)\} \;\; .
\label{enyeqn}
\ee
This has the solution (cf. Paragraph 2.1.2.B's first example of 
quadrature, eq. \ref{autonmsfoode})
\be
\pm \int  \frac{dx}{ \surd\{ 2E - V(x)\}} \; = \; \int dt \equiv t 
\;\; ;
\label{enyeqnintgted}
\ee
which we then invert  so as to give $x$ as a function of $t$.\\
\indent To spell out the illustration of the three merits:--- The 
conservation of energy reduces the problem from two dimensions 
(variables)---from needing to both know $x$ and ${\dot x}$---to one: 
(Reduce). A single function $V$ describes the system: (Fewer). And our 
method solves, in the sense of quadrature, an entire class of 
problems: (Wider).  

\indent (Separate): The ability to change to variables that are  
``de-coupled'', in that one has to solve, either:\\
\indent  (a) ideally, independent rather than coupled equations; or \\
\indent (b) much more commonly, equations that are coupled less 
strongly (at least not pairwise!) to one another than was the 
originally given set.\\
\indent This merit often occurs in association with (Reduce). Reducing 
a problem from $n$ variables to $n-1$ will typically leave us, after 
we solve the $(n-1)$-variable problem, having to find the $n$th 
variable, $q_n$ say, from a single equation $\frac{d q_n}{dt} = 
f(q_1,\dots,q_{n-1})$ where the right-hand side gives $\frac{d 
q_n}{dt}$ as a function of the other variables, which are now 
themselves given as functions of time. So $q_n$ can be found by 
quadrature.\\
\indent More specifically, it also occurs in the method of separation 
of variables. Paragraph 2.1.2.B's example (ii) gave the simplest 
example of this method; but it will come to the fore in 
Hamilton-Jacobi theory. 

I submit that this is an impressive list of merits, especially since 
all three schemes have all of them. In any case, it fills out my claim 
that analytical mechanics is devoted to providing general schemes for 
representing and solving mechanical problems.

In Section \ref{Meth}'s opening statement of  my main moral about 
method, (Scheme), I also claimed that the provision of such schemes is 
a significant topic in the analysis of scientific theories---not least 
because it falls between two topics often emphasised by philosophers, 
``laws of nature'' and ``models''.\\
\indent  Assessing my claim must largely rest with the reader. But I 
think there are two reasons why philosophers  should take note of 
(Scheme)---apart from the simple fact that one science, viz. 
analytical  mechanics, is mostly devoted to such schemes, and has 
succeeded most impressively. These reasons concern how the development 
of schemes for treating ``any'' problem, is ignored in the 
philosophical literature; (or so it seems to me). I think this happens 
in two ways. The first relates to observations; the second relates to 
theory, and will be taken up in the next moral. 

As to observations: the literature emphasizes the ``opposite'' idea. 
That is, it emphasizes what it takes to account for, or give a model 
of, given observations (or a given phenomenon): and in particular, the 
nature of the approximations involved. Indeed, one influential vein of 
literature emphasizes the limitations of theory. Namely, it stresses 
that one needs to choose a model that one can solve in some strong 
sense (as people say: exactly, or nearly exactly), and then exploit it 
as much as possible to deal with other problems. (Here `model' and 
`solve' have various senses, for various examples and various authors:  
for example, contrast Kuhn's emphasis on exemplars in the Postscript 
(1970) of his (1962), with Cartwright (1999).) I agree this  happens 
often, maybe ``all the time'', in science.\footnote{But I think this 
literature also misses a point about the frequent exploitation of a 
single model: namely that there are sometimes good theoretical reasons 
for the selection of the preferred model, and for the success of its 
application to other problems. The obvious example in physics concerns 
analysing problems in terms of harmonic oscillators. Not only is $V 
\propto x^2$ the simplest polynomial way to specify a spatially 
varying force; also, Taylor's theorem implies that locally it provides 
the dominant contribution to a generic smoothly spatially varying 
force. Of course, such considerations are the springboard for a wealth 
of theory: the obvious example is the analysis of small oscillations 
about equilibrium; and catastrophe theory (Butterfield 2004b) provides 
a very advanced example.} But---my moral again---we should also note 
the ``opposite'', i.e. schemes for treating any problem in a large 
class.

My second reason for noting (Scheme), relating to theory rather than 
observations, is covered in my second, minor, moral about method.

\subsubsection{Reformulating and Restricting a Theory: (Reformulate) 
and (Restrict)}\label{Refo}
 Most of the philosophical  literature conceives the development of a 
theory as a matter of increasing the theory's logical strength 
(information content). The two main ways in which this can happen are 
taken to be: deepening the theory's account of the phenomena in its 
domain (especially by invoking a more detailed causal and/or 
microstructural account); and extending the theory to new phenomena, 
outside its domain.
Again, I agree this happens often, and maybe ``all the time''. But one 
should also notice the ``opposite'' idea: 
theory development without an increase in logical strength. There are 
two main cases to consider. The new formulation might be equivalent 
(theoretically, not just empirically) to the old; I call it 
(Reformulate). Or it might be logically weaker; I call it (Restrict).

Agreed: the first case, (Reformulate), is old news. That is, it is not 
controversial that providing equivalent formulations of a theory can 
be very significant, both methodologically and ontologically. For one 
of a pair of theoretically equivalent formulations might extend better 
than the other to another domain of phenomena; or deal better, in some 
sense, with the given domain. This can even be so when theoretical 
equivalence is construed strongly enough that theoretical equivalence 
implies, or near enough implies, that the two formulations have the 
same ontology.\footnote{Here are two examples from non-relativistic 
quantum mechanics of a fixed number of particles: (i) the 
Schr\"{o}dinger and Heisenberg pictures; (ii) the wave-mechanical and 
path-integral formulations of the position representation. The members 
of each pair are provably equivalent; and though the ontology of 
quantum mechanics is a murky business, I think no one sees a relevant 
difference between the members of a pair. But when we turn to 
relativistic quantum theories, in particular quantum field theory, 
each pair's symmetry is broken: in various ways, the Heisenberg 
picture and path-integral formulation ``win''.} 

But analytical mechanics provides several examples of equivalent 
formulations, for which this old news is worth reading; for three 
reasons. The first reason relates back to (Scheme): one of the 
equivalent formulations might have the specific methodological 
advantage of providing such a scheme. Second, as to ontology: such 
equivalent formulations help rebut the false idea that classical 
mechanics gives us a single matter-in-motion picture. Third, these 
equivalences are subtler than is suggested by textbook impressions, 
and folklore slogans like `Lagrangian and Newtonian mechanics are 
equivalent'.  

The second case, (Restrict), at first seems paradoxical: how can one 
develop a theory by restricting it, i.e. by decreasing its logical 
strength? My answer to this again relates to (Scheme), but I can state 
it in general terms. For it reflects the usual trade-off in scientific 
enquiry between the aims of covering (i.e. describing and explaining) 
a wide domain of phenomena, and covering phenomena in detail---between 
width and depth, as people say. So imagine a case where a theory 
admits, for just a subset of its domain of phenomena, a formulation 
which offers, for just that subset, advantages of ``depth''. In such a 
case, it could be best to restrict one's attention to the subset, and 
pursue just the special formulation.

Analytical  mechanics provides several major examples of this; we will 
see some as early as Sections 3.1 and 3.2. Besides, in analytical 
mechanics  the  advantage of ``depth'' offered by the special 
formulation is {\em not} a more detailed causal and/or microstructural 
account, or even the possibility of adding such an account. It is 
rather the provision of a general scheme for solving problems; (or 
adding further merits to a given scheme). So in these examples, 
(Restrict) is not simply a preliminary to adding logical strength.  
Furthermore, in these examples, the special formulation and the 
original general theory are often logically equivalent, as regards 
what they say about the given subset of phenomena. So in this way, 
(Restrict) and (Reformulate) will  often be exemplified together.

\subsection{Ontology}
\label{Onto}
In Section \ref{Intr}, I stressed that the ontology of classical 
mechanics  is a subtler affair than the matter-in-motion picture 
suggests. This general point will be borne out in two morals, which 
concern respectively: modality and objects.

\subsubsection{Grades of modal involvement: (Modality)}\label{Moda}
As I said at the start of this Section, the starting-point of each of 
the schemes---Lagrangian, Hamiltonian and Hamilton-Jacobi---is to 
postulate the {\em state-space}: the set of all possible states of the 
system it is concerned with; (though the structure of this set varies 
between the different schemes).

At first sight, the philosophical import of this would seem to be  at 
most some uncontroversial version of the idea that laws support 
counterfactuals. That is: whether or not one believes in a firm 
distinction between laws of nature and accidental generalizations, and 
whatever one's preferred account of counterfactuals, a theory (or 
``model'') that states `All As are Bs' surely in some sense warrants 
counterfactuals like `If any object were an A, it would be a B'. And 
so when analytic mechanics postulates state-space and then specifies 
e.g. laws of motion on it, it seems at first that this just 
corresponds to the passage from `All actual systems of this kind 
(having such-and-such initial states---usually a ``small'' proper 
subset of state-space) evolve thus-and-so' to `If any system of this 
kind were in any of its possible initial states, it would evolve 
thus-and-so'.\footnote{Here, and in all that follows, I of course set 
aside the (apparent!) fact that the actual world is quantum, not 
classical; so that I can talk about e.g. an actual system obeying 
Hamilton's Principle. Since my business throughout is the philosophy 
of classical mechanics, it is unnecessary to encumber my argument, 
from time to time, with antecedents like `If the world were not 
quantum': I leave you to take them in your stride. Cf. also footnote 
10 in Section \ref{Solv}.}\label{stride}   

But this first impression is deceptive. The structures with which 
state-space is equipped by analytical mechanics, and the constructions 
in which it is involved, make for a much more varied and nuanced 
involvement with modality than is suggested by just the idea that laws 
support counterfactuals. This is my third moral, which I call 
(Modality).

I propose to delineate, in Quinean fashion,  three grades of modal 
involvement; so I shall write (Modality;1st) etc. Like Quine's three 
grades, the first is intuitively the mildest grade, and the third the 
strongest. But this order will not correspond to any ordering of the 
three schemes, Lagrangian, Hamiltonian and Hamilton-Jacobi. In 
particular, this paper's scheme, the Lagrangian one, was historically 
the first, and is in various respects the most elementary, of the 
three: but it exhibits the {\em third} grade of modal involvement.

The grades are defined in terms of which kind of actual matters of 
fact they allow to vary counterfactually. One kind is the given 
initial conditions, and/or final and/or boundary  conditions; roughly 
speaking, this is the given initial state of the system. Another kind 
is the given physical {\em problem}: which I here take as specified by 
a number of degrees of freedom, and a Lagrangian or Hamiltonian. (In 
elementary terms, this means: specified by the number of particles 
involved, and the forces between them.) A third kind is the laws of 
motion, e.g. as specified by Newton's or Hamilton's equations. Thus I 
propose the following grades.\footnote{I don't claim that these three 
grades are the best way to classify the modal involvements of 
analytical mechanics. But they have the merit of being obvious, and of 
showing clearly the variety of modal involvement that occurs. I also 
think:\\
\indent (i): the grades could be sub-divided in various ways, for 
example using Section \ref{ConstrtsGenCoords}'s classification of 
kinds of constraints (or finer classifications made in the 
literature);\\
\indent (ii): similar grades can be discerned in other physical 
theories.\\
But I shall not develop (i) or (ii) here.}

(Modality;1st): The first i.e. mildest grade keeps fixed the given 
actual physical problem and laws of motion. But it considers different 
initial conditions, and/or final and/or boundary conditions, than the 
actual given ones; roughly speaking, different initial states of the 
system. And so it also considers  counterfactual histories of the 
system. (Under determinism, a different initial state implies a 
different history, i.e. trajectory in state-space.) 

So this grade includes the idea above, that laws support 
counterfactuals. But it will also include subtler modal involvements. 
Perhaps the most striking case occurs in Hamilton-Jacobi theory. For 
details, cf. Butterfield (2004d: Sec. 4, 2004e: Sec. 4) or the 
companion paper. But in short: one solves a problem, as it might be an 
actual one, by introducing an ensemble of systems, i.e. a set of 
possible systems, of which the actual system is just one member. 
Furthermore, the ensemble can be chosen in such a way that the problem 
is solved {\em without} performing integrations, i.e. just by 
differentiation and elimination: a remarkable---one might well say 
`amazing'---technique. 

(Modality;2nd): The second grade keeps fixed the laws of motion, but 
considers different problems than the actual one (and thereby in 
general, different initial states). Such cases include considering a 
counterfactual number of degrees of freedom, or a counterfactual 
potential function. Maybe no actual system is, nor even is well 
modelled as, a Lagrangian system with 5,217 coordinates; and maybe no 
actual system has a potential given (in certain units) by the 
polynomial $13x^7 + 5x^3 + 42$.
But analytical mechanics (in all three schemes) considers such 
counterfactual cases. And one can have good reason to do so, the 
obvious reason being that the counterfactual case provides an 
idealization or approximation needed to get understanding of an actual 
system.

However, this second grade also includes more ambitious cases of 
considering counterfactual problems: namely, cases where one makes a 
generalization about a whole class of problems. An elementary example 
is the conservation of energy theorem, which we will see in Lagrangian 
mechanics.\footnote{Again: an advanced, indeed spectacular, example is 
the classification of catastrophes by catastrophe theory.} 

(Modality;3rd): The third grade allows different laws, even for a 
given problem. Again, this can happen  even in Lagrangian mechanics. 
Here one does not explicitly formulate non-actual laws (much less 
calculate with them). Instead, one states the actual law as a 
condition that compares the actual history of the system  with 
counterfactual histories of it that do not obey the law (in 
philosophers' jargon: are {\em contralegal}). That is, the 
counterfactual histories share the initial (and final) conditions, but 
do not obey the given deterministic laws of motion, with the given 
forces. (Agreed, for any sufficiently smoothly varying counterfactual 
history, there {\em could} be forces which in conjunction with the 
actual laws and initial and final conditions, would yield the 
counterfactual history. But this does not matter, in the sense that it 
is not appealed to in the formulation of the actual law.)

This is at first sight surprising, even mysterious. How can it be 
possible to state the actual law by a comparison of the actual history 
with possible histories that do not obey it? Besides, metaphysicians 
will recognize that this seems to contradict a Humean  view of laws, 
and in particular  Lewis'  doctrine of Humean supervenience (Lewis 
1986, pp.ix-x). I  address this issue at length elsewhere (2004e: 
Section 5). For this paper, it suffices to note (and thereby reassure 
Humeans) that this third grade of modal involvement can be reconciled 
with Humeanism. 

To sum up this moral: the detail of analytical  mechanics  reveals it 
to have a varied and nuanced involvement with modality.

\subsubsection{Accepting Variety: (Accept)}\label{Acce}
My second moral is about ontology in a more obvious way than was 
(Modality): it is about what are the basic objects of analytical 
mechanics. In general terms, it is that this is a subtler and more 
varied affair than suggested by the matter-in-motion 
picture.\footnote{This general idea is already suggested by the moral 
(Reformulate) of Section \ref{Refo}. For given that the world is in 
fact quantum, we can only interpret a phrase like `the basic objects 
of analytical mechanics' as about the objects postulated by analytical 
mechanics. And we already know from (Reformulate) that different 
approaches or theories within analytical mechanics might have 
heuristic, and even ontological, differences (and might do so, even if 
they are in some sense theoretically equivalent). So unless some 
single approach or theory is favoured, we already expect that 
analytical mechanics might be pluralist about ontology.} More 
specifically, the ontology need not be ``just point-particles''. I 
shall develop this moral in three more specific points.

\paragraph{2.2.2.A Infinite as finite, and conversely}\label{222A}
First, analytical mechanics is much more flexible about its basic 
ontology than one might think---especially if one thinks of classical 
mechanics as requiring {\em au fond} point-particles. In particular, 
each of my three schemes can treat both finite and infinite systems. 
(By this I mean, respectively, systems with a finite, or infinite, 
number of degrees of freedom.) Furthermore, each scheme can, starting 
with a system given to it as finite/infinite, treat it as 
infinite/finite---and under appropriate circumstances, can justify 
doing so.

This point returns us to (Ideal), discussed at the end of Section 
\ref{Intr}. I said there that analytical mechanics typically describes 
bulk matter (bodies) in terms of a {\em small} finite number of 
variables (degrees of freedom). But this idealization  can be 
justified, both empirically, and theoretically---by proving theorems 
to the effect that collective variables of many-dimensional systems 
(i.e. systems  with many, even infinitely many, degrees of freedom) 
would behave as described by a low-dimensional analysis.  So here we 
see analytical mechanics treating an infinite or ``large-finite'' 
system as ``small-finite''---and justifying its doing so.  

But the ``converse'' can also happen. Analytical mechanics has (on all 
three approaches) a rigorous formalism for describing infinite systems 
(aka: continuous systems); though as mentioned in Section \ref{Intr}, 
I'll say nothing about these formalisms. But the use of such a 
formalism does not commit one to the described system being really 
infinite. When for example, analytical mechanics describes a string or 
a gas as a continuous system (so as to describe e.g. sound waves in 
it), it is not committed to the string or gas really having infinitely 
many degrees of freedom. For a finite system can have so many degrees 
of freedom as to justify a model that treats it as continuous. For 
example, one can treat the density or pressure in a gas as a 
continuous function of spatial position, but take this function to 
represent, in an idealized way, an average over a macroscopically 
small volume, of very many microscopic degrees of freedom. And again, 
the justification for this kind of treatment can be both empirical and 
theoretical. (Continuous models of discrete phenomena are of course 
not special to analytical mechanics: they are endemic in physics.)  

To sum up: when analytical mechanics successfully describes a system 
as finite/infinite, the system could really (i.e. in a classical 
world!) be the ``opposite'', i.e infinite/finite: a salutary lesson in 
flexibility, and an antidote to the matter-in-motion picture (whether 
it takes matter as continuous bodies or as point-particles).

\paragraph{2.2.2.B Beware micro-reductionism}\label{222B} Second, 
analytical mechanics is less micro-reductionist than one might 
think---especially if one focusses on the matter-in-motion picture. In 
particular, if a given approach to analytical mechanics is applied to 
a problem or range of problems, and some objects (or  quantities i.e. 
variables) are defined from the fundamental objects etc. postulated by 
the approach, then we should accept the derived objects etc. as no 
less real than the fundamental ones.

Here, I say `accept as no less real than the fundamental objects', 
rather than `accept as real, like the fundamental objects', in order 
not to presuppose, nor favour, some version of scientific 
realism---either in general or about analytical mechanics. Like most 
others including anti-realists like van Fraassen,  I  take 
interpreting a physical theory to be a matter of describing what the 
world would be like if it were true: an endeavour that makes sense 
even if as an anti-realist, one is agnostic, or even atheistic, about 
theories' assertions about unobservables.

 I admit that this point raises issues about reductionism etc. which I 
cannot address here.\footnote{Section 2 of Butterfield and Isham 
(1999) is a brief statement of my general views. It stresses how 
widespread explicit definability is within physics. Wilson  stresses 
such definability in analytical mechanics, and puts it to work in a  
critique of currently popular doctrines in the metaphysics of mind and 
of properties; (1985: 230-238; 1993: 75-80).} Suffice it to make two 
remarks in defence of the point.\\
\indent (a): Since it concerns {\em defined} objects and quantities, 
it is surely plausible: why deny reality to what is defined in terms 
of the real?\\
\indent (b): This point is of course related to Paragraph 2.2.2.A  
above. For many if not most examples of defined quantities are 
collective variables, typically of infinite or ``large-finite'' 
systems. And as in Paragraph 2.2.2.A, there are two main cases: the 
defined variable might be used in a ``small-finite'' description of 
the system; or it might be used in a continuous description (for 
example, taking a continuous function of spatial position to represent 
a macroscopically local average of very many microscopic degrees of 
freedom). So for collective variables as examples of defined 
quantities, my second point is in effect an application of my first to 
the topic of reductionism. 

Here is an example combining these points. Consider a circular wave 
spreading on a still pond. An analytical mechanical treatment will no 
doubt identify a few variables, such as the radius and height of the 
wave as relevant; and so describe the crest as propagating radially, 
though no water molecules do (except briefly). But apart from these 
few variables, analytical mechanics can treat the problem in three 
main ways. Either it uses just these few variables; or it takes the 
water to be composed of a vast number of particles; or it takes the 
water to be a continuum:(Paragraph 2.2.2.A). According to the last two 
ways, the few selected variables are collective variables of a very 
complex system; but they are no less real than they are on the first 
treatment: (this Paragraph).

To put these points in a slogan: we should accept that analytical 
mechanics suggests a varied ontology. Hence my label, (Accept). This 
slogan will be sharpened in the next Paragraph. Though it mostly 
concerns infinite systems i.e. continua, while my subsequent Sections 
are confined to finite systems, it is worth expounding it here---not 
least because it exposes a common ``micro-reductionist'' error.

\paragraph{2.2.2.C Beware the particles-in-motion picture}\label{222C} 
One might reply to Paragraphs 2.2.2.A and 2.2.2.B that nevertheless 
point-particles are the ``basic micro-ontology'' of analytical 
mechanics. For when analytical mechanics conceives a body or a fluid 
as a continuum, it thereby postulates point-sized ``bits of stuff'', 
``cheek by jowl'' with one another. Agreed, these are not 
point-particles in the sense of mass-points separated from each other 
by a void; (the sense usually associated with Boscovitch---though in 
fact the notion was  introduced two decades earlier by Euler). But 
they are close cousins, and deserve the name `point-particle'. For 
each has a definite position and so also its time-derivatives 
(velocity, acceleration etc.), mass-density---and other properties, 
such as pressure, depending on the details of the continuum being 
treated. 

I reply: fair comment. I am willing to agree that point-particles, in 
this weak sense including point-sized bits of matter in a continuum as 
well as Boscovitchean point-particles,  are the ``basic 
micro-ontology'' of analytical mechanics. Accordingly, I will from 
Section 3 onwards adopt the common habit of saying `particle' to mean, 
according to context, either:\\
\indent (i) a point-particle (Boscovitchean or not); or following 
(Ideal),\\
\indent (ii): a small solid (maybe rigid) body, which is assigned just 
one position vector ${\bf r}$, i.e. treated as having no internal 
structure or orientation.

But beware! Point-particles being the basic objects does {\em not} 
mean that analytical mechanics  should be, or even can be, understood 
in a particle-by-particle way. And in fact, it cannot be. Here, we 
meet what  Section \ref{AgstMiM} called the second error in the idea 
that classical mechanics is unproblematic: what I call the 
`particles-in-motion picture'. This picture claims that analytical 
mechanics can and should be understood in a particle-by-particle way: 
i.e. that analytical mechanics not only takes matter to be composed of 
point-particles (in the above weak sense), but also analyses all the 
physics of matter's behaviour in terms of particle-to-particle 
relations.

It is this second claim that is false. Agreed, some parts of 
analytical mechanics conform to it. The main example is of course  the 
analytical mechanics of point-particles in the stricter i.e. 
Boscovitchean sense, with action-at-a-distance forces. Here the main 
illustration of the claim is that the total force on each particle is 
the sum of the forces exerted on it by each other particle. That is, 
all the fundamental forces are from one particle to another; in 
physics jargon, all interactions are two-body, not many-body. The 
standard example is of course  Newtonian gravity.\footnote{The way I 
have stated this illustration assumes that the component forces are 
just as real as the total force on a particle. Though I fully accept 
this assumption,  it has been denied. No matter: the illustration 
could be restated in more cumbersome language so as to be independent 
of this controversy.} 
   
Agreed also, some parts of the analytical mechanics of continua 
conform to it. For example, the forms of the terms in the Lagrangian 
and Hamiltonian densities for continua are in many cases deduced by 
conceiving the continua as an infinite limit of a large finite 
assembly of point-particles. For example, one assumes that each 
point-particle in the assembly interacts only with its nearest 
neighbours, and does so by a quadratic potential (which might be 
modelled by a spring).\footnote{This kind of deduction (which goes 
back to at least Green (1855) and Thomson (1863)) raises an 
interesting methodological and historical point. Namely, if the 
deduced continuum model is empirically successful, it is tempting to 
think the assumptions are not ``merely heuristic'', a ladder to be 
thrown away {\em \`{a} la} Wittgenstein, once the continuum treatment 
is written down. That would represent ``too great a coincidence'': we 
should expect a body or fluid, which is well modelled as a continuum 
with such Lagrangian and Hamiltonian densities, to  consist in fact of 
a vast number of point-particles of some sort, that (to a good 
approximation) interact as assumed, e.g. with a nearest-neighbour 
quadratic potential. In other words (with less connotation of 
scientific realism---and of our acquaintance with quantum theory!): we 
should {\em interpret} analytical mechanics, when it successfully 
describes a body or fluid as a continuum with such densities, as 
idealizing in the way discussed in Paragraph 2.2.2.A---as taking the 
body or fluid to be a swarm of such interacting point-particles. 

Indeed, this line of thought was pursued and contested in the 
nineteenth century debates about the foundations  of classical 
mechanics; but I shall not go into this. In any case, it is only in 
many, not in all, cases that the Lagrangian or Hamiltonian densities 
are deduced by taking such an infinite limit.}

Nevertheless, the claim is false. In fact, the analytical mechanics  
of continua has to be formulated in terms of {\em spatially extended} 
regions and their properties and relations. Two examples will suffice: 
one is kinematical, the other dynamical.

\indent (i): I mentioned above that a point-sized bit of stuff in a 
continuum (a point-particle, in the weaker sense) has a mass-density. 
At first sight, that seems to support the particles-in-motion picture: 
that the mass of a finite volume is the integral of the mass-density 
seems to be a case, albeit a very  simple one, of 
``micro-reductionism'' or the ``supervenience of the global on the 
local''. But in fact a rigourous formulation proceeds in the opposite 
direction. It takes as primitive the attribution of masses to finite 
volumes, and defines mass-density as a limit of ratios of mass to 
volume, as the volume tends to zero; (the details are taken over from 
measure theory). Besides this opposite procedure is necessary, in 
order to avoid various conundrums; (unsurprisingly, these conundrums  
are  not really specific to mass---they have analogues in general 
measure theory.)

\indent (ii): One cannot understand the forces operating in continua 
(whether solids or fluids) as particle-to-particle. In particular, 
returning to Section \ref{AgstMiM}'s example of two continuous bodies 
that touch, at a point or over a finite region: it is wrong to think 
each point-sized bit of matter exerts a force on some (maybe 
``nearby'') particles, or on all other particles. Rather, one needs to 
consider, for each arbitrary finite (i.e. not infinitesimal) portion 
of matter: (a) a force exerted on its entirety by matter outside it, 
and (b) a force exerted at each point of its  surface by matter 
outside it. (Agreed, this strategy, of describing the forces on all 
the countless overlapping extended sub-regions of a continuum, is 
highly ``redundant'', in the sense that each sub-region is described 
countless times, viz. as a part of the description of a larger region 
in which it is included. Nevertheless, analytical mechanics 
adopts---and needs to adopt---this strategy.)

To sum up: (i) and (ii) both show that in the mechanics of continua, 
one cannot analyse the physics wholly in terms of particle-to-particle 
relations: the particles-in-motion picture is false. Rather, 
one needs to take spatially extended regions as 
primitives.\footnote{My 2004a gives a more detailed critique of the 
particles-in-motion picture; it also connects the picture to 
philosophical concerns about intrinsic properties, and Humean 
supervenience. My 2004 applies this critique to the philosophical 
debate about the nature of objects' persistence through time.

 For more technical details about (i) and (ii), cf. e.g.: Truesdell 
(1991, Sections II.2, III.1, III.5); and for (ii), Marsden and Hughes 
(1983), Section A.2 and Chapter 2. Topic (ii) has a rich history. It 
was Euler who in 1775 first realized continuum mechanics' need for (a) 
and (b); but the road to general acceptance was long and complex. For 
a glimpse of this history (emphasising the relation to the principle 
of rigidification), cf. Casey (1991: especially 333, 362-369).}

\newpage

\section{Analytical mechanics introduced}
\label{Anal}

\paragraph{3.0 Difficulties of the vectorial approach to 
mechanics}\label{diffiesveclapproach}
In expounding the analytical mechanics of finite-dimensional systems, 
one can of course take various routes. In this Section, my route will 
be based on those of Goldstein et al. (2002) and Lanczos (1986). 

First of all, I will motivate analytical mechanics  by considering two 
difficulties faced by  the more elementary approach to mechanics 
familiar from high school (mentioned in Section \ref{AgstMiM}).  
Roughly speaking, this approach takes the motion of each body to be 
determined by the vector sum of all forces on it. More precisely, a 
body is taken to be either:\\
\indent (i) a Boscovitchean point-particle; or \\
\indent (ii) (following (Ideal), the idealization discussed at the end 
of Section \ref{Intr}): a body which is small and rigid enough to be 
assigned just one position vector ${\bf r}$, i.e. to be treated as 
having no internal structure or orientation; or\\
\indent (iii) composed of particles in sense (i) or (ii).\\
I shall from now on use `particle' in this way, i.e. to mean (i) or 
(ii). 
Then according to this approach, the motion of each particle is to be 
determined by the vector sum of all forces on it. So to determine the 
motion of a system of particles labelled by $i$ is in principle to 
solve
\be
m_i{\ddot {\bf r}}_i = {\bf F}_i = \Sigma_j {\bf F}_{ij} + {\bf 
F}^{(e)}_i \label{eqn;newt}
\ee
where ${\bf F}_{ij}$ is the force on particle $i$ due to particle $j$, 
and ${\bf F}^{(e)}_i$ is the external force on particle $i$. (Here I 
will wholly set aside the topic, familiar in philosophy of physics, of 
the need to identify inertial frames with respect to which the 
quantities in eq. \ref{eqn;newt} are to be measured.) This is often 
called the {\em vectorial approach} to mechanics.\footnote{It is also 
called `Newtonian'; eq. \ref{eqn;newt}, or `${\bf F}=m{\bf a}$', being 
the most familiar form of Newton's second law. But this name is 
anachronistic. In particular: (i) it was Euler in 1749 who first wrote 
down ${\bf F}=m\frac{d^2{\bf r}}{dt^2}$, though in cartesian 
coordinates rather than vector notation (which was developed only in 
the nineteenth century e.g. by Heaviside and Gibbs (Crowe 1985)); (ii) 
it was Boscovitch (1758) who advocated an ontology of point-particles 
moving in a void.}

In  general, solving eq. \ref{eqn;newt} is so complex as to be utterly 
intractable, since there are countless particles in any macroscopic 
body. But the vectorial approach has two tactics, one empirical and 
one theoretical, for simplifying the problem. They are both aspects of 
(Ideal).

(i):  It is often empirically adequate to treat the body in question 
as a particle (e.g. a bob on a pendulum), or as a small set of 
particles; perhaps  with other simplifications, such as all the forces  
acting at the body's centre of mass.

(ii): Some problems allow an assumption about, and-or an analysis of, 
the many inter-particle forces, that greatly simplifies the problem. 
The simplification may even allow the problem to be solved in some 
``medium sense'' such as quadrature. The standard example is the rigid 
body: conceived as composed of particles, it is defined as having 
constant inter-particle distances. The configuration of the body in 
space can then be specified by just six numbers.\footnote{Here is an 
argument  for the number six (which works even if the body is 
continuous). (1):  Rigidity implies that the positions of all the 
particles are fixed once we fix the position of just three of them. 
For imagine: if the tips of three of your fingers were placed at  
three given positions within a rigid brick, and someone specified the 
exact positions of the finger-tips, then they would have implicitly 
specified the positions of all the brick's constituent parts. (2): the 
position of three particles given as forming a certain triangle can be 
specified by six numbers.} And other assumptions may further simplify 
the problem. For example, the assumption that internal forces lie 
along the lines between particles implies that internal forces do no 
work; which further implies that if these forces are derived from a 
potential, then the internal potential energy is constant.  

But these tactics are of limited value when one considers {\em 
constrained} systems, i.e. systems that are required to be placed, or 
to move, in certain limited ways.  For constraints lead to two 
difficulties, which tactics (i) and (ii) cannot in general  
overcome---and which analytical mechanics, in particular Lagrangian 
mechanics, {\em does} overcome. Namely:---

\indent (Dependent): Constraints imply that the ${\bf r}_i$ are not 
{\em independent}: i.e. one cannot in imagination vary each of them 
while leaving all  the others fixed. (This implication holds good, 
whether ${\bf r}_i$ represents the position of a body, using (Ideal), 
or the position of a point-particle.) Note that independence and 
dependence of the ${\bf r}_i$ is a modal notion: I shall return to 
this in Section \ref{ssecGene}.  

\indent (Unknown): In general, the forces that maintain the 
constraints ({\em forces of constraint}; aka: forces of reaction) are 
not known. Thus we may suppose that we know what are often called the 
{\em applied forces} (aka: impressed, or given, or active, forces) on 
each particle: forces like gravity, spring forces etc. that apply to 
the parts of the system whether constrained or not. (So `applied' does 
not mean `external': the source of an applied force can be internal to 
the system.) Even so, we will in general not know the constraint 
forces on the system, e.g. from the surface on which it rests and 
across which it is constrained to move. 
And so the use of eq. \ref{eqn;newt} is forestalled.\footnote{This  
point is not affected by the vagueness of my distinction between 
applied and constraint forces. For if one wishes, one can {\em define} 
the applied forces as the known ones (e.g. Desloge 1982: 528), so that 
it is merely usual that unknown forces maintain constraints.}

Lagrangian  mechanics overcomes these difficulties; (as do Hamiltonian 
and Hamilton-Jacobi mechanics). In short: for (Dependent), one 
eliminates variables so as to work with a smaller set of variables 
which {\em are} independent. As for (Unknown): under remarkably 
general conditions, one can solve problems without knowing the 
constraint forces! Furthermore, after  solving the problem, one can 
then calculate the constraint forces: in effect, they have the values 
they need to have so as to maintain the constraints.\\
\indent So for both difficulties, Lagrangian  mechanics will 
illustrate the strategy of reducing the number of variables that 
describe a problem---i.e. the merit (Reduce).
 
Section \ref{ssecGene} will describe how to overcome (Dependent). 
Section \ref{ssecPrin} will begin on the topic of how to overcome 
(Unknown): it describes  the principle of virtual work and Lagrange's 
method of multipliers. The last Subsection, Section \ref{ssecDale}, 
introduces d'Alembert's principle, and from that deduces Lagrange's 
equations. These lie at the centre of Lagrangian mechanics, which is 
further developed in Section \ref{Lagr}.\\
\indent  Here I should stress a qualification about this order of 
exposition: a qualification which will become clearer in Section 
\ref{ssecDale}. To get to Lagrange's equations, it is {\em not} 
necessary to proceed as I do, {\em via} the principle of virtual work 
and d'Alembert's principle. One could go ``straight'' from eq. 
\ref{eqn;newt} to Lagrange's equations; and some fine expositions do. 
But as John Bell said, about how to teach special relativity: `the 
longer road sometimes gives more familiarity with the country' (1987: 
77). 

We will see all four of my morals illustrated in this Section. I 
already mentioned the merit (Reduce). By and large, the minor morals, 
(Reformulate), (Restrict) and (Accept) will be a bit more prominent 
than the main ones, (Scheme) and (Modality). But these two will come 
to dominate in Section \ref{Lagr} onwards.

\subsection{Configuration space}\label{ssecGene}
The key to overcoming the first difficulty, (Dependent), lies in what 
can claim to be Lagrangian  mechanics' leading idea: configuration 
space. It will be clearest to first describe how this idea addresses 
(Dependent) (Section \ref{ConstrtsGenCoords}), and then turn to 
describing dynamics in terms of configuration space (Section 
\ref{KEandwork}).

\subsubsection{Constraints and generalized 
coordinates}\label{ConstrtsGenCoords}
The idea is to represent the configuration of the entire system by a 
point in a higher-dimensional space. This is the basic version of the 
idea of state-space, which, as discussed in Section \ref{Mora}, 
underlies all my morals and all three schemes of analytical mechanics. 
Indeed, one might expect this, simply in view of determinism---which 
implies that for a given system and given forces on it,``all 
problems'' are fixed by all initial conditions, and so by the 
state-space: (Modality;1st). But we can already be more specific.

Namely, we can state two important  advantages of the idea of state 
space.\\
\indent (i): Such a space can be described using many different 
coordinate systems (in ways analysed in detail in differential 
geometry). This yields a striking generalization of the idea of 
changing variables the better to solve a problem. Stated thus, the 
idea is endemic to science and hardly remarkable. But allowing 
arbitrary coordinates on a state-space suggests that one develop a 
scheme that encompasses all choices of variables, with a view to 
obtaining the most tractable representation (if one is lucky: a 
solution, at least in Section \ref{Solv}'s ``medium sense'') of any 
problem: the merit (Wider).\\
\indent (ii) Using a state-space allows us to relate dynamics to the 
geometry of higher-dimensional spaces, in particular their curvature. 
(These two advantages were first emphasized by Jacobi.)

Returning to the difficulty (Dependent), the main ingredient for 
overcoming it is the idea of allowing arbitrary coordinates on 
configuration space, i.e. (i) above. To spell this out, I need to 
first state two independent distinctions between constraints. I will 
also need these distinctions throughout what follows; in fact most of 
my discussion will concern the first half of each distinction. 

 {\em Holonomic} constraints are expressible in the form of say $k$ 
equations 
\be
f_j({\bf r}_1, {\bf r}_2, \dots, t) = 0; \mbox{ where } j = 1,2,\dots 
k \label{eq;holo}
\ee
governing the coordinates ${\bf r}_i$. The obvious example is any 
rigid body: the constraints are expressed by the equations stating the 
fixed inter-particle distances. (The term ``holonomic'' is due to 
Hertz.) On the other hand, non-holonomic constraints are not thus 
expressible; though they might be expressible by inequalities, or by 
equations governing differentials of coordinates.  For example:--\\
\indent (i): confinement of a particle to one side of a surface (e.g. 
one particle confined to the region beyond a sphere of radius $a$ 
centred at the origin---expressed by the inequality $\mid {\bf r}_1 
\mid^2 \, > a^2$);\\ 
\indent (ii): a rigid body rolling without slipping on a surface; the 
condition of rolling is a condition on the velocities (i.e. that the 
point of contact is stationary), a differential condition which can be 
given an integrated form only once the problem is solved.

{\em Scleronomous} (respectively: {\em rheonomous)} constraints are 
independent of (dependent on) time. (The terminology is due to 
Boltzmann: ``scleronomic'' and ``rheonomic'' are also used.) So the 
time-argument in eq. \ref{eq;holo} (and in corresponding equations for 
non-holonomic constraints) is to allow for rheonomous constraints. 

As to overcoming the difficulty (Dependent), the situation is clearest 
for holonomic constraints. The idea is to use our free choice of 
coordinate systems on configuration space so as to describe the system 
by appropriate independent variables, called  {\em generalized 
coordinates}. So consider a system of $N$ particles (i.e. 
point-particles or bodies small and rigid enough to be described by a 
single position vector $\bf r$), with $k$ holonomic constraints. The 
system's configuration $({\bf r}_1, {\bf r}_2, \dots,{\bf r}_N)$ is at 
each time confined to a hypersurface in $\mathR^{3N}$ which will in 
general be $(3N - k)$-dimensional. (Rheonomous constraints just mean 
that the hypersurface in question varies with time.)\\
\indent More precisely: we say the $k$ constraints eq. \ref{eq;holo}, 
$f_j = 0$ are {\em (functionally) independent} if at each point on the 
hypersurface, the $3N$-dimensional gradients $\nabla f_j$ (i.e. with 
cartesian coordinates $(\frac{\pl f_j}{\pl x_1}, \dots, \frac{\pl 
f_j}{\pl z_N})$) are linearly independent: i.e. if at each point, the 
$k \times 3N$ matrix with entries $\frac{\pl f_j}{\pl x_1}, \dots, 
\frac{\pl f_j}{\pl z_N}$ has maximal rank $k$.\footnote{Cf. the 
definition of functional independence in Paragraph 2.1.3.A, (iii).}\\
\indent If the constraints are independent, then the hypersurface to 
which the system is confined (for rheonomous constraints: at a given 
time) is $(3N - k)$-dimensional, and there is a system of coordinates 
on $\mathR^{3N}$  in which the constraints become
\be
q_{3N - k + 1} = 0, \; \dots,\; q_{3N} = 0 \;\; .
\ee 
(These generalized  coordinates need not have the dimension of 
position.) 
 In other words, there are $3N - k$ independent variables 
$q_1,\dots,q_{3N-k}$ coordinatizing the hypersurface, such that (again 
allowing for rheonomous constraints)
\be
{\bf r}_i = {\bf r}_i(q_1,\dots,q_{3N-k},t), \;\;\; i = 1,2,\dots, N.
\label{eq;rfromq}
\ee
\indent Such coordinates are said to be {\em adapted} to the 
constraints. I shall usually focus on the hypersurface, not the 
ambient $3N$-dimensional configuration space; and I shall write $n$ 
for $3N - k$, and therefore write the generalized coordinates as 
$q_1,\dots,q_n$. The $n$-dimensional space is sometimes called the 
{\em constraint surface} in the $3N$-dimensional space; `configuration 
space' is used for either of these spaces.

Configuration space, and the way it underlies (Scheme), will be 
centre-stage in this Section and Section \ref{Lagr}.

First of all, let me address two doubts which can be raised about this 
strategy of analysing the system wholly within the lower-dimensional 
constraint surface. (The second is more important in what follows.)

\indent (A): In everyday problems the constraints are often 
non-holonomic; as in  examples (i) and (ii) above. Here we see our 
first example of the moral (Restrict); and a main one to boot. In 
fact, many of the methods and results of analytical mechanics depend 
on the constraints, if any, being holonomic; and most of my discussion 
will assume this.  

\indent (B): Since forces of constraint are in fact finite, a 
constrained system  is not rigorously confined to the constraint 
surface---it can depart slightly from it; (more precisely, its 
configuration can depart slightly).\\
\indent There are three replies to this. The simplest and dullest is  
that the forces of constraint are often strong enough that this 
strategy is an empirically adequate idealization. (Incidentally, it is 
an idealization in Section \ref{schemes}'s proposed usage of 
neglecting the negligible while posing, not while solving, the 
problem.)\\
\indent A more interesting reply is that there are theorems proving 
that in the limit as the forces of constraint become infinitely 
strong, the system's dynamics in the full configuration space becomes 
as analytical mechanics describes it, on the constraint surface. Such 
theorems illustrate (Ideal) and (Accept). I will report such a theorem  
in Paragraph 3.3.2.A.\\
\indent Finally, the most interesting reply is one of the triumphs of 
Lagrangian mechanics---and a prime example of the merit (Reduce). 
Namely, suppose we maintain---say on the strength of the first two 
replies---that the constraint equations hold, i.e. the system is 
confined to the constraint surface. Then: under some very general 
conditions, Lagrangian mechanics enables us to rigorously solve the 
mechanical problem, i.e. to find the generalized coordinates  
$q_1,\dots,q_n$ as functions of time (at least in a ``medium sense'' 
such as quadrature), without ever knowing the forces of 
constraint!\footnote{To anticipate a little: it is sufficient for this 
ability that the constraints are holonomic and ideal, and the applied 
forces are monogenic.} Furthermore, after solving the problem in this 
way, we can come back and calculate what the constraint forces were 
(as a function of time).

\subsubsection{Kinetic energy and work}\label{KEandwork}
In this Subsection, I  describe how to represent on configuration 
space, notions which are close cousins of the two sides of Newton's 
second law---the $m{\bf a}$ representing a body's inertia and the $\bf 
F$ representing the force on it. For in analytical mechanics, these 
Newtonian notions are displaced as central concepts by these cousins: 
respectively, the kinetic energy, and the work done by the forces 
(which latter is in many cases the derivative of a certain function, 
the work function, i.e. the negative of the potential energy).  
Thus:--

 (a): {\em The analogue of inertia}:-- The kinetic energy $T := 
\Sigma_i \frac{1}{2} m_i {\bf v}^2_i$ defines a line-element (in 
modern jargon: a metric) in the $3N$-dimensional configuration space 
by
\be
(ds)^2 \equiv ds^2 := 2Tdt^2 = \Sigma_i m_i{\bf v}^2_i dt^2 = \Sigma_i 
m_i(dx_i^2 + dy_i^2 + dz_i^2)  \;\; .\label{eq;Tdefinesline}
\ee     
Incidentally, this implies that
\be
 T = \frac{1}{2}m \left(\frac{ds}{dt} \right)^2 \;\; \mbox{   with    
}\;\;  m = 1  \;\; ; \label{eq;Tonconfigmass1}
\ee
i.e. the system's kinetic energy is represented by the kinetic energy 
(relative to the new line-element) of a single particle of mass 1.

 (b): {\em The analogue of force}:-- To explain this, I need the idea, 
which will be crucial in all that follows, of a {\em virtual 
displacement}. This idea also provides a main example of the moral 
(Modality;1st). That is, it considers counterfactual states, but not 
counterfactual problems or laws---indeed, so far laws are not in play. 

A virtual displacement, $\delta{\bf r}_i$, of particle $i$ is defined 
as a possible displacement of $i$ that is consistent with both the 
applied force ${\bf F}^{a}_i$, and the force of constraint ${\bf 
f}_i$, on particle $i$, at the given time. Here, it will be 
consistency with the constraints {\em at the given time} that matters. 
We similarly define a virtual displacement of the system as possible 
displacements of its particles that are jointly consistent with the 
constraints at the time. We will usually be concerned with arbitrarily 
small virtual displacements, often called `infinitesimal'; and I will 
usually omit this word. 

 We can express this more precisely by supposing the constraints yield  
differential conditions on the coordinates. ({\em Warning}: This 
paragraph is  more precise than I will need: it is not used later on.) 
If the constraints are holonomic, as in eq \ref{eq;holo}, we can 
differentiate to get the differential  conditions; if the constraints 
are rheonomous, at least one of the conditions will be time-dependent. 
So if the conditions are, in terms of $n$ generalized coordinates (and 
with $k$ constraints),
\be
\Sigma_j A_{ij}(q_1,\dots,q_n,t)dq_j + A_{it}(q_1,\dots,q_n,t)dt = 0 
\;\; ; \;\;\; i = 1,\dots,k ,
\ee
then a virtual displacement of the system at time $t$ is defined to be 
any solution $(\dd q_1,\dots,\dd q_n)$ of the $k$ equations  
\be
\Sigma_j A_{ij}(q_1,\dots,q_n,t)\dd q_j = 0 \;\; ; \;\;\; i = 
1,\dots,k.
\ee

Obviously, virtual displacements need not be actual. But beware: the 
converse also fails: actual displacements need not be virtual, because 
the forces and constraints might vary with time (the rheonomous case), 
and virtual displacements must be consistent with the forces and 
constraints {\em at the given time}.\footnote{So there is no real 
conflict with the venerable principle of modal logic, that the actual 
is possible. But the point is important: it underpins the derivation 
of the conservation  of energy  from D'Alembert's principle, cf. 
Paragraph 3.3.1.A.} Note also that `virtual' will be used in this 
sense, not just for displacements, but more generally. We distinguish 
between virtual and actual variations of any quantity; and use the 
respective notations (introduced by Lagrange), $\dd$ and $d$, for 
them. Also, I shall  sometimes use $\dd$, either\\
\indent (i): for a small (not infinitesimal) variation; or\\
\indent  (ii): to indicate that the differential is not exact.

If the total applied force on particle $i$ has cartesian components 
$X_i,Y_i,Z_i$, the total work done by this applied force in an 
infinitesimal virtual  displacement can be written as a differential 
form in the cartesian coordinates, i.e. as a sum over the $3N$ 
cartesian coordinates 
\be
\dd w = \Sigma^{N}_{i=1} X_i \dd x_i + Y_i \dd y_i + Z_i \dd 
z_i;\label{eq;dwcart}
\ee  
or, transforming to generalized coordinates, say $q_1,\dots,q_n$, as a 
differential form in the generalized coordinates
\be
\dd w = \Sigma^n_{j=1} F_j \dd q_j \label{eq;dwgnlzd} 
\ee
where the transformation eq. \ref{eq;rfromq} determines the $F_j$ in 
terms of the cartesian components of force $X_i,Y_i,Z_i$.

In general, $\dd w$ is not integrable. But if it is, then its integral 
is called the {\em work function} $U$. Lanczos (1986, p. 30) calls 
applied forces of this kind {\em monogenic} (`monogenic' for `single 
origin') as against {\em polygenic}. Goldstein et al (2002) follow 
him; and I shall also adopt these terms.\footnote{But beware: Lanczos' 
text sometimes suggests `monogenic' is also defined for constraint 
forces. Agreed, in statics at equilibrium, the total applied force is 
the negative of the total constraint force:  so if the former are 
monogenic with work function $U$, one can also call the latter 
monogenic with ``work function'' $-U$. But in dynamics (Section 
\ref{ssecDale}), this correspondence breaks down: even with monogenic 
applied forces, most constraint forces will not be derivable from a 
work function. Thanks to Oliver Johns for this point.}

 In many of the simpler cases of monogenic forces, $U$ is independent 
of both the time, and the generalized  
velocities ${\dot q}_i$, so that $U = U(q_1,\dots,q_n)$. If so, we say 
the system is {\em conservative}; and we have
\be
\Sigma_j F_j \dd q_j = \frac{\partial U}{\partial q_j}\dd q_j
\ee 
so that by the mutual independence of the $q$s, we can set all but one 
of the $\dd q_j$ equal to 0, and so infer 
\be
F_j = \frac{\partial U}{\partial q_j}.
\ee  
Writing $V:= -U$, we have $F_j = -\frac{\partial V}{\partial q_j}$ and 
we interpret $V$ as a {\em potential energy}. This corresponds to the 
``cartesian definition'' of conservative systems, viz. that the force 
on the $i$th particle $\F_i$ is derivable ($\forall i$) from a 
time-independent scalar function $V$ on configuration space, i.e.
\be
\F_i = - \nabla_i V
\label{FgradV}
\ee
where the $i$ indicates that the gradient is to be taken with particle 
$i$'s coordinates. (Incidentally: Here we meet the prototypical case 
of an integrability condition. For each $i$, eq. \ref{FgradV}, with 
given $\F_i$, has a solution $V$ only if in the domain considered $\F$ 
is curl-free, i.e. $\nabla \wedge \F = 0$: or equivalently, any closed 
loop integral $\oint \F_i \cdot d{\bf s}$ vanishes.)

When we add to the assumption of conservativity (i.e. $V$ depending 
only on the generalized coordinates, not on the time, nor on the 
generalized  
velocities), the assumption that constraints, if any, are 
scleronomous, we get the conservation of energy, i.e. the constancy in 
time of $T + V$. This can be deduced in various ways, e.g. from 
d'Alembert's principle in Section 3.3 below. But however the theorem 
is derived, its being a generalization over many problems means it 
illustrates (Modality;2nd).\footnote{Two remarks about the more 
general case where $U$ depends on the velocities and time, so that $U 
= U(q_j,{\dot q}_j,t)$. (1) As we shall see, analytical mechanics and 
its variational principles apply in this case. (2) In particular, for 
scleronomous systems with monogenic forces---i.e. systems with a 
work-function $U$ that has no explicit dependence on time, but may be 
dependent on velocities $U = U(q_j,{\dot q}_j)$---there is a 
conservation of energy theorem: $T + V$ is conserved but with $V$ 
defined by $V := \Sigma_j \frac{\partial U}{\partial q_j}{\dot q}_j - 
U$.}

\subsection{The Principle of Virtual Work}\label{ssecPrin}
\subsubsection{The principle introduced}\label{PVWIntrodd} 
I turn to the second difficulty, (Unknown), faced by vectorial 
mechanics at the start of this Section: that the forces that maintain 
the constraints (`forces of constraint', `forces of reaction') are not 
known. So we ask: can we somehow  formulate mechanics in such a way 
that we do not need to know the forces of constraint? In fact for many 
problems, we can: viz. problems in which these forces would  do no 
work in a virtual displacement. Such constraints are called {\em 
ideal}. The prototypical case is a rigid body; where indeed the work 
done by the internal forces (with ${\bf F}_{ij}$ assumed to lie along 
the line between particles $i$ and $j$) is zero. But this condition is 
quite common, even for non-holonomic constraints; though to be sure, 
it also excludes very many cases e.g. friction.

Restricting ourselves to ideal constraints  (characterized, note, by a 
counterfactual!) is a crucial example of the moral (Restrict). It is 
even more important an example than my previous one (viz. analytical 
mechanics' frequent restriction to holonomic constraints). For most of 
analytical mechanics depends on this restriction. In particular, those 
of its great principles  that I will discuss in this paper---the 
principle of virtual work, d'Alembert's principle, the principle of 
least action, and Hamilton's principle---do so.   Accordingly, I (like 
many authors) will often {\em not} repeat that constraints are being 
assumed to be ideal.\\
\indent Besides, as envisaged at the end of Section \ref{Refo}, the 
principles are often closely related to each other, and in some cases 
equivalent under the assumption of ideal and/or holonomic constraints; 
so that they exemplify (Reformulate) as well as (Restrict). 

The first of the principles governing such problems is the {\em 
principle of virtual work}. Again we need  Section \ref{ssecGene}'s 
idea of 
an (infinitesimal) virtual displacement (and so (Modality;1st)). For 
the principle of virtual work concerns the total work done in such a 
displacement in the special case of equilibrium. 

So let us assume that the system is in {\em equilibrium}. This means 
that for each particle $i$ the total force ${\bf F}_i$ on it vanishes. 
Then for any virtual displacement $\delta{\bf r}_i$, ${\bf F}_i \cdot 
\delta{\bf r}_i = 0$ and so, summing, $\Sigma_i {\bf F}_i \cdot 
\delta{\bf r}_i = 0$. Let us split $\F_i$ in to the applied force 
(impressed force) ${\bf F}^{(a)}_i$, and the force of constraint ${\bf 
f}_i$. And let us make our restricting assumption that the constraints 
are ideal, i.e. the virtual work of the ${\bf f}_i$ is 0. Then we have 
\be
\Sigma_i \, \F^{(a)}_i \cdot \dd\rr_i = 0. \label{eq;pvwcart}
\ee
That is:  a system is in equilibrium only if the total virtual work of 
all the applied forces vanishes.

Under certain conditions the converse of this statement also holds; as 
follows. First note that we cannot conclude from eq \ref{eq;pvwcart} 
that each $\F^{(a)}_i = 0$, because the $\dd\rr_i$ are not linearly 
independent vectors---recall that we are considering virtual 
displacements. 
But it does represent a statement of orthogonality; as does the 
corresponding statement in generalized coordinates (cf. the transition 
from eq. \ref{eq;dwcart} to eq. \ref{eq;dwgnlzd})
\be
\Sigma^n_{j=1} F_j \dd q_j = 0, \label{eq;pvwgnlzd} 
\ee    
which says that the ``vector'' of the $F_j$ must be ``orthogonal'' to 
the surface of allowed variations in the coordinates $q_j$.

On the other hand, suppose the following two conditions hold.\\
\indent (i): The coordinates ---whether $\rr_i$  
or $q_j$---{\em are} indeed independent; i.e. in the case of $q_j$: 
the constraints are holonomic so that we work in their $n$-dimensional 
space, the constraint surface; {\em and}\\
\indent (ii): the displacements are {\em reversible} in the sense that 
if $\dd q_j$ is allowed by the constraints, so is $-\dd q_j$).\\
\indent Then each $F_j = 0$. For only the zero vector can be 
orthogonal to all vectors. And so we have the converse of the above 
statement. That is, we have  {\em the principle of virtual work}:
\begin{quote} A system (subject to our restrictions) is in equilibrium 
if and only if for any virtual displacement  the total virtual work of 
all the applied forces vanishes. (It is of course the `if' half of the 
principle that is substantive.)
\end{quote}
(This is an example of what Section \ref{Refo} envisaged: (Restrict) 
and (Reformulate) together.)

Note that if the applied forces are monogenic, the total virtual work 
of these forces is the variation of the work-function $U$. So in this 
case equilibrium means: $\dd U = - \dd V = 0$. So the topic of 
equilibrium with holonomic constraints leads to the topic of  a 
function being stationary, subject to other functions taking 
prescribed values. And similarly, the topic of equilibrium with 
non-holonomic constraints leads to the topic of  a function being 
stationary, subject to conditions other than prescribed value(s) of  
function(s)---conditions that might be expressed as equations relating 
some functions' values or functions' differentials. (Here, `being 
stationary' means, as in elementary calculus, having a zero 
derivative: details below. But again as in calculus, our interest in 
stationary points of functions is often that they are extrema, i.e. 
maxima or minima. And so I will often talk of `extremizing a function' 
etc., to avoid cumbersome phrases like `find a point at which a 
function is stationary'---there is no word `stationarize'! )

 This is one of the several reasons for analytical mechanics' endemic 
use of the method of {\em Lagrange multipliers}, to analyse 
extremizations of a function subject to constraints. I finish this 
Subsection with a brief introduction to this method. It will lead us 
back to the topic of overcoming the difficulty (Unknown), that we do 
not know the forces of constraint.

\subsubsection{Lagrange's undetermined multipliers}\label{parLagr} 
This method has two significant advantages over the obvious method of 
eliminating as many variables as there are constraint equations, and 
then using differential calculus to perform an unconstrained 
extremization in fewer variables.\\
\indent (i): In many cases, there is no natural choice of variables to 
be eliminated: either because of the symmetrical, or nearly 
symmetrical, way that the variables occur; or because any choice makes 
for cumbersome algebra.\\
\indent (ii): The Lagrange method is more powerful. It can handle 
constraints given by differential conditions (in mechanical terms: 
non-holonomic constraints); which the elimination method cannot. 

Apart from its advantages (i) and (ii), it is also worth noting 
that:---\\
\indent (a): The method is not confined to the context in which it is 
often met, viz. variational principles (where the function to be 
extremized is an integral).\\
\indent (b): In mechanics, the method has a physical interpretation, 
which provides a way to calculate the  forces that maintain the 
constraints; more details in Paragraph 3.2.2.B.

\paragraph{3.2.2.A Lagrange's method}\label{322ALagMethod}
 The idea is clearest in a visualizable elementary setting. Suppose we 
want to extremize $f(x,y,z)$ i.e. $f:\mathR^3 \rightarrow \mathR$ 
subject to two constraints $g_1(x,y,z) = 0$ and $g_2(x,y,z) = 0$. 
Generically, the constraint surfaces meet in a line, and at the  
solution point $(x_0,y_0,z_0)$ the gradient $\nabla f$ must be 
orthogonal to the line's tangent vector ${\bf v}$; (otherwise $f$ 
could be increased or decreased by a displacement along the line). But 
${\bf v}$ lies in the intersection of the two tangent planes of the 
constraint surfaces, and these planes are normal to $\nabla g_1$ and 
$\nabla g_2$ respectively. So at the solution point $(x_0,y_0,z_0)$ 
the gradient $\nabla f$ must lie in the plane defined by $\nabla g_1$ 
and $\nabla g_2$, i.e. it must be a linear combination of $\nabla g_1$ 
and $\nabla g_2$;
\be
\nabla f = \lambda_1\nabla g_1 + \lambda_2\nabla g_2.
\label{3dlagmult}
\ee
This argument generalizes to higher dimensions (say $n$), and an 
arbitrary number (say $m$) of constraints. So using $\nabla$ for the 
$n$-dimensional gradient,
\be
\nabla f = \Sigma_j \lambda_j \nabla g_j.
\label{manydlagmult}
\ee   
   
We now put the argument (in higher dimensions) algebraically. We will 
use $x_i$ not $q_i$, even though the argument makes no use of 
cartesian coordinates; this has the merit of indicating that the 
equations, eq.  \ref{3dlagmult} and \ref{manydlagmult}, and eq. 
\ref{eq;LagMultBasic} below, refer to the `larger' configuration 
space, i.e. whose dimension, $n$ say, exceeds the dimension of the 
constraint surface by $m$, where $m$ is the number of constraints. We 
are to find the point $x_0 := (x_{0_1},\dots,x_{0_n})$ at which $\dd f 
= 0$ for small variations $x' - x_0$ such that $g_j(x') := 
g_j(x'_1,\dots,x'_i,\dots,x'_n) = 0$ for $j = 1,2,\dots,m$. So eq. 
\ref{manydlagmult} requires that there are $\lambda_j$ such that 
defining $h(x) := f(x) + \Sigma_j \lambda_j g_j(x)$, we have:
\be
\dd h = \dd f + \Sigma_j \lambda_j \dd g_j = 0 \;\; \mbox{, i.e.  } 
\;\; \frac{\partial f}{\partial x_i} + \Sigma_j \lambda_j 
\frac{\partial g_j}{\partial x_i} = 0, \;\;\; \forall i=1,\dots,n 
\label{eq;LagMultBasic}
\ee 
We use these $n$ equations, together with the $m$ equations $g_j(x_i) 
= 0 $ to find the $n + m$ unknowns (the $n \;\; x_i$ and the $m \;\; 
\lambda_j$).

Thus the fundamental idea is to replace a constrained extremization in 
$n$ variables subject to $m$ constraints by an unconstrained 
extremization in $n+m$ variables. I will make three brief comments, 
(1)-(3), about further developments, before I turn to the physical 
interpretation.

 (1): Thinking of the $\lambda_j$ as variables, and so of $h$ as a 
function of the $n+m$ variables $(x_i,\lambda_j)$ we can ask that it 
be stationary. This variation problem gives eq. \ref{eq;LagMultBasic} 
again, if we vary with respect to $x_i$; and the constraint  equations 
$g_j = 0$, if we vary with respect to $\lambda_j$. In short: Variation 
of the $\lambda_j$ gives back the constraint equations {\em a 
posteriori}. 
 
(2): Lagrange's method also applies to constraints expressed not by 
equations $g_j = 0$ but only by conditions on differentials, i.e. a 
set of equations
\be
\dd g_j := G_{j1}\dd x_1 + \dots + G_{jn}\dd x_n = 0; \;\;\; j = 
1,2,\dots,m \label{eq;varynonholoconstrt}
\ee  
where the left-hand side uses the $\dd$ to indicate that it is not an 
exact differential, i.e. $G_{ji}$ is not the $i$th partial derivative 
of a function $g_j$. Lagrange's method again applies and we get the 
condition
\be
\dd f + \lambda_1 \dd g_1 + \dots + \lambda_m \dd g_m = 0,
\ee
where we are to treat all the $x_i, i = 1,\dots,n$ as independent  
variables; and into this equation the expressions for $\dd g_j$ from 
eq. \ref{eq;varynonholoconstrt} can be substituted. (It is just that 
the left-hand side of eq. \ref{eq;varynonholoconstrt} is not the 
differential of a function $g_j$, as it was above).  

(3): Lagrange's method (including the above two comments) also applies 
to the central idea of calculus of variations---the extremization of 
an integral, which we will meet in Section \ref{Lagr}; (details in 
Section \ref{ConstrExtr}).

\paragraph{3.2.2.B Physical interpretation: the determination of the 
constraint forces}\label{322BInterpnLagMethodFindConstraintForces}
When Lagrange's multiplier method is applied to mechanics, it  has a 
physical  interpretation. The interpretation is easily seen for our 
present topic, equilibria for monogenic applied forces. As we have 
seen, for such forces, there is a $V$ which, once added to some linear 
combination of constraints, is to be extremized. In short, the 
physical interpretation is that the Lagrange multipliers give the 
forces of constraint. I shall develop this interpretation only for the 
special case of holonomic constraints and constraint forces that are 
derivable from a potential (unusual though this is: cf. footnote 35). 
But the interpretation holds much more generally.

So suppose that the constraints are holonomic (as well as ideal), so 
that we are to extremize $V$ subject to the constraints that all the 
$g_j = 0$, i.e. to extremize $V + \lambda_jg_j$. Suppose also that 
each force of constraint is ``monogenic'', i.e. is derivable from a 
potential energy. Then, two results follow. First, for each $j$, 
$\lambda_jg_j$ represents the potential energy of the $j$th force of 
constraint. Second, the fact that each $\lambda_j$ is known only at 
the solution-point $x_0$ reflects our scanty knowledge about the 
forces of constraint. For in forming the gradient of the $j$th 
additional potential energy $\lambda_jg_j$, we get as the contribution 
$F_{ji}$ to the $i$ cartesian component of the force
\be
F_{ji} := -\frac{\partial}{\partial x_i}(\lambda_jg_j) = - \lambda_j 
\frac{\partial g_j}{\partial x_i} - \frac{\partial \lambda_j}{\partial 
x_i}g_j  \;\;\;\; \mbox{(no summation on $j$)};
\label{lamdaforceofconstt}
\ee   
at the solution-point $x_0$, $g_j$ vanishes, i.e. $g_j(x_0) = 0$, so 
that at $x_0$
\be
F_{ji} = - \lambda_j \frac{\partial g_j}{\partial x_i} \;\;\; 
\mbox{(no summation on $j$)}.
\label{lamdaforceofconstt2}
\ee

Remarkably, this physical interpretation carries over to the case of 
non-holonomic (but ideal!) constraints, and to the case where  the 
ideal constraint  forces do not have a work-function, i.e. no 
potential energy $\lambda_jg_j$; (as Lanczos might say, the case of 
polygenic constraint  forces). One proceeds as in comment (2) of 
Paragraph 3.2.2.A; the forces are again given by the $\lambda$-method. 
Furthermore, this physical interpretation carries over to the case of 
non-equilibrium, i.e. dynamics, for both holonomic and non-holonomic 
constraints. I will discuss this a little more in Sections 
\ref{ApplicMechs} and \ref{ssecTime}; but for more details, cf. 
Desloge (1982: 532-534) and Johns (2005: Chapter 3.4).

This physical interpretation underpins the striking way in which 
Lagrangian mechanics  enables one to solve problems without knowing 
the forces of constraint. Again, I will not go into details: not even 
in Section \ref{ssecDale}'s discussion of dynamics---since there I use 
ideal constraints and d'Alembert's principle to reduce mechanical 
problems very directly to a description on the constraint surface 
which does not even {\em mention} the constraint forces: vividly 
illustrating the merit (Reduce).\\
\indent But in short, the idea is that, similar to just above: the 
$i$th generalized component ($i = 1,\dots,n$) of the $j$th constraint 
force ($j = 1,\dots,m$) is $- \lambda_j {\pl g_j}/{\pl q_i}$. This 
means that knowing the constraint equations as functions of the 
generalized coordinates $g_j(q_1,\dots,q_n) = 0$ is enough. For as in 
Paragraph 3.2.2.A, there are enough equations to determine not just 
the system's motion $q_i(t)$, but also the $\lambda$s, and thereby the 
forces of constraint.

\indent So to sum up: under some widespread conditions, we can, after 
we solve the problem (i.e. find the motion of the system) without even 
knowing the constraint forces, go back and calculate the constraint 
forces. In effect, the idea of this calculation is that the constraint 
forces have whatever values they need to have so as to maintain the 
constraints on the previously calculated motion. In this way we can 
overcome the difficulty (Unknown) in the best possible way. (For 
details, cf. Desloge (1982: 545-546, 549-552, 555), Johns (2005: 
Chapter 3-5,3-8,3-11).)

\subsection{D'Alembert's Principle and Lagrange's 
Equations}\label{ssecDale}

\subsubsection{From D'Alembert to Lagrange}\label{FromDale}
To sum up Section \ref{ssecPrin}: it described how the principle of 
virtual work eliminates the force of constraint ${\bf f}_i$ on each 
particle $i$ for the case of equilibrium (thus overcoming the 
difficulty (Unknown) for that case). The idea of 
D'Alembert's principle is to eliminate the ${\bf f}_i$ also for 
non-equilibrium situations, by the simple and ingenious device of 
treating the negative of the mass-acceleration, $-{\dot{p}_i }$, as a 
force; as follows.\footnote{As usual, the history of the principle is 
much more complicated than modern formulations suggest. For 
d'Alembert's original formulation, cf. Fraser (1985a).}  

Newton's second law $\F_i = {\dot{p}_i }$ ``reduces to statics'' if we 
rearrange it as if there were a ``reversed effective force'' 
$-{\dot{p}_i}$; i.e. if we write
\be
\Sigma_i \, (\F_i - {\dot{p}_i})\cdot \dd\rr_i = \Sigma_i \, 
(\F^{(a)}_i - {\dot{p}_i})\cdot \dd\rr_i + \Sigma_i \, {\bf f}_i \cdot 
\dd\rr_i = 0.  
 \ee
Again we assume that the virtual work of the forces of constraint 
${\bf f}_i$ is 0, so that:
\be
\Sigma_i \, (\F^{(a)}_i - {\dot{p}_i})\cdot \dd\rr_i = 
0.\label{eqn;dalem}  
\ee
This is {\em d'Alembert's Principle}. Since the forces of constraint 
${\bf f}_i$ are now eliminated, I will now drop the superscript 
$^{(a)}$ for `applied'.

This prompts three immediate comments: technical, philosophical and 
strategic.\\
\indent  (i): Given d'Alembert's Principle, we can argue, as we did 
after eq \ref{eq;pvwgnlzd}. Namely, suppose that the coordinates are 
independent (so that constraints, if present, are holonomic and we 
focus on the constraint surface), and that the displacements 
reversible. Then each $\F_i - {\dot{p}_i} = 0$. This difference of the 
applied force and the inertial force, $\F_i - {\dot{p}_i}$, is 
sometimes called the `effective force' on particle $i$; (and also 
sometimes the `constraint force', despite equalling $-{\bf f}_i$!). So 
d'Alembert's principle eq. \ref{eqn;dalem}  implies: the total virtual 
work done by the effective forces is zero. This is sometimes written 
as: $\dd w^e = 0$, where $\dd w^e$ is a non-exact differential; 
(non-exact because in general ${\bf f}_i$ is not derived from a work 
function).

\indent  (ii): Again, we see (Restrict) and (Modality;1st) at work.

\indent  (iii): A warning about my chosen route for expounding 
analytical mechanics. On this route, d'Alembert's principle figures 
large (despite having so simple a deduction from the principle of 
virtual work). Thus I shall report in the Paragraphs just below how it 
implies a form of the conservation of energy, and a form of Lagrange's 
equations. It also underlies the other central principles of 
analytical mechanics, including  the most important  one, Hamilton's 
Principle---which I will discuss in Section \ref{Lagr}. Besides, 
`underlies' here is logically strong: it means `implies when taken 
together with just pure mathematical apparatus, and (for some 
implications) some general physical assumptions; and in some cases, 
the converse implication holds'.\\
\indent But I admit: many fine expositions  adopt other routes, on 
which d'Alembert's principle hardly figures. In particular, one can 
proceed directly from Newton's equations to Lagrange's equations. For 
example, cf. Woodhouse (1987: 31-34,41-47), Johns (2005: Chapter 
2-2,2-7); or in more detail, for successively less straightforward 
systems, e.g. first for holonomic, then for  non-holonomic, 
constraints, Desloge (1982: 522-523, 542-545, 554-557, 564-565).

\indent As it stands, d'Alembert's principle has a significant 
disadvantage; (so that it does not itself supply a general scheme, on 
a par with the Lagrangian or Hamiltonian one). Though the virtual work 
of the applied forces is often an exact differential (i.e. the forces 
are monogenic, there is a work-function) there is no such function for 
the virtual work of the inertial forces.\\
\indent It is one of the key insights of Lagrangian  mechanics that 
this disadvantage can be overcome, by expressing d'Alembert's 
principle in terms of configuration space. In particular, by 
integrating d'Alembert's principle with respect to time, we can derive 
Hamilton's principle itself. More precisely: by integrating the total 
virtual work of the effective force $\dd w^e$ along the system's 
history (trajectory in configuration space) with time as the 
integration variable, the inertial forces become monogenic: for 
details cf. Section \ref{ssecHami}.

Besides, we can similarly derive from d'Alembert's principle other 
principles of Lagrangian  mechanics. (In some cases, this is done via 
Hamilton's principle, i.e. by first deriving Hamilton's principle from 
d'Alembert's principle.) For reasons of space, I shall not report such 
derivations, though they illustrate well my moral (Reformulate). For 
some details, cf. Lanczos (1986: 106-110). He discusses in order:\\
\indent (a): how Gauss' principle of least constraint expresses 
d'Alembert's principle as a minimum principle;\\
\indent (b): the merits and demerits of Gauss' principle; and\\
\indent  (c): Hertz' interpretation of Gauss' principle as requiring  
the system's path in configuration space be of minimal curvature---an 
idea developed by Jacobi's principle, which I will discuss in Section 
4.6.\\
\indent I emphasise that these derivations and discussion are all 
conducted under the restriction we imposed at the start of  Section 
\ref{ssecPrin}; viz. that the constraints are ideal, i.e. the virtual 
work of the forces of constraint is 0. Thus we again see 
(Restrict).\footnote{Incidentally, Lanczos raises this restriction to 
a postulate, called Postulate A (1986: 76).} 

But I shall report: (A) how d'Alembert's principle implies the 
conservation of energy (under appropriate conditions); and (B) how it 
also implies Lagrange's equations.  These equations, set in the 
context of Hamilton's principle, will be centre-stage in the next 
Section; so it is also worth seeing that they are implied by 
d'Alembert's principle directly, i.e. not via Hamilton's principle.\\
\indent Then (in the next Subsection) I will end this Section by 
discussing how Lagrange's equations, regardless of how they are 
deduced, represent mechanical problems and illustrate (Scheme). 

\paragraph{3.3.1.A  Conservation of energy}\label{331A}
 Integrating D'Alembert's Principle, under certain assumptions, yields 
as a result the conservation of energy; as follows. If the applied 
forces are monogenic, then the Principle becomes
\be
\dd V + \Sigma_i m_i {\ddot {\bf r}_i}\cdot \dd\rr_i = 0.
\label{Dalemono}
\ee
Then: If (and only if!) the system is not only holonomic, but also 
scleronomous in work-function and constraints (i.e. the work-function 
$U$ can be $U(q_i,{\dot q}_i)$ but $U$ cannot be an explicit function 
of $t$, and the constraints are $g_j(q_i) = 0$ but not 
$g_j(q_i,t)=0$), then we can choose the $\dd \rr_i$ to be the {\em 
actual} changes $d \rr_i$ in an infinitesimal time $dt$; (Lanczos 
1986: 92-94; cf. also footnote 33). This implies:
\be
\Sigma_i m_i {\ddot {\bf r}_i}\cdot \dd\rr_i = \Sigma_i m_i {\ddot 
{\bf r}_i}\cdot d\rr_i = dT 
\ee
so that D'Alembert's Principle gives:
\be
dV + dT = d(T + V) = 0.
\label{consenyfromdale}
\ee 
This result, being a generalization across a whole class of problems, 
illustrates my moral (Modality;2nd).

\paragraph{3.3.1.B  Deducing Lagrange's equations}\label{331B} The 
deduction of Lagrange's equations from d'Alembert's Principle 
illustrates (Reformulate) and (Restrict). In the course of the 
derivation, one makes two restrictions in addition to constraints 
being ideal (which is implicit in d'Alembert's principle): first to 
holonomic constraints, and then to monogenic systems. (Conservativity, 
which is used in some expositions, is not necessary.) {\em Warning:} 
The details of this derivation are not used later on. 

The idea will be to transform d'Alembert's principle eq. 
\ref{eqn;dalem} to generalized coordinates $q_j$, of which there are 
say $n$ (e.g. above we had $n = 3N - k$, with $N$ particles and $k$ 
constraints). So the transformed equation will concern virtual 
displacements $\dd q_j$. Then we will assume the constraints are 
holonomic, i.e. the $q_j$ are independent, so that each coefficient of 
$\dd q_j$ in the transformed equation must vanish.

 We begin by noting that the transformation ($i$ again labelling 
particles)
\be
\rr_i = \rr_i(q_1,q_2,\dots,q_n,t)\label{equati
on;rfromq}
\ee
yields
\be
{\bf v}_i := \frac{d \rr_i}{dt} = \Sigma_j \frac{\partial 
\rr_i}{\partial q_j}{\dot{q}_j} + \frac{\partial \rr_i}{\partial t}    
.\label{equation;vfromq}
\ee
This implies ``cancellation of the dots'', i.e.
\be
\frac{\partial {\bf v}_i}{\partial {\dot{q}}_j} = \frac{\partial 
\rr_i}{\partial q_j};
\label{canceldots}
\ee
and also commutation of differentiation with respect to $t$ and $q_j$, 
i.e.
\be
\frac{d}{dt}\frac{\partial \rr_i}{\partial q_j} = \frac{\partial {\bf 
v}_i}{\partial q_j}.
\ee
Besides, note that 
\be
\dd \rr_i = \Sigma_j \;\; \frac{\partial \rr_i}{\partial q_j}\dd q_j 
\ee
implies
\be
\Sigma_i \;\; \F_i \cdot \dd \rr_i = \Sigma_j \;\; Q_j \dd q_j \;\;\; 
\mbox{   with  }\;\;\; Q_j := \Sigma_i \; \F_i \cdot 
\left(\frac{\partial \rr_i}{\partial q_j} 
\right).\label{equation;defineQj}
\ee
The $Q_j$ are the components of {\em generalized force}. Though the 
$Q_j$ need not have the dimensions of force (and will not if the $q_j$ 
do not have the dimensions of length), $Q_j \dd q_j$ must have the 
dimensions of work.

Applying these results to the second term of d'Alembert's Principle, 
eqn \ref{eqn;dalem}, i.e. to
\be
\Sigma_i \;\; {\dot{p}_i} \cdot \dd\rr_i = \left(\Sigma_i \; \; m_i 
{\dot{\bf v}_i}\right) \cdot\left(\Sigma_j \;\; \frac{\partial 
\rr_i}{\partial q_j}\dd q_j \right)  
\ee
and using the definition of total kinetic energy $T := \Sigma_i 
\frac{1}{2}m_i{\bf v}^{2}_i$, d'Alembert's Principle becomes:
\be
\Sigma_j[\{\frac{d}{dt}(\frac{\partial T}{\partial \dot{q}_j}) - 
\frac{\partial T}{\partial q_j}\} - Q_j]\dd q_j = 0.
\ee
Now let us assume the constraints are holonomic, so that the $q_j$ are 
independent. Then we can conclude that for each $j$
\be
\frac{d}{dt}(\frac{\partial T}{\partial \dot{q}_j}) - \frac{\partial 
T}{\partial q_j} = Q_j.\label{eqn;lagforT}
\ee  

Equations \ref{eqn;lagforT} are sometimes called {\em Lagrange's 
equations}. But this name is more often reserved for the form these 
equations  take for a system that is not just holonomic, but also 
monogenic with a velocity-independent work function. That is: If each 
applied force $\F_i$ (the force on the $i$th particle) is a gradient 
(with respect to $i$'s coordinates) of a (possibly time-dependent) 
scalar function $V$ on configuration space, i.e.
\be
\F_i = - \nabla_i V
\ee
then the definition of $Q_j$, eq. \ref{equation;defineQj}, immediately 
yields
\be
Q_j = - \frac{\partial V}{\partial q_j}
\ee
so that defining the {\em Lagrangian} $L := T - V$, we get from eqn 
\ref{eqn;lagforT}:
\be
\frac{d}{dt}(\frac{\partial L}{\partial \dot{q}_j}) - \frac{\partial 
L}{\partial q_j} = 0.\label{eqn;lag}
\ee
(Furthermore, we can get this same form for the equations (again with 
$L= T - V$) if there is velocity-dependence, provided the generalized 
forces $Q_j$ are then obtained by
\be
Q_j = -\frac{\pl V}{\pl q_j} + \frac{d}{dt}\left(\frac{\pl V}{\pl 
{\dot q}_j}\right).\label{QfromvelydepdtV}
\ee   
This formula applies in electromagnetism; (cf. e.g. Goldstein et al 
(2002: 22).)

This is a good point at which to note the form of the kinetic energy 
$T$ in terms of the generalized coordinates; and in particular, the 
form $T$ takes when the constraints are scleronomous---which is the 
case I will mostly consider.  
We transform between cartesian and generalized coordinates by 
equations \ref{equation;rfromq} and \ref{equation;vfromq}. Note in 
particular that 
\be
T = \Sigma_i \; \frac{1}{2}m_iv^2_i = \Sigma_i \;\; \frac{1}{2}m_i 
\left(\Sigma_j \frac{\partial \rr_i}{\partial q_j}{\dot{q}_j} + 
\frac{\partial \rr_i}{\partial t} \right)^2 
\ee 
and that expanding this expression, we can express $T$ in terms of the 
generalized coordinates as
\be
T = a + \Sigma_j a_j{\dot q_j} + \Sigma_{j,k}a_{jk}{\dot q_j}{\dot 
q_k}\label{equation;Tingeneral}
\ee  
where $a,a_j,a_{jk}$ are definite functions of the $\rr$'s and $t$, 
and hence of the $q$'s and $t$. Besides, if the transformation 
equations \ref{equation;rfromq} and \ref{equation;vfromq} do not 
contain time explicitly (i.e. the constraints are scleronomous), then 
only the last term of eq. \ref{equation;Tingeneral} is non-zero: i.e. 
$T$ is a homogeneous quadratic form in the generalized velocities.\\
\indent This result has a geometric significance. For we saw in eq. 
\ref{eq;Tdefinesline} that $T$ defines a metric (a line-element) in 
the $3N$-dimensional configuration space of $N$ particles; we now see 
that for scleronomous constraints  it  defines a metric on the 
constraint surface. This geometric viewpoint will be developed in 
Paragraph 3.3.2.E.     

\subsubsection{Lagrange's equations: (Accept), (Scheme) and 
geometry}\label{LagEqAccScheme}
 Lagrange's equations (especially in the form of eq \ref{eqn;lag}) are 
the centre-piece of Lagrangian mechanics; and since Lagrangian 
mechanics is the basis of other approaches to analytical mechanics, 
such as Hamiltonian mechanics, these equations can fairly claim to be 
the crux of the subject. I end this Section with five comments (in 
five Paragraphs) about these equations.\\
\indent The first comment is foundational: it concerns using 
Lagrange's equations to justify analysing a system with holonomic 
constraints wholly in terms of the constraint surface.  This comment 
illustrates my moral (Accept). The second, third and fourth comments  
are about solving Lagrange's equations, and the practical advantages 
of using them, i.e. my moral (Scheme). These comments lead in to the 
fifth comment, about the modern geometric description of Lagrangian 
mechanics, and the representation it provides of the solution of a 
mechanical problem.

\paragraph{3.3.2.A  Confinement to the constraint surface}\label{332A}
I said at the end of Section 3.1 when I first introduced the idea of 
the constraint  surface, that there were theorems proving that in the 
limit as the forces of constraint become infinitely strong, the 
system's dynamics in the full configuration space becomes as 
analytical mechanics describes it, on the constraint surface. With 
Lagrange's equations eq. \ref{eqn;lag} in hand, we can state such a 
theorem (cf. Arnold 1989: 75-77).

Suppose again we are given $N$ particles, and so a $3N$-dimensional 
configuration space $M$, which we equip with the line-element eq 
\ref{eq;Tdefinesline}. Let $S$ be an $n$-dimensional hypersurface of 
$M$; (so we imagine there are $k := £N - n$ holonomic constraints); 
let ${\bf q}_1$ be $n$ coordinates on $S$ and let ${\bf q}_2$ be $k$ 
coordinates in directions orthogonal to $S$. Let the potential energy 
have the form $V = V_0({\bf q}_1, {\bf q}_2) + C{\bf q}^2_2$. The idea 
is that we will let $C$ tend to infinity, to represent a stronger and 
stronger force constraining the system to stay on $S$. So consider the 
motion in $M$ according to eq \ref{eqn;lag} (with $3N$ coordinates) of 
a system with initial conditions at $t = 0$
\be
{\bf q}_1(0) = {\bf q}^0_1 \;\;\; {\dot {\bf q}}_1(0) = {\dot {\bf 
q}}^0_1
\;\;\; {\bf q}_2(0) = {\dot {\bf q}}_2(0) = 0 
\label{icsArnoldslimit}
\ee
Then as $C \rightarrow \infty$, a motion on $S$ is defined with the 
Lagrangian 
\be
L_* = T\mid_{{\bf q}_1 = {\bf q}_2 = 0} - V_0 \mid_{{\bf q}_2 = 0} .
\ee 
This result illustrates (Accept) and (Ideal).  This conception of a 
constrained system as a limit also plays a role in the equivalence of 
some analytical mechanical principles; cf. Section 4.2 and Arnold 
(1989: 91f.).

\paragraph{3.3.2.B  Integrating Lagrange's equations}\label{332B} I 
begin with three general comments about solving Lagrange's equations, 
in the usual form, eq \ref{eqn;lag}.

(1): {\em Differential equations on a manifold: velocity  phase 
space}:---\\
 The first point to stress is that the $n$ second-order differential 
equations eq. \ref{eqn;lag} are (despite the appearance of the partial 
differentials!) ordinary differential equations; equations  which are 
defined on a differential manifold, the constraint surface.\\
\indent In Paragraph 2.1.3.A's brief review of the theory of ordinary 
differential equations, I mentioned that the basic theorem about the 
local existence and uniqueness of solutions of (and local constants of 
the motion  for) first-order ordinary differential equations carried 
over to higher-order equations defined on a differential manifold. And 
Paragraph 3.3.2.E will give more details about the description of 
Lagrangian mechanics in terms of modern geometry, i.e. manifolds. But 
there are two important  points to make about this, which do not 
require any modern geometry.

\indent \indent (i): It is worth introducing jargon for the 
$2n$-dimensional space coordinatized by the $q$s and ${\dot q}$s taken 
together ($n$ is the number of configurational degrees of freedom). 
After all, the crucial function, the Lagrangian $L(q,{\dot q})$ is 
defined on this space. It is called {\em velocity phase space}. 
(Sometimes, it is called `phase space'; but this last term is more 
often used for the {\em momentum phase space} of Hamiltonian 
mechanics.) It is often denoted by  $TQ$: here $T$ stands for 
`tangent', not `time', for reasons given in Paragraph 3.3.2.E.\\
\indent Incidentally: In writing $L(q,{\dot q})$ and saying $L$ is 
defined on $TQ$, I have simplified by setting aside time-dependent 
potentials and rheonomous constraints. For treating them, there is 
again useful jargon. If $Q$ is a configuration space given 
independently of time, then the space $Q \times \mathR$, with points 
$(q,t)$, $t \in \mathR$ representing a time, is often called the {\em 
extended configuration space}. And the treatment of time-dependent 
potentials and-or rheonomous constraints might then proceed in  what 
can be called {\em extended velocity phase space} $TQ \times \mathR$.

\indent \indent (ii): As regards integrating Lagrange's equations:--- 
Recall  the idea from elementary calculus that $n$ second-order 
ordinary differential equations have a (locally) unique solution, once 
we are given $2n$ arbitrary constants.  This idea holds good for 
Lagrange's equations, even in the ``fancy setting'' of a manifold $TQ$ 
or $TQ \times \mathR$. And the $2n$ arbitrary constants can be given 
just as one would expect: as the initial configuration and generalized 
velocities $q_j(t_0), {\dot q}_j(t_0)$ at time $t_0$. Comments (2) and 
(3) expand a little on this.

(2): {\em The Hessian condition}:---\\
 Expanding the time derivatives in eq. \ref{eqn;lag} gives 
\be
\frac{\pl^2 L}{{\pl {\dot q}_k}{\pl {\dot q}_j}}{\ddot q}_k =
- \frac{\pl^2 L}{{\pl {q}_k}{\pl {\dot q}_j}}{\dot q}_k -
\frac{\pl^2 L}{{\pl t}{\pl {\dot q}_j}} + 
\frac{\pl L}{\pl {\dot q}_j} \;\; .
\label{elpara=texpand}
\ee
So the condition for being able to solve these equations to find the 
generalized accelerations at some initial time $t_0$, ${\ddot 
q}_j(t_0)$, in terms of $q_j(t_0), {\dot q}_j(t_0)$ is that the {\em 
Hessian} matrix $\frac{\pl^2 L}{{\pl {\dot q}_j}{\pl {\dot q}_k}}$ be 
nonsingular. Writing the determinant as $\mid \;\; \mid$, and partial 
derivatives as subscripts, the condition is that:
\be
\mid \frac{\pl^2 L}{{\pl {\dot q}_j}{\pl {\dot q}_k}} \mid \;\;\; 
\equiv \;\;\;
\mid L_{{\dot q}_j{\dot q}_k}\mid \;\;\; \neq \;\;\; 0 \;\;\;\;\; ;
\label{nonzerohessian}
\ee
This {\em Hessian condition} holds in very many mechanical problems; 
and henceforth, we impose it. Indeed it underpins most of what 
follows: it will also be the condition needed to define the {\em 
Legendre transformation}, by which we will pass from Lagrangian to 
Hamiltonian mechanics.

But I should also stress that the Hessian condition can fail, and does 
fail in important problems. The point  has been recognized since the 
time of Lagrange and Hamilton; though it was only in the mid-twentieth 
century, that Dirac, Bergmann and others developed a general framework 
for mechanics that avoided the Hessian condition. I shall make just 
three remarks about this: one mathematical, one physical and one 
terminological.\\
\indent (i): It is easy to show that the Hessian condition implies 
that $L$ cannot be homogeneous of the first degree in the ${\dot 
q}_i$. That is, $L$ cannot obey, for all $\lambda \in \mathR$: 
$L(q_i,\lambda{\dot q}_i,t) = \lambda L(q_i,{\dot q}_i,t)$. It is also 
easy to show that homogeneity of the first degree in the ${\dot q}_i$ 
for positive $\lambda$ is equivalent to an integral of $L$ (viz.,  the 
integral that is central to the calculus of variations: cf. Section 
\ref{ssecHami}) being independent of the choice of its integration 
variable (called being `parameter-independent').  (For details, cf. 
e.g. Lovelock and Rund (1975: Section 6.1).)     \\
\indent (ii): Some problems are naturally analysed using an $L$ that 
is homogeneous in this sense, and so has a parameter-independent 
integral.\\
\indent \indent Perhaps the best-known case occurs in Fermat's 
principle in geometric optics. It says, roughly speaking, that a light 
ray between spatial points $P_1$ and $P_2$ travels by the path that 
minimizes the time taken. If one expresses this  principle as 
minimizing an integral with time as the integration  variable, one is 
led to an integrand that is in general, e.g. for isotropic media, 
homogeneous of degree 1 in the velocities ${\dot q}_i$---conflicting 
with the Hessian condition eq. \ref{nonzerohessian}. So geometric 
optics usually proceeds by taking a spatial coordinate as the 
integration variable, i.e. the parameter along the light path. For 
details and references, cf. e.g. (Butterfield 2004c: Sections 5,7).\\
\indent \indent  But also in mechanics as against optics, there are 
cases of homogeneous $L$ and parameter-independence. This is 
especially true in relativistic theories---beyond this paper's scope! 
(Cf. Johns (2005: Part II) for a beautifully thorough account.)\\
\indent (iii): Beware of jargon. The framework of Dirac et al.  is 
called `constrained dynamics''---so be warned that this is a very 
different sense of `constraint' than ours.  

Of course, even with eq. \ref{nonzerohessian}, it is still usually 
hard {\em in practice} to solve for the ${\ddot q}_j(t_0)$: they are 
buried in the left hand side  of eq. \ref{elpara=texpand}.  This 
circumstance prompts the move to Hamiltonian mechanics, taken up in 
the companion paper. Meanwhile, the topic of the practical difficulty 
of solving equations prompts (3).

(3): {\em The ineluctable}:---\\
 I admit that from a very general viewpoint, Lagrange's equations 
represent no advance over the vectorial approach to mechanics. Namely: 
the dynamical problem of $n$ degrees of freedom is still expressed by 
$n$ second-order differential equations. Broadly speaking, this 
``size'' of the dynamical problem is an ineluctable consequence of 
Newton's second law being second-order in time. By and large, the most 
that a general scheme  can hope to do to reduce this ``size'' is:\\
\indent \indent (i) to provide help in finding and-or using new 
variables that simplify the problem; especially by reducing the number 
of equations to be solved, by some of the new variables dropping out 
(cf. (Reduce) and (Separate));\\
\indent \indent (ii) to make a useful trade-in of second-order 
equations for first-order equations.\\
\indent \indent As we shall see, Lagrangian mechanics does (i). 
(Hamiltonian mechanics does both (i) and (ii).) 
    
So much by way of general remarks about integrating Lagrange's 
equations.
I now turn to the practical advantages of using them to solve 
problems.  We can already see two substantial advantages---advantages 
that are valid for all holonomic systems (eq. \ref{eqn;lagforT}), not 
just those holonomic   systems which are monogenic with a 
velocity-independent work function (eq. \ref{eqn;lag}).

\paragraph{3.3.2.C Covariance; (Wider)}\label{332CCovarce}
The above deduction of Lagrange's equations, eq. \ref{eqn;lagforT} and 
\ref{eqn;lag}, makes it clear that they are covariant under any 
coordinate transformations (aka: {\em point-transformations}) $q_j 
\rightarrow q'_j$. (Of course, one can also prove this covariance 
directly i.e. assume the equations hold for the $q_j$, and assume some 
transformation $q_j \rightarrow q'_j$, and then prove they also hold 
for the $q'_j$.) This covariance means we can analyse a problem in 
whichever generalized coordinates we find convenient: whichever 
coordinates   we choose, we just  write down the Lagrangian in those 
coordinates  and then solve Lagrange's equations in the form eq. 
\ref{eqn;lagforT} or eq. \ref{eqn;lag}. This is one of our main 
illustrations of (Scheme), and its merit (Wider). (We will later see 
this covariance as an automatic consequence of a variational 
principle; cf. the end of Section \ref{ssecHami}.)

\paragraph{3.3.2.D One function; (Fewer)}\label{332DOnefunction}
In any such generalized coordinates, and for any number of particles 
(or generalized coordinates), the solution of the problem is encoded 
in a smaller number of functions than the number of degrees of freedom 
immediately suggests. In eq. \ref{eqn;lagforT}, the inertia is encoded 
in one function $T$ (cf. (a) in Section \ref{KEandwork}). And more 
remarkably, eq. \ref{eqn;lag} encodes the forces in one function $V$; 
besides, it encodes the solution to the problem in just one function, 
viz. $L := T - V$. This illustrates merit (Fewer) of (Scheme). This 
situation prompts a technical comment, and some philosophical remarks.

(1): {\em Equivalent Lagrangians}:---
The technical comment is that my phrase `one function' needs 
clarifying. To explain this,  let us consider just holonomic  
conservative systems, described by eq. \ref{eqn;lag}. It is easy to 
show that two Lagrangians $L_1$ and $L_2$ determine the very same 
equations eq. \ref{eqn;lag} if they differ by the time derivative of a 
function $G(q_j(t))$ of the generalized coordinates, i.e. if 
$L_1(q,{\dot q}) - L_2(q,{\dot q}) = \frac{d}{dt}\; G(q(t))$. Such 
Lagrangians are called {\em equivalent}.

\indent The converse is false: two inequivalent Lagrangians can yield 
the same equations of motion.  A two-dimensional harmonic oscillator 
gives an example. We met this system in Paragraph 2.1.3.B, with 
different frequencies in the two dimensions. Now we need only the 
special case of a common frequency. So the usual Lagrangian and its 
Lagrange equations are (with cartesian coordinates written as $q$s):
\be
L_1 =  \frac{1}{2}\left[{\dot q_1}^2 + {\dot q_2}^2 \; - \; \omega^2 
(q^2_1 + q^2_2) \right] \;\; ; \;\; {\ddot q}_i + \omega^2 q_i = 0 
\;\;, i = 1,2.
\label{Lag2DHO;eqns}
\ee
But the same Lagrange equations, i.e. the same dynamics, is given by
\be
L_2 = {\dot q_1}{\dot q_2} - \omega^2 q_1 q_2 \;\; ,
\label{OddLag2DHO}
\ee
which is not equivalent to $L_1$. This example is given by Jos\'{e} 
and Saletan (1998: 68, 103, 145), together with a proof that the 
converse results holds {\em locally}.\footnote{Thanks to Harvey Brown 
for alerting me to this example and Jos\'{e} and Saletan's discussion; 
and to two other uses of this example in connection Noether's theorem 
(Section \ref{Noetsubsubsec}).}

(2): {\em How the Lagrangian controls the motion}:---\\
Turning to philosophy: it is at first sight puzzling that the motion 
in 3-dimensional space of an arbitrary number of particles can be 
controlled by fewer functions than there are degrees of freedom: how 
so?\\
\indent Part of the answer is of course that the functions are defined 
not on physical space. In particular, $V$ is defined on configuration 
space $Q$ (or for a time-dependent potential on extended configuration 
space $Q \times \mathR$). And $L$ and $T$ are defined on the 
$2n$-dimensional velocity phase space $TQ$ with points $(q,{\dot q})$ 
(or again: on the extended velocity phase space $TQ \times \mathR$). 
So these functions encode properties of the entire system's 
configuration, or of the configuration taken together with the $n$ 
generalized velocities. And there is no difficulty in general about 
how a single function  on a higher-dimensional space might determine a 
motion in the space. After all, one could take the function's 
(higher-dimensional) gradient.\\
\indent But this is of course {\em not} how these functions determine 
the motion; (though incidentally, it is in effect how another 
function, the Hamiltonian, determines the motion in Hamiltonian 
mechanics). So the question arises how they do so, i.e. whether there 
is some notion---perhaps a geometric one, like taking the 
gradient---that underlies how these functions determine the motion, 
via eq.s \ref{eqn;lagforT} and \ref{eqn;lag}.   

The short answer is that there {\em is} such a notion: these equations 
reflect the fact that the dynamical laws (the determination of the 
motion) can be given a {\em variational formulation}. In particular, 
for holonomic  conservative systems (cf. eq. \ref{eqn;lag}): it turns 
out that when we consider $L$'s values not just at various times for 
the actual motion, but also for suitably similar {\em possible} 
motions, then the collection of all these values encodes the physical 
information that determines the motion.

But this short answer immediately invites two further questions, one 
philosophical and one technical.  The philosophical question is: `how 
can it be that {\em possible} values of a function such as $L$ 
determine {\em actual} motions?' As I mentioned at the end of  Section 
\ref{Moda}, I  address this issue at length elsewhere (2004e: Section 
5) and will not pursue it here. The technical question is (again, 
stated just for holonomic conservative systems): `how do $L$'s values 
for some merely possible motions determine the actual motion?' I will 
answer that in detail in Section \ref{Lagr}; (and Hamiltonian 
mechanics gives a deeper perspective on the answer).

\paragraph{3.3.2.E Geometric formulation}\label{332ELagEqGeom}
I turn to give a brief description of the elements of Lagrangian 
mechanics in terms of modern differential geometry. ({\em Warning: 
This Paragraph is not used later on.}) Here `elements' indicates 
that:\\
\indent (i): As this paper mostly eschews modern geometry, I will here 
assume without explanation various geometric notions, in particular: 
manifold, vector, one-form, metric, Lie derivative and tangent bundle. 
(But a reassurance: Section \ref{VecfieldsSymmies} gives some 
explanations of manifold, vector field and tangent bundle, which apply 
equally here.)\\
\indent  (ii): I make the simplifying assumptions that led to the 
usual form of Lagrange's equations eq. \ref{eqn;lag}: in particular, 
that the constraints are holonomic, scleronomous and ideal, and that 
the system is monogenic with a velocity-independent work-function. But 
much of the description below can be generalized in various ways to 
avoid these assumptions.\\
\indent (iii): I will also simplify by speaking ``globally, not 
locally''. For example, I will speak as if the relevant scalar 
functions, and vector fields and their integral curves, are defined on 
a whole manifold; when in fact all that Lagrangian mechanics can claim 
in application to most systems is a corresponding local statement---as 
we already know from Paragraph 2.1.3.A's report that differential 
equations are guaranteed  the existence and uniqueness only of a {\em 
local} solution.

Finally, a warning:--- Hitherto I have written $q_j, q_k$ etc. for the 
generalized coordinates. But in this Paragraph, I need to respect the 
distinction between contravariant and covariant indices (in more 
modern jargon: vectors and forms). So I write the coordinates as $q^i, 
q^j$ etc. Similarly, I will in this Paragraph, though  not elsewhere 
in the paper, adopt the  convention that repeated indices are summed 
over.

We begin by assuming that the configuration space (i.e. the constraint 
surface) is a manifold $Q$. So the kinetic energy $T$, being a 
homogeneous quadratic form in the generalized velocities (cf. 
discussion of eq. \ref{equation;Tingeneral}), defines a metric on 
$Q$.\\
\indent The physical state of the system, taken as a pair of 
configuration and generalized velocities, is represented by a point in 
the tangent bundle $TQ$. That is, writing $T_x$ for the tangent space 
at $x \in Q$, $TQ$ has points $(x, \tau), x \in Q, \tau \in T_x$;  so 
$TQ$ is a $2n$-dimensional manifold. As I said in Paragraph 3.3.2.B, 
$TQ$ is sometimes called {\em velocity phase space}. We will of course 
work with the natural coordinate systems on $TQ$ induced by  
coordinate systems $q$ on $Q$; i.e. with the $2n$ coordinates 
$(q,{\dot q}) \equiv (q^i, {\dot q}^i)$.   

The fundamental idea  is now that this tangent bundle is the arena for 
the geometric description of Lagrangian mechanics: in particular, the 
Lagrangian is a scalar function $L:TQ \rightarrow \mathR$ which 
``determines everything''. But I must admit at the outset that this 
involves limiting our discussion to  Lagrangians, and coordinate 
transformations, that are time-independent.\\
\indent More precisely: recall, first, the simplifying assumptions in 
(ii) above. Velocity-dependent potentials and-or  rheonomous 
constraints would  prompt one to use  the  extended configuration 
space $Q \times \mathR$, and-or the extended velocity phase space $TQ 
\times \mathR$.\\
\indent  So would  time-dependent coordinate  
transformations.\footnote{Thanks to Harvey Brown for urging this last 
limitation.} I admit that this last is  a considerable limitation from 
a philosophical viewpoint, since it excludes boosts, i.e. 
transformations to a coordinate system moving at constant velocity 
with respect to another; and boosts are central to the philosophical 
discussion of  spacetime symmetry groups, and especially of  
relativity principles. To give the simplest example: the Lagrangian of 
a free particle in one spatial dimension is just its kinetic energy, 
i.e. in cartesian coordinates $\frac{1}{2}m{\dot x}^2$. Under a boost 
with velocity $v$ in the $x$-direction to another cartesian coordinate 
system, $x \mapsto x' := x - vt$; i.e. the point labelled $x$ in the 
first system is labelled by $x - vt$ in the second system (assuming 
the two spatial coordinate systems coincide when $t=0$). For example, 
if $v = 5$ metres per second, the point first labelled $x = 10$ metres 
is labelled by the second system at time $t = 2$ seconds  as $10 - 5.2 
= 0$, i.e. as the origin. So ${\dot x}' = {\dot x} - v$, and the  
Lagrangian, i.e. the kinetic energy, is not invariant under the boost: 
by choosing $v$ equal to the velocity of the particle in the 
$x$-direction, one can even make the particle have zero energy. (I 
shall return to the topic of transformations under which the 
Lagrangian  is invariant, though again with my limitation to $TQ$, 
when presenting Noether's theorem; Section \ref{NoetherLag}, cf. 
especially Section \ref{VecfieldsSymmies}).

But setting aside these {\em caveats}, I now describe Lagrangian 
mechanics on $TQ$, with four comments.

\indent  (1): {\em  $2n$ first order equations; the Hessian 
again}:---\\
The Lagrangian equations  of motion (in the natural coordinates $(q, 
{\dot q})$) are now $2n$ {\em first-order} equations for the functions 
$q^i(t), {\dot q}^i(t)$, determined by the scalar function $L:TQ 
\rightarrow \mathR$. The $2n$ equations fall in to two groups: 
namely\\
\indent \indent (a) the $n$ equations eq. \ref{elpara=texpand}, with 
the ${\ddot q}^i$ taken as the time derivatives of ${\dot q}^i$ with 
respect to $t$; i.e. we envisage using the Hessian condition eq. 
\ref{nonzerohessian} to solve eq. \ref{elpara=texpand} for the ${\ddot 
q}^i$, hard though this usually is to do in practice;\\
\indent \indent (b) the $n$ equations ${\dot q}^i = \frac{d q^i}{dt}$.

(2): {\em Vector fields and solutions}:---\\
\indent \indent (a): These $2n$ first-order equations are equivalent 
to a vector field on $TQ$. This vector field is called the `dynamical 
vector field', or for short the `dynamics'. I write it as $D$ (to 
distinguish it from the generic vector field $X,Y,...$). So the 
solutions are integral curves of $D$.\\  
\indent\indent (b): In the natural coordinates $ (q^i, {\dot q}^i)$, 
the vector field $D$ is expressed as
\be
D = {\dot q}^i \frac{\pl}{\pl q^i} + {\ddot q}^i \frac{\pl}{\pl {\dot 
q}^i} \;\; ;
\label{LagDelta}
\ee
and the rate of change of any dynamical variable $f$, taken as a 
scalar function on $TQ$, $f(q,{\dot q}) \in \mathR$ is given by
\be
 \frac{df}{dt} = {\dot q}^i \frac{\pl f}{\pl q^i} + {\ddot q}^i 
\frac{\pl f}{\pl {\dot q}^i} = D(f) .
\label{LagDeltaforf}
\ee   
\indent\indent (c): Again, the fundamental idea of the Lagrangian  
framework is that  the Lagrangian $L$ ``determines everything''. In 
particular, it determines: the dynamical vector field $D$, and so (for 
given initial $q,{\dot q}$) a solution, a trajectory in $TQ$, $2n$ 
functions of time $q(t), {\dot q}(t)$ (with the first $n$ functions 
determining the latter).\\
\indent\indent  (d): The (local) existence and uniqueness of solutions 
to sets of first-order equations means not just that initial 
conditions $q^i(t_0), {\dot q}^i(t_0)$ determine a unique solution; 
but this solution is now a curve (parametrized by time) in $TQ$. This 
separation of solutions/trajectories within $TQ$ is important for the 
visual and qualitative understanding of solutions.\\
\indent\indent  (e): I stress again that if the Hessian condition eq. 
\ref{nonzerohessian} fails, the (local) existence and uniqueness of 
solutions can fail: there can be points in $TQ$ through which no 
solution, or more than one, passes.  

(3): {\em Geometric formulation of Lagrange's equations}:---\\
We can formulate Lagrange's equations in a coordinate-independent way, 
by using three ingredients. Namely: $L$ itself (a scalar, so 
coordinate-independent);  the vector field $D$ that $L$ defines; and 
the one-form on $TQ$ defined by $L$ (locally, and in terms of the 
natural coordinates $(q^i,{\dot q}^i)$) by
\be
\theta_L := \frac{\pl L}{\pl {\dot q}^i}dq^i \;\;.
\label{defpdq}
\ee
(This one-form takes a central role in Hamiltonian mechanics, where it 
is called the {\em canonical one-form}.)

The  Lie derivative of $\theta_L$ along the vector field  $D$ on $TQ$ 
defined by $L$ is, by the Leibniz rule:
\be
{\cal L}_{D} \theta_L =   ({\cal L}_{D} \frac{\pl L}{\pl {\dot q}^i} ) 
dq^i +  \frac{\pl L}{\pl {\dot q}^i}{\cal L}_{D}(dq^i) \;\;.
\label{LieOfpdq}
\ee
But the Lie derivative of any scalar function $f:TQ \rightarrow 
\mathR$ along any vector field $X$ is just $X(f)$; and for the 
dynamical vector field $D$, this is just ${\dot f} = \frac{\pl f}{\pl 
q^i}{\dot q}^i + \frac{\pl f}{\pl {\dot q}^i}{\ddot q}^i$. So we have
\be
{\cal L}_{D} \theta_L =   (\frac{d}{dt} \frac{\pl L}{\pl {\dot q}^i} ) 
d q^i +  \frac{\pl L}{\pl {\dot q}^i} d {\dot q}^i  \;\; .
\label{LieOfpdqdbydt}
\ee
Rewriting the first term by the Lagrange equations, we get
\be
{\cal L}_{D} \theta_L =   (\frac{\pl L}{\pl {q}^i} ) dq^i +  \frac{\pl 
L}{\pl {\dot q}^i}d{\dot q}^i \equiv dL \;\; .
\label{coordindpdtLag}
\ee
We can conversely deduce the familiar Lagrange equations from eq. 
\ref{coordindpdtLag}, by taking coordinates. So we conclude that these 
equations'  coordinate-independent form is:\\
\be
{\cal L}_{D} \theta_L =  dL \;\; .
\label{coordindpdtLag2}
\ee

\indent  (4): {\em Limitations}:---\\
 Finally, a comment about the  Lagrangian framework's limitations as 
regards solving problems, and how they prompt the transition to 
Hamiltonian mechanics.\\
\indent Recall the remark at the end of Paragraph 3.3.2.B (2), that 
the $n$ equations  eq. \ref{elpara=texpand} are in general hard to 
solve for the ${\ddot q}^i(t_0)$: they lie buried in the left hand 
side  of eq. \ref{elpara=texpand}. On the other hand, the $n$ 
equations ${\dot q}^i = \frac{d q^i}{dt}$ (the second group of $n$ 
equations  in (1) above) are as simple as can be. \\
\indent This makes it natural to seek another $2n$-dimensional  space 
of variables, $\xi^{\al}$ say ($\al=1,...,2n$), in which:\\
\indent \indent (i): a motion is described by first-order equations, 
so that we have the same advantage as in $TQ$ that a unique trajectory 
passes through each point of the space; but in which\\
\indent \indent  (ii): all $2n$ equations have the simple form 
$\frac{d \xi^{\al}}{dt} = f_{\al}(\xi^1,...\xi^{2n})$ for some set of 
functions $f_{\al} (\al=1,...,2n)$.\\
\indent \indent Indeed, Hamiltonian mechanics provides exactly such a 
space: viz.,  the cotangent bundle of the configuration manifold, 
instead of its tangent bundle. 

\section{Lagrangian mechanics: variational principles and reduction of 
problems}
\label{Lagr}
\paragraph{4.0 Preamble}\label{PreambleLagMechs}
This Section will begin with conceptual discussion, and then move to 
more  technical matters. I will first introduce the two main 
variational principles of Lagrangian mechanics: the principle of least 
action (understood as it was by Lagrange and Euler), and Hamilton's 
Principle; (Section \ref{ssecPrel}).\\
\indent Beware: Hamilton's Principle is often called a (or even: the) 
least action principle. Indeed, a more general warning is in order. 
`Action' has, unfortunately, various meanings; there is no agreed and 
exact usage, though it always has the dimension of momentum $\times$ 
length = energy $\times$ time. In this Section, action will tend to 
mean the integral with respect to time, along a possible  history or 
trajectory of the system in configuration space, of a quantity with 
the dimension of energy: $\int E \; dt$.\\
\indent (But the companion paper will give increasing prominence to:\\
$\;\;$ (i) the integral (not always along a possible  history of the 
system!) of a momentum $p$ with respect to length $\int p \; dq$; or 
more generally, summing over degrees of freedom $\int \Sigma_i \; p_i 
\; dq_i$;\footnote{$\int \Sigma_i \; p_i \; dq_i$ is the canonical 
one-form, which is central to Hamiltonian mechanics.} and\\
$\;\;$  (ii) the integral (again, not always along a history) of the 
{\em difference}, $\int \Sigma_i \; p_i \; dq_i - E \; dt$.)

After Section \ref{ssecPrel}, Hamilton's Principle takes centre-stage. 
Its technical features are reported in  Sections \ref{ssecHami}, 
\ref{ssecExte}. Then the rest of the Section is dominated by the theme 
of symmetry: especially,  how a symmetry can help reduce the number of 
variables of a problem---again, the merit (Reduce).\\
\indent First, I introduce generalized momenta in the context of the 
conservation of energy  (Section \ref{ssecGeneMom}). Section 
\ref{ssecCycl} begins with the simple but important result that the 
generalized  momentum of any cyclic coordinate is a constant of the 
motion, and so reduces the dimension of the dynamical system by one. 
The rest of the Section develops this result in two ways.\\
\indent (1): First, I describe the method of Routhian reduction. This 
leads to Section \ref{ssecTime}'s explanation of how, starting from 
Hamilton's Principle,  Routhian reduction applied to time as a cyclic 
coordinate  recovers Euler's and Lagrange's principle of least action. 
The discussion will also cover another famous variational  principle 
of Lagrangian  mechanics, Jacobi's principle.\\
\indent (2): Finally in Section \ref{NoetherLag}, I describe Noether's 
theorem, which provides a  powerful general perspective on symmetry.\\
\indent (All these aspects of this Section's discussion of symmetry 
will have analogues, and further developments, in Hamiltonian 
mechanics.) 

As regards my four morals, this Section will illustrate all of them. 
But the  main morals will be prominent:\\
\indent (i): the four merits of (Scheme); i.e. (Fewer), (Wider), 
(Reduce) and (Separate); as just discussed, (Reduce) will be 
especially prominent in connection with symmetries.\\
\indent (ii): all three grades of (Modality); but especially the third 
grade, (Modality;3rd), which involves considering possibilities that 
violate the actual laws.

\subsection{Two variational principles introduced}\label{ssecPrel} 
Analytical mechanics contains many variational principles, which are 
closely related (and in some cases equivalent) to one another. 
But I will focus on just two principles, and their relationship to 
each other: Euler's and Lagrange's ``principle of least action''; and 
Hamilton's Principle.\footnote{The history of the principles' 
discovery and evolution is fascinating: in particular, Lagrange 
himself worked with Hamilton's Principle---the name was coined by 
Jacobi, and only became prevalent in the 20th century. But I will not 
go into this history.}  In this Subsection, I introduce them without 
technicalities. In particular, I will present them as using two 
different kinds of `variation of a path': a distinction that I will 
gloss  philosophically in terms of (Modality)'s three grades of modal 
involvement.

\subsubsection{Euler and Lagrange}\label{ELsPLA}
 For simplicity, let us consider a single  point-particle in a 
time-independent potential $V= V(\rr)$; in short, a conservative 
one-particle system. (We will remark later that the principle in fact 
applies much more widely.) Suppose one is given the initial conditions 
that the particle is at spatial point $P_1$ at time $t_1$, with a 
given total energy $E = T + V(P_1)$ compatible with the value of 
$V(P_1)$, i.e. $E \geq V(P_1)$. Then for any spatial path $\gamma$ 
starting from $P_1$, the initial conditions, together with the 
conservation of energy, determine the particle's motion over the next 
time-step, if we assume that it must start out travelling along 
$\gamma$. For  the initial value of $T$ and the specification of 
$\gamma$ determine an initial velocity. And so they determine at which 
point $P'$ along $\gamma$ the particle will be at time $t_1 + dt$ i.e. 
an infinitesimal time-step later. Furthermore, the conservation of 
energy, and value of $V(P')$ determine (by $T = E - V(P')$) what the 
speed of the particle  is at the time $t_1 + dt$. The argument can be 
iterated. That is: if we  assume that also at time $t_1 + dt$ the 
particle must continue to travel along our chosen path $\gamma$ , then 
its motion over the next time-step is determined---and so 
on.\footnote{By the way: the argument so far clearly also works for 
$V$ a prescribed [i.e. independent of the particle] function of time. 
But the principle of least action , to follow, requires $V$ 
independent of time.}  

Of course, we chose $\gamma$ arbitrarily; and so (since $V$ is given) 
we were almost certainly {\em wrong} to suppose that the particle must 
follow $\gamma$, even assuming it starts out along $\gamma$ at time 
$t_1$. That is: the imagined motion is not just counterfactual but 
contralegal: it violates the dynamical laws (i.e. Newton's equations).  
But Euler and Lagrange discovered that:  for this system, these laws 
are equivalent to a statement about a whole class of possible paths 
through $P_1$. To formulate this statement, first note that the 
previous paragraph also shows: The initial conditions and the 
requirement of energy conservation at all times also determine, for 
any time-interval $[t_1,t_2]$, the time-integral of $T$ along the path 
$\gamma$. And similarly, for any other possible path, $\eta$ say, 
through $P_1$: the initial conditions, in particular the initial 
energy, and the requirement of energy conservation at all times 
determine, what the time-integral of $T$ along $\eta$ would be if the 
particle were to traverse $\eta$.

We can now state Euler's and Lagrange's principle of least action, for 
a single particle. The idea is as follows: Given\\
\indent  (i): the initial conditions that the particle is at spatial 
point $P_1$ at time $t_1$, with a given velocity ${\bf v}$ (which 
fixes the total energy $E = T + V(P_1)$); and given also\\
\indent (ii): the particle later passes through the point $P_2$ (i.e. 
one assumes that $P_2$ lies on the particle's actual spatial path,, as 
determined by the dynamical laws and the given potential $V$): it 
follows that\\
\indent (iii): the actual path traversed will be that path among all 
possible paths connecting $P_1$ to $P_2$, motion along which, with a 
common fixed initial energy $E$, makes the time-integral of $T$ along 
the path, a minimum.

That is the idea. But (iii) needs amendment. For the  actual path 
might not make the integral a minimum: even in comparison with just 
the class of all paths close to the actual one, rather than all 
possible paths. The precise statement of the principle is rather that 
the actual path makes the integral {\em stationary} in comparison with  
all sufficiently close paths.\\
\indent Here `stationarity' means that a derivative is zero. The 
details are made precise in the calculus of variations, and reviewed 
in Section \ref{ssecHami}. For the moment, we only need the idea  
that, as in elementary calculus, a zero  derivative is compatible, not 
only with a minimum of the function in question, but also with a 
maximum or a turning-point of it. (We similarly replaced minimality by 
stationarity at the end of Section \ref{PVWIntrodd}.)

\indent This distinction, between minimization and stationarity of a 
function, has both a historical and a terminological 
significance---and both points will apply just as much to Hamilton's 
Principle as to the principle of least action.\\
\indent Historically, some early advocates of the principle of least 
action asserted that the actual path minimized the integral. Besides, 
this was regarded  as a remarkable ``efficiency'' or ``economy'' on 
the part of Nature---and as suggesting a proof of the existence of 
God. The main example of this tendency is Maupertuis, who announced 
(an obscure form of) the principle, claiming minimization, in 1744. 
But already in the same year, Euler published his ground-breaking 
treatise on the calculus of variations, in an addendum of which the 
principle is expounded as a precise theorem. (For details, cf. Fraser 
(1994); or more briefly, Kline (1972: 577-582), Yourgrau and 
Mandelstam (1979: 19-29).) In due course, it became clear that only 
the stationarity version of the principle held good; and similarly, 
that Hamilton's Principle (Section \ref{HamsPrinc}) was a matter of 
stationarity, not of minimization.  Thus for example, Hamilton in 1833 
criticized the idea of minimization, noting that `the quantity 
pretended to be economized is in fact often lavishly expended': so he 
preferred to speak of a `principle of stationary action'.\\
\indent This distinction  also raises a significant mathematical 
question, regardless of mechanics: what are the necessary  or 
sufficient conditions, for problems in the calculus of variations, of 
securing a minimum rather than just stationarity? Though the 
investigation of this question has a long and distinguished history 
(starting essentially with Legendre in 1786), and a good deal is now 
known about it, we will not need any of these details. (For the 
history, cf. Kline (1972: 589-590, 745-749); for the technicalities, 
cf. e.g. Courant and D. Hilbert (1953: 214-216), Fox (1987: Chapters 
2, 9).)   \\
\indent Finally, this distinction also has a terminological aspect. It 
is easier to say `minimize a function' or `extremize a function' 
(where `extremize' means `minimize or maximize'), than to say `render 
a function stationary': there is no English word `stationarize'! So 
despite the remarks above,  I shall from now on (for Hamilton's 
Principle as well as the principle of least action) usually say 
`minimize' or `extremize': in fact, this is a widespread practice in 
the textbooks. But `stationarize' is to be understood!

This principle of least action for a single particle is a remarkable 
principle---and a fine example of (Reformulate). But much more is 
true: the principle can be extended to systems with an arbitrary 
number $N$ of particles. Here, I do not just mean the trivial $N$-fold 
conjunction of the one-particle principle, which follows immediately 
if we assume no particle-particle interactions (and no collisions in 
the time-period).\footnote{That is:--- Given  a system of  $N$ 
point-particles, at spatial points $P_{1_1},\dots,P_{1_N}$ at time 
$t_1$, subject to an external potential $V({\bf r})$, and with given 
initial velocities ${\bf v}_1,\dots, {\bf v}_N$ (and so initial 
kinetic energies $T_1(t_1),\dots,T_N(t_1)$); and given $N$ points 
$P_{2_1},\dots,P_{2_N}$ (mutually distinct and in general distinct 
from $P_{1_1},\dots,P_{1_N}$), through which, respectively, the $N$ 
particles later pass (with no collisions):
 for each particle $i$, the actual path traversed will be the path  
connecting $P_{i_1}$ to $P_{i_2}$, motion along which, with a fixed 
initial energy $E_i = T_i(t_1) - V(P_{i_1})$ common to all comparison 
paths, makes the time-integral of $T_i$, along the path, stationary.} 
I mean, rather, that for any system (i) which is conservative, and 
(ii) for which the constraints, if any, are ideal, holonomic and 
scleronomous (so that we can work in the constraint surface): a 
corresponding principle  holds.

 This will be stated very precisely in Section \ref{ssecTime}. For the 
moment, we only need the main idea, that:
\begin{quote} 
For such a system, whatever the details of the interactions between 
its parts (encoded in $V$), the representative point in configuration 
space moves along the curve between given initial and final 
configurations that makes stationary (in comparision with neighbouring 
curves) the time-integral along the curve of the total kinetic energy 
$T$.
\end{quote}
This is a very striking, even amazing, principle; both technically and 
philosophically. Technically, one can apply the calculus of variations 
to deduce the corresponding Euler-Lagrange equations; (Section 
\ref{ssecHami} gives more explanation). Indeed, Euler and Lagrange did 
just this, obtaining the correct equations of motion for conservative 
systems that, if constrained, have ideal, holonomic and scleronomous 
constraints.

As regards philosophy, there are three immediate comments.  The first 
concerns my morals; the other two are more general, and have a 
historical aspect.

(1): {\em Morals}:---\\
 Clearly, the principle is a fine illustration of (Scheme), and the 
merit (Fewer). (Once Section \ref{ssecHami} connects it with the 
Euler-Lagrange equations, we shall also see the other merits, (Wider) 
etc.; and the moral (Reformulate).)

\indent The principle is also a fine illustration of (Modality). 
Recall that Section \ref{Moda} distinguished three broad grades of 
modal involvement:  (Modality;1st) to (Modality;3rd).  In 
(Modality;1st) we keep fixed the problem and laws of motion, but vary 
the initial and/or final conditions;\footnote{In some usages of the 
word `problem', varying the initial conditions would count as varying 
the problem. But not mine: in Section \ref{Moda}, I stipulated that a 
problem is specified by the number of degrees of freedom and the 
forces involved, here coded in the potential  function; (and more 
generally in a Lagrangian or Hamiltonian).} while in (Modality;2nd) we 
consider various problems but again keep fixed the laws of motion; and 
in (Modality;3rd) we vary the laws of motion in the sense that we 
consider histories that violate the actual laws (for the given forces, 
i.e. problem).

Broadly speaking, the principle of least action illustrates all three 
grades, though  there is a minor wrinkle about (Modality;1st).\\
\indent Applying the principle  to a given problem obviously involves 
(Modality3rd): most of the various counterfactual histories, along the 
paths not traversed, are contralegal. And the principle itself clearly 
involves (Modality;2nd), since it generalizes across a whole class of 
problems.

\indent The wrinkle about (Modality;1st) is that, despite the variety 
of positions and speeds at intermediate times (i.e. times after $t_1$ 
but before arrival at $P_2$), the principle does {\em not} vary the 
initial or final conditions in the sense of position and {\em speed}. 
The reason is that\\
\indent \indent (i):  the given initial position $P_1$, and so 
$V(P_1)$, and velocity ${\bf v}$ determine a total energy $E$; 
(indeed, to determine $E$, one needs only $V(P_1)$ and the speed $v$); 
and\\
\indent \indent (ii): the conservation of the energy $E$, together 
with the given final position $P_2$, and so $V(P_2)$, determines  the 
speed with which the particle would arrive at $P_2$ along any 
path---whether the unique dynamically allowed one, or another one.\\
\indent Nevertheless the principle illustrates (Modality;1st). For the 
variety of paths makes  for a variety of (both initial and final) 
velocity or momentum, though not of speed---and that variety counts as 
varying the initial and final conditions.

(2): {\em Other formulations}:---\\
 Historically, the principle for a single particle was of course  
formulated first; and it was often formulated in terms of minimizing 
the integral along the path of the particle of the momentum $m{\bf 
v}$, with distance $s$ as the integration variable. (Thus wrote Euler 
in 1744.) It was also formulated in terms of the integral of twice the 
kinetic energy $2T$, with time as the integration variable. These 
alternative formulations are trivially equivalent to the 
single-particle principle  above. But they are historically important, 
because of the role they played in discussions of the relative  
dynamical importance of momentum and kinetic energy; in particular, in 
the controversy about {\em vis viva}, which was in effect defined as 
$2T$. I shall not need the first alternative, using $\int m{\bf v}\; 
ds$, at all. But in Section \ref{ssecTime}, I shall recover the second 
(with $2T$) from Hamilton's Principle. 

(3): {\em Teleology foresworn}:---\\
 The principle's reference to the final configuration suggests  
teleology and final causes: that the values attained by the integral 
at the end of various possible trajectories through configuration  
space somehow determines (or even: causes or explains) {\em from the 
start} which trajectory is traversed. (This is clearest for our first 
case, the single particle: it looks as if the particle's final 
position determines which path it takes to that position.)\\
\indent This suggestion is of course  not peculiar to this principle. 
It arises for any variational principle that fixes a final condition 
(in time), and so it is endemic in analytical mechanics. In 
particular, we will see that Hamilton's Principle similarly refers to 
the final configuration of the system. Accordingly, this aspect of 
analytical mechanics has been much discussed. Indeed, teleology was 
the dominant topic in  philosophical  discussion of variational 
principles, from the beginnings in the eighteenth century (with 
Maupertuis' theological arguments, mentioned above), to Planck's 
advocacy of them in the late nineteenth century: a dominance which 
helps explain the logical empiricists' strong rejection of variational 
principles. (For discussion and references, cf. Yourgrau and  
Mandelstam (1979: 163-165, 173-175) and St\"{o}ltzner (2003).) 

But I shall not pursue this topic, for two reasons. First, it would 
involve discussion of causation and explanation---large subjects 
beyond my scope. Second and more important: there is a strong reason 
to reject an interpretation of the principle, and of other 
final-condition variational  principles in analytical mechanics, in 
terms of final causes---as against one in terms of efficient causes. 
This reason is based on the formalism of mechanics, and does not 
depend on any general philosophical  objections to final causes.\\
\indent Namely: given such a principle, the calculus of variations 
deduces a set of differential equations, the corresponding 
Euler-Lagrange equations, which are (in the cases we will consider) 
{\em equivalent} to the principle; and these equations suggest an 
interpretation in terms of efficient causes. In particular, we will 
see that Lagrange's equations in their most familiar form, eq. 
\ref{eqn;lag}, are equivalent to Hamilton's Principle for a holonomic 
and monogenic system. Recalling that these equations are $n$ 
second-order equations (for $n$ degrees of freedom), and so need $n$ 
positions and velocities as initial conditions (just like Newton's 
equations), we can surely regard the initial conditions, together with 
the equations, as determining (or causing or explaining) which 
trajectory in configuration space  is traversed; (and so in 
particular, the value of the integral of $T$).\\
\indent This interpretation is surely just as good as the one in terms 
of final causes. Besides, we can similarly defend an efficient-cause 
interpretation for principles other than the principle of least action 
and Hamilton's; (though I will not go into details).\footnote{I claim 
no originality for appealing to the Euler-Lagrange equations to 
suggest an efficient-cause interpretation of variational principles, 
as a reply to the proposed final-cause interpretations. This appeal is 
common enough; e.g. Torretti (1999: 92). But so far as I know, my 
emphasis  on modality, here and in (2004e), is novel.}

To sum up this Subsection: Euler and Lagrange discovered that for any 
conservative system that, if constrained, has ideal, holonomic and 
scleronomous constraints: two functions, $T$ and $V$, determine the 
motion of the system, by a principle that selects the actual motion 
from a whole class of conceivable motions. Note that the motions in 
the class have a common fixed start-time $t_1$, common initial and 
final configurations, and a common fixed energy (and so a common 
$T(t_1)$). So the times of arrival at the final configuration {\em 
vary}. This will {\em not} be so for Hamilton's Principle. 

\subsubsection{Hamilton}\label{HamsPrinc}
 Hamilton's Principle replaces Euler and Lagrange's ``common-energy, 
varying arrival times'' variation, with a variation that has ``varying 
energies, common arrival time''. But as for Euler and Lagrange, the 
variations have fixed initial and final configurations (i.e. spatial 
positions $P_1, P_2$ for one particle, and $\{P_{i_1}; P_{i_2}\}$ for 
$N$ particles).

With this kind of variation, it turns out that for many kinds of 
system, just {\em one} function  determines the motion, again {\em 
via} a  variational principle that selects the actual motion by 
comparison with a class of ``nearby'' motions. In particular, for a 
conservative system that, if constrained, has ideal and holonomic 
constraints, the function is the Lagrangian, the  difference $L := T - 
V$ of the total kinetic and potential energies of the system.\\
\indent That is: Hamilton's Principle for such a system, consisting of 
$N$ particles, says:
\begin{quote}
The actual motion between a configuration $\{P_{i_1}\}$ at time $t_1$ 
and $\{P_{i_2}\}$ at time $t_2$ will be the motion that makes 
stationary the time-integral, along the trajectory in configuration 
space, of $L := T - V$.
\end{quote}
This calls for two immediate comments. The first concerns my morals; 
the second is about Hamilton's Principle's advantages over the 
principle of least action.  

(1): {\em Morals}:---\\
 Like the principle of least action, Hamilton's principle is a fine 
illustration of (Scheme), and (Fewer); and we will later see the other 
merits, (Wider) etc. \\
\indent And again like the principle of least action, Hamilton's 
Principle illustrates all three grades of modal involvement. But the 
details about (Modality;1st) and (Modality;3rd) are a bit different 
from the case of the principle of least action, because now the 
various histories have varying energies and a common arrival time. 
There are two points here.\\
\indent First: even for a single problem (specified by a number of 
degrees of freedom and the forces), the counterfactual histories have 
to have  different energies one from another, in order for them to 
have a common arrival time; e.g. a constituent particle could have 
different initial speeds in two histories. This illustrates 
(Modality;1st) and (Modality;3rd), as the principle of least action 
did. But in so far as one finds it unnatural in counterfactual 
suppositions to fix some future actual fact (arrival time), and 
thereby have to counterfactually vary present facts (energy, 
speed)---rather than {\em vice versa}---one will find Hamilton's 
Principle's  (Modality;1st) more ``radical'' than the principle of 
least action's. \\
\indent Second: for Hamilton's Principle, the energy is in general  
not preserved (constant) {\em within} a counterfactual history---even 
for a system that actually obeys the conservation of energy, i.e. a 
conservative scleronomous system. Indeed, for any problem with 
prescribed initial and final conditions: almost all (in a natural 
measure) of the histories considered will violate energy-conservation. 
This is because Hamilton's principle considers all smooth curves in 
configuration space that are close to the actual trajectory; (details 
in Section \ref{ssecHami}).
 
(1): {\em Advantages}:---\\
 I emphasise the three main advantages of Hamilton's Principle over 
Euler and Lagrange's principle of least action. In ascending order of 
importance, they are:--\\
\indent (i): It encompasses the principle of least action, in two 
senses. First, it immediately yields the Euler-Lagrange equations, eq. 
\ref{eqn;lag}---describing conservative systems, whose constraints, if 
any, are ideal, holonomic and scleronomous---that the principle of 
least action also obtains; cf. Section \ref{ssecHami}. Second, it {\em 
explains} how the principle of least action obtains those equations; 
cf. Section \ref{ssecTime}.\\  
\indent (ii) It can be extended to other kinds of system, even some 
non-holonomic systems. For some of these kinds, it again uses as its 
integrand $L = T - V$ (Section \ref{ssecHami}); for others, it uses an 
analogous integrand; (cf. Section \ref{ssecExte}). \\
\indent (iii) It leads to Hamilton's equations, and thereby to 
Hamiltonian mechanics; which have various advantages over Lagrange's 
equations, and indeed Lagrangian mechanics; cf. the companion paper.

The first two advantages will be spelt out in the sequel. I begin with 
an exact statement of Hamilton's Principle. 

\subsection{Hamilton's Principle for monogenic holonomic  
systems}\label{ssecHami}
{\em Beware}: this Section's title is somewhat misleading, for two 
reasons. First, the discussion is as usual restricted to ideal 
constraints, not just to monogenic and holonomic systems.\\
\indent Second and more important: `monogenic' is a slight misnomer. 
For one of my main points will be that Hamilton's Principle with $L = 
T - V$ as integrand is equivalent to Lagrange's equations in the 
familiar form eq \ref{eqn;lag}. As discussed in Section 
\ref{ssecDale}, these equations are physically correct, i.e. follow 
from d'Alembert's principle, not only when the potential $V$ is  
time-independent and velocity-independent (i.e. conservative systems), 
or when $V$ is time-dependent but velocity-independent, but {\em also} 
in some cases of velocity-dependence, e.g. when eq. 
\ref{QfromvelydepdtV} holds. Such conditions are a mouthful to say; 
and this Section's title abbreviates that mouthful in the somewhat 
more general word `monogenic'. But this inaccuracy will be harmless.

Thus understood, Hamilton's Principle for a monogenic holonomic system 
says:--- 
\begin{quote}
The motion in configuration space between prescribed configurations
at time $t_1$ and time $t_2$ makes stationary the line integral
\be
I = \int^{t_2}_{t_1} L(q_1,\dots,q_n,{\dot q_1},\dots,{\dot q_n}, t) 
\; dt
\ee 
of the Lagrangian $L:= T - V$. (Note the inclusion of time $t$ as an 
argument to allow $V$ to be time-dependent.)
\end{quote}
Hamilton's Principle (in this form) is a necessary and sufficient 
condition for Lagrange's equations, i.e. eq. \ref{eqn;lag}: ((Scheme) 
with merits (Fewer) and (Wider), again). I will not prove necessity; 
(for this, cf. e.g. Whittaker (1959: Section 99: 245-247) and Lanczos 
(1986: 58-59, 116)). But I show sufficiency, since the argument:\\
\indent (i) simply applies the basic result of the calculus of 
variations, that the unconstrained stationarity of an integral 
requires the Euler-Lagrange equations; and\\
\indent  (ii) makes clear that the Principle involves (Modality;3rd), 
as announced in Section \ref{HamsPrinc}.

{\em The basic result of the calculus of variations}:---\\
 The variation, with fixed end-points, of an integral 
\be
J =  \int^{x_2}_{x_1} f(y_1(x),\dots,y_n(x),{\dot y_1(x)},\dots,{\dot 
y_n}(x),x) \; dx
\ee
(where the dot ${\dot {}}$ indicates differentiation with respect to 
$x$) is obtained by considering $J$ as a function of a parameter 
$\alpha$ which labels the possible curves $y_i(x,\alpha)$. We take 
$y_1(x,0), y_2(x,0), \dots$ as the solutions of the stationarity 
problem; and we let $\eta_1(x), \eta_2(x),\dots$ be arbitrary 
functions except that they vanish at the end-points, i.e. $\eta_i(x_1) 
= \eta_i(x_2) = 0 \;\; \forall i$. Then we write:
\be
y_i(x,\alpha) = y_i(x,0) + \alpha\eta_i(x) \;\; \forall i.
\ee
(Such variations can be analysed using the idea of a functional 
derivative, and for infinite i.e. continuous systems they need to be; 
but I shall not need that idea.)

The condition for stationarity is then that
\be
\left(\frac{\partial J}{\partial \alpha}\right)_{\alpha = 0} = \; 0;
\ee  
and the variation of $J$ is given in terms of that of $\alpha$ by
\be
\dd J = \frac{\partial J}{\partial \alpha}d\alpha = \int^{x_2}_{x_1} 
\Sigma_i \left(\frac{\partial f}{\partial y_i}\frac{\partial 
y_i}{\partial \alpha}d\alpha + \frac{\partial f}{\partial {\dot 
y_i}}\frac{\partial {\dot y_i} }{\partial \alpha}d\alpha \right) \; 
dx.
\label{eq;basiccalcvar}
\ee
Integrating by parts, for each $i$, the second term in the integrand, 
and using  the fact that the variations $\dd y_i = (\frac{\partial 
y_i}{\partial \alpha})_0d\alpha$ are independent, we get that $\dd J = 
0$ only if 
\be
\frac{\partial f}{\partial y_i} = \frac{d }{dx}\frac{\partial 
f}{\partial {\dot y_i}} \;\;\;\; \forall i = 1,\dots,n; 
\label{eq;ELgeneral}
\ee   
which are called the {\em Euler-Lagrange} equations. (They first occur 
in a 1736 paper of Euler's. But the $\dd$-notation and this neat  
deduction is due to Lagrange: he developed his approach in letters to 
Euler from 1754, but first published it in 1760: Kline (1972: 
582-589), Fraser (1983, 1985).)

We remark that for a variational principle that uses fixed end-points, 
the integrand is undetermined up to the total derivative with respect 
to  the independent variable $x$ of a function $g(y)$.   That is: 
suppose we are given a variational principle $\dd J := \dd \; \int f  
dx = 0$, and accordingly its Euler-Lagrange equations. Then exactly 
the same Euler-Lagrange equations would arise from requiring instead 
$\dd J' := \dd \; \int [f + \frac{dg}{dx}] dx = 0$. For the end-points 
being fixed means that $\dd \; \int \frac{dg}{dx}  dx = \dd [g(x_2) - 
g(x_1)] \equiv 0$; so that $\dd J = 0$ iff $\dd J' = 0$.\\
\indent Though simple, this result is important: it corresponds to the 
result in Paragraph 3.3.2.D that Lagrangians differing by a total time 
derivative determine identical Lagrange's equations. (It also 
underlies the idea of generating functions for canonical 
transformations---developed in the companion paper.)

Now I return to mechanics. Consider a monogenic holonomic system: 
Hamilton's system implies, by the above argument but with the 
substitutions
\be
x \rightarrow t\;\; ;\;\; y_i \rightarrow q_i\;\; ; \;\; f \rightarrow 
L,
\ee
Lagrange's equations (\ref{eqn;lag}), i.e.:
\be
\frac{\partial L}{\partial q_i} = \frac{d }{dt}\frac{\partial 
L}{\partial {\dot q_i}} \;\; \forall i = 1,\dots,n.
\label{eq;lag2}
\ee 
The use of arbitrary functions $\eta$ in the variation problem means 
that the Principle mentions contralegal histories, illustrating 
(Modality;3rd). 

Note that the assumption of holonomic constraints is {\em used} in the 
argument; for we appeal to independent variations $\dd q_i$ to get eq. 
\ref{eq;lag2}, in the way we got eq. \ref{eq;ELgeneral} from eq. 
\ref{eq;basiccalcvar}. (For the modification of Hamilton's Principle 
to cover non-holonomic systems, see the next Subsection.)\\
\indent On the other hand, the argument proceeds independently of how 
$L$ is defined, and so of the assumption of monogenicity. The role of 
this  assumption is, rather, to limit of the scope of Hamilton's 
Principle, for the sake of empirical correctness (cf. Lanczos 1986: 
114). 

This deduction of Lagrange's equations from Hamilton's Principle 
implies that they have an important property. Namely, they are 
covariant under coordinate transformations (point-transformations) 
$q_j \rightarrow q'_j$; (the merit (Wider)).  For since the 
stationarity of a definite integral is {\em ipso facto} independent of 
a change of the independent variable, $q_j \rightarrow q'_j$, deducing 
Lagrange's equations from Hamilton's Principle implies the covariance 
of the equations (a covariance that holds even if the 
point-transformation $q_j \rightarrow q'_j$ is time-dependent). (This 
property also followed from the deduction from d'Alembert's principle; 
cf. Paragraph 3.3.2.C.)

I end this Section by discussing the relation between Hamilton's 
Principle and d'Alembert's principle. As I mentioned at the start of 
Section  \ref{FromDale}, one can deduce Hamilton's Principle from 
d'Alembert's principle. More precisely, one can deduce Hamilton's 
Principle in the above form---for a holonomic system whose applied 
forces are monogenic with their work function $U$ independent of 
velocities---from d'Alembert's principle for such a system. Besides, 
the deduction can be reversed: this is an equivalence, illustrating 
(Reformulate).

Lanczos (1986: 111-113) gives details of this. So does Arnold (1989: 
91-95). Arnold's discussion has the merit that it also formulates an 
equivalence with the conception of a constrained system as a limit 
(cf. eq. \ref{icsArnoldslimit} in Paragraph 3.3.2.A)---(Reformulate) 
again! But Arnold also assumes conservative systems, and  makes some 
use of modern geometry. I will just summarize Lanczos' deduction in 
the first direction, i.e. from d'Alembert's principle to Hamilton's 
Principle.

The idea is to overcome the intractable because polygenic character of 
the inertial forces, by representing them by the kinetic energy $T$ 
(and by boundary terms that, by the variation in Hamilton's principle 
having fixed end-points, are equal to zero). One integrates, with 
respect to time, the total virtual work  $\dd w$ done by what Section 
\ref{FromDale}  called the `effective forces': i.e. the work done by 
the difference between the  applied force on particle $i$, $\F_i$ and 
the rate of change of $i$'s momentum, ${\dot p}_i$, summed over $i$.\\
\indent d'Alembert's Principle sets this total virtual work  $\dd w$ 
equal to 0. So we write, summing over particles $i$ and working in 
cartesian coordinates:
\be
\int \dd w \; dt = \int \Sigma \left(\F_i - \frac{d}{dt}(m_i{\bf v}_i) 
\right)\cdot \dd \rr_i \; dt = 0 \label{eq;intergrateddAlemb} 
\ee  
Assuming that the $\F_i$ are monogenic with their $U$ independent of 
velocities, and setting $V = - U$, we deduce (by integrating by parts) 
that
\be
\int \dd w \; dt = \dd \int L \; dt -\left[ \Sigma \; m_i{\bf v}_i 
\cdot \dd \rr_i \right]^{t_2}_{t_1} = 0 \mbox{   where }L := T - V. 
\label{eq;integrateddAlembwithbdary} 
\ee 
Then requiring that the $\dd \rr_i$ vanish at the end-points of the 
integration makes the right-hand side the variation of a definite 
integral: i.e.
\be
\int \dd w dt = \dd \int L \; dt = 0 \mbox{   with }L := T - V.
\ee
Finally, the assumption that the system is holonomic (i.e. $q_j$ 
freely variable) means that the variations in this last equation match 
those of Hamilton's principle.\footnote{By the way: the boundary term 
that is here zero will later be  very important in Hamiltonian 
mechanics and Hamilton-Jacobi theory. It is essentially the canonical 
one-form I mentioned before.}

\subsection{Extending Hamilton's Principle}\label{ssecExte}
In this Subsection, I discuss (simplifying Goldstein et al (2002: 
46f.)) how Hamilton's Principle can be extended to some kinds of 
non-holonomic system. But {\em warning}: this Section can be skipped, 
in that its material is not central to the rest of this paper. 
However,  the topic does illustrate the merit (Reduce); and it 
involves an important application of Lagrange's method of undetermined 
multipliers, viz. its application to extremizing integrals. I begin 
with an interlude about this; (for more details, cf. Lanczos (1986: 
62-66)).\footnote{Also beware: Lanczos remarks (1986: 92, 114) that 
Hamilton's principle applies only to holonomic systems. That is  
wrong. Lanczos seems best interpreted as mis-reporting the {\em true} 
requirement (also stated by him, p. 112) on Hamilton's principle in 
the form using $L: = T - V$,  that the applied forces have a work 
function---so that we can write $V = - U$. The confusion arises 
because  Lanczos also  sometimes (e.g. p. 85) takes holonomic 
(respectively:  non-holonomic) constraints to be maintained by 
monogenic (polygenic) forces. I deny that (cf. footnote 34 in Section 
\ref{KEandwork}). But even if it were true, it would be a point about 
the forces of constraint, not about the applied forces. So it would 
not make the true requirement above, that the applied forces have a 
work function, imply that Hamilton's Principle is restricted to 
holonomic systems.}

\subsubsection{Constrained extremization of 
integrals}\label{ConstrExtr}
 Recall from Section \ref{parLagr} the main idea of Lagrange's method. 
We are to find the point $x_0 := (x_{0_1},\dots,x_{0_n})$ at which 
$\dd f = 0$ for small variations $x' - x_0$ such that $g_j(x') = 
g_j(x'_1,\dots,x'_i,\dots,x'_n) = 0$ for $j = 1,2,\dots,m$. So 
defining $h(x) = f(x) + \Sigma_j \lambda_j g_j(x)$, we have:
\be
\dd h = \dd f + \Sigma_j \; \lambda_j \dd g_j = 0 \;\; \mbox{, i.e.  
}\;\; \frac{\partial f}{\partial x_i} + \Sigma_j \; \lambda_j 
\frac{\partial g_j}{\partial x_i} = 0, \;\;\;\;\; i=1,\dots,n
\ee 
We use these $n$ equations, together with the $m$ equations $g_j(x_i) 
= 0 $ to find the $n + m$ unknowns (the $n \;\; x_i$ and the $m \;\; 
\lambda_j$).

I now show how to apply this technique to the case where $f$ is an 
integral; treating first (A) holonomic, then (B) non-holonomic, 
constraints. With an eye on applications to Hamilton's Principle, I 
assume the independent variable of the integral is time $t$, i.e. we 
are to extremize $f = \int F(q_i,\dot{q}_i,t)\;dt$, subject to some 
constraints. Also:\\
\indent (1): I will now use $q$s not $x$s to emphasise that the 
coordinates need not be cartesian.\\
\indent (2) Because of the constraints, the $q$s in this Subsection 
(unlike Section \ref{ssecHami}) are {\em not} independent.

 (A): {\em Holonomic constraints}:--- Assume that the constraints are 
holonomic, so that we have equations $g_j(q_i) = 0$ for $j = 
1,\dots,m$. 
Then variation of the constraint equations gives
\be
\dd g_j = \Sigma_i \; \frac{\partial g_j}{\partial q_i}\dd q_i \mbox{      
for each }j = 1,\dots,m \;\; .
\ee
Lagrange's method is to multiply each of these equations by an 
undetermined multiplier $\lambda_j$. But since these constraint 
equations are to hold for each $t$, each $\lambda_j$ becomes an 
undetermined function of $t$, and an integral over $t$ is added to the 
summation over $j$.  So the condition for extremization (the 
variational principle) is:
\be
\dd \int F \; dt + \int (\lambda_1 \dd g_1 +  \dots + \lambda_m \dd 
g_m) \; dt = 0.
\ee
Then the usual calculus of variations argument, using the fact that $F 
= F(q_i,\dot{q}_i,t)$ while each $g_j$ is a function only of the 
$q_i$, leads to the Euler-Lagrange equations, for each $i = 
1,\dots,n$:
\be
\frac{\partial F}{\partial q_i} - \frac{d }{dt }\left( \frac{\partial 
F }{\partial \dot{q}_i}\right) + \lambda_1 \frac{\partial 
g_1}{\partial q_i} + \dots + \lambda_m \frac{\partial g_m}{\partial 
q_i} = 0. \label{eq;EulLagwithholoconstrt}
\ee  
In other words: the original variational problem of extremizing $\int 
F \;dt$ subject to $g_j(q_i) = 0$ is replaced by the equivalent 
problem of extremizing (subject to {\em no} constraints)
\be
\int F + \lambda_1 g_1 + \dots + \lambda_m g_m .
\ee
We have so far considered the $\lambda_j$ as constants of the 
variation problem. But, as in Section 3.2.1, comment (1), after eq. 
\ref{eq;LagMultBasic}: we do not have to do so; and if we vary the 
$\lambda_j$, we get back the constraint equations {\em a posteriori}.

(B): {\em Non-holonomic constraints}:--- We now suppose that the 
constraints are non-holonomic, so that we do not have equations 
$g_j(q_i) = 0$ for $j = 1,\dots,m$. But suppose that, as in Section 
3.2.1, comment (2) and its eq. \ref{eq;varynonholoconstrt},
we have differential constraint equations
\be
\dd g_j = \Sigma_i \; G_{ji} \dd q_i = 0 \mbox{      for each }j = 
1,\dots,m.
\ee
Lagrange's method again applies. We get again eq. 
\ref{eq;EulLagwithholoconstrt}, but with factors $G_{ji}$ (as in eq. 
\ref{eq;varynonholoconstrt}) replacing the $\frac{\partial 
g_j}{\partial q_i}$ of eq. \ref{eq;EulLagwithholoconstrt}. That is, we 
get:-
\be
\frac{\partial F}{\partial q_i} - \frac{d }{dt }\left( \frac{\partial 
F }{\partial \dot{q}_i}\right) + \lambda_1 G_{1i} + \dots + \lambda_m 
G_{mi} = 0. \label{eq;EulLagwithnonholoconstrt}
\ee  
We will see this in more detail with eq. \ref{eq;nonholohpstart} et 
seq. below.

\subsubsection{Application to mechanics}\label{ApplicMechs}
It is  the second case, (B), that we need when formulating Hamilton's 
Principle for  non-holonomic systems. I will consider only those 
non-holonomic systems whose equations of constraint can be expressed 
as a relation between the differentials of the $q$'s, i.e. can be put 
in the form of say $m$ equations 
\be
\Sigma_i \; a_{ji}dq_i + a_{jt}dt = 0, \;\; j = 1,\dots,m .
\label{eq;nonholohpstart}
\ee
The variation considered in Hamilton's Principle holds constant the 
time, so that virtual displacements $\dd q_i$ must satisfy
\be
\Sigma_i \; a_{ji}\dd q_i = 0.
\ee
This implies that for any undetermined functions of time 
$\lambda_j(t)$, we have $\lambda_j \Sigma_i a_{ji}\dd q_i = 0$. We sum 
over $j$, and integrate over an arbitrary time interval, to get
\be
\int^{t_2}_{t_1} \Sigma_{i,j} \; \lambda_j a_{ji}\dd q_i \;\; dt = 0.
\ee
We now assume the integrated statement of Hamilton's Principle in the 
form
\be
\int^{t_2}_{t_1} dt\;\; \Sigma_i \left(\frac{\partial L}{\partial q_i} 
- \frac{d}{dt}\frac{\partial L}{\partial {\dot q_i}} \right) \dd q_i = 
0;
\ee
and add the equations, so that
\be
\int^{t_2}_{t_1} dt\;\; \Sigma_i \left(\frac{\partial L}{\partial q_i} 
- \frac{d}{dt}\frac{\partial L}{\partial {\dot q_i}} + \Sigma_j \; 
\lambda_j a_{ji} \right) \dd q_i = 0.
\label{equation;integHPnonhol}
\ee
We can then choose the first $n-m$ of the $\dd q_i$ freely (so that 
the last $m$ are thereby fixed); and we can choose the $\lambda_j$ so 
that:
\be
\left(\frac{\partial L}{\partial q_i} - \frac{d}{dt}\frac{\partial 
L}{\partial {\dot q_i}} + \Sigma_j \; \lambda_j a_{ji} \right) = 0, 
\mbox{   for } i = n-m+1,\dots,n;
\label{equation;chooselambdaintegrand0}
\ee
(these are in effect equations of motion for the last $m$ of the 
$q_i$). Then eq. \ref{equation;integHPnonhol} becomes
\be
\int^{t_2}_{t_1} dt\;\; \Sigma^{n-m}_i \left(\frac{\partial 
L}{\partial q_i} - \frac{d}{dt}\frac{\partial L}{\partial {\dot q_i}} 
+ \Sigma_j \; \lambda_j a_{ji} \right) \dd q_i = 0,
\label{equation;integHPnonholton-m}
\ee
an equation which involves only the independent $\dd q_i$, so that we 
can deduce by the usual calculus of variations argument
\be
\frac{\partial L}{\partial q_i} - \frac{d}{dt}\frac{\partial 
L}{\partial {\dot q_i}} + \Sigma_j \; \lambda_j a_{ji} = 0 \mbox{   
for } i = 1,\dots,n-m . 
\ee
Putting this together with eq \ref{equation;chooselambdaintegrand0} 
(and with $\lambda \rightarrow - \lambda$), we have $n$ equations (cf. 
eq. \ref{eq;EulLagwithnonholoconstrt})
\be
\frac{\partial L}{\partial q_i} - \frac{d}{dt}\frac{\partial 
L}{\partial {\dot q_i}} = \Sigma_j \; \lambda_j a_{ji} \;\; i = 
1,\dots,n .
\ee 
There are altogether $n + m$ unknowns (the $q_i$ and $\lambda_j$). The 
other $m$ equations are the equations of constraint relating the 
$q_i$'s, now considered as differential equations
\be
\Sigma_i \; \; a_{ji}{\dot q_i} + a_{jt} = 0, \;\; j = 1,\dots,m .
\ee

Finally, I note that the $\lambda_j$ again have the physical 
interpretation discussed in Paragraph 3.2.2.B; (but now with $\lambda 
\rightarrow - \lambda$, and without the assumption of cartesian 
coordinates). This interpretation is related to the issue of 
justifying treating motion wholly within the constraint surface, first 
raised after eq \ref{eq;rfromq}---and so to (Reduce), and also  
(Ideal) and (Accept). For discussion, cf. the references given in 
Paragraph 3.2.2.B and Lanczos (1986: 141-145) who relates the 
interpretation to the fact that the constraints are microscopically 
violated. 

So much by way of extending Hamilton's Principle to non-holonomic 
systems. NB: The rest of this Section will assume that the 
constraints, if any, are holonomic---and as usual, ideal.  

\subsection{Generalized momenta and the conservation of energy}
\label{ssecGeneMom} 
The rest of this Section is dominated by one idea: {\em using a  
symmetry to reduce the number of variables of a problem}. First, in 
this Subsection, I introduce generalized momenta and discuss the 
conservation of energy. This is a preliminary to Section 
\ref{ssecCycl}'s  result that the generalized  momentum of any cyclic 
coordinate is a constant of the motion. Though very simple, that 
result is important: for it is the basis of both  Routhian reduction 
and Noether's theorem---which will take up the rest of the Section.

In Section \ref{ssecDale} we deduced the conservation of energy, for a 
(ideal and holonomic)  system that is scleronomous in both work 
function and constraints, from d'Alembert's principle---by choosing 
virtual displacements $\dd \rr_i$ equal to the actual displacements 
$d\rr_i$ in an infinitesimal time $dt$. We can similarly deduce the 
conservation of energy for such a system from Hamilton's principle, by 
considering such a variation for the generalized coordinates $q_j$, 
i.e. with $dt = \epsilon$
\be
\dd q_j = dq_j = \epsilon {\dot q}_j .
\ee  
This deduction is important in that it introduces two notions which 
will be important in what follows.

Since this variation does not fix the configurations at the 
end-points, we get a boundary term when we perform the integration by 
parts in the usual derivation of the Euler-Lagrange equations (cf. eq. 
\ref{eq;basiccalcvar} in Section 4.2.). That is, we get from 
Hamilton's principle and an integration by parts
\be
\dd \int^{t_2}_{t_1} L \; dt = \left[\Sigma_j \frac{\partial 
L}{\partial {\dot q}_j} \dd q_j 
\right]^{t_2}_{t_1}\label{eq;bdaryterminH
amPrinc}
\ee
Here we see for the first time two notions that will be central in the 
sequel.

\indent \indent (i): Generalized momenta:---\\
Elementary examples prompt the definition of the {\em generalized 
momentum}, $p_j$,  {\em conjugate} to a coordinate $q_j$ as: 
$\frac{\partial L}{\partial {\dot q_j}}$; (Poisson 1809). 
So Lagrange's equations for a holonomic monogenic system, eq. 
\ref{eq;lag2}, can be written:
\be
\frac{d}{dt} p_j = \frac{\pl L}{\pl {q_j}} \;\; ;
\label{LagWithpj}
\ee
and we can write eq. \ref{eq;bdaryterminHamPrinc} as:
\be
\dd \int^{t_2}_{t_1} L \; dt = \left[\Sigma_j p_j \dd q_j 
\right]^{t_2}_{t_1}\label{eq;HPwithbdaryterm}
\ee
Note that $p_j$ need not have the dimensions of momentum: it will not 
if $q_j$ does not have the dimension length. And even if $q_j$ is a 
cartesian coordinate, a velocity-dependent potential will mean $p_j$ 
is not the usual mechanical momentum. 

\indent \indent  (ii): The differential $\Sigma_j p_j \dd q_j$, as in 
the right-hand side  of eq. \ref{eq;HPwithbdaryterm}:---\\
As I have mentioned, this will be important in Hamiltonian mechanics.

If $L$ does not depend on time explicitly, i.e. $L = L(q_j;{\dot 
q}_j)$, then using $dq_j = \epsilon {\dot q}_j \Rightarrow d{\dot q}_j 
= \epsilon {\ddot q}_j$, we get:
\be
\dd L = dL = \epsilon {\dot L}
\ee 
so that 
\be
\dd \int L \; dt = \int \dd L \; dt = \int \epsilon {\dot L}\; dt = 
\epsilon \left[L \right]^{t_2}_{t_1}.
\ee
Then eq. \ref{eq;HPwithbdaryterm} implies 
\be
\left[\Sigma_j p_j{\dot q}_j - L \right]^{t_2}_{t_1} = 0.
\ee 
Since the limits $t_1, t_2$ are arbitrary, we get
\be
H : = \Sigma_j p_j{\dot q}_j - L = \mbox{ constant }
\label{eq;defineH}
\ee 
 This function  $H$ (`$H$' for `Hamiltonian') can be called the `total 
energy' of the system; (though as we shall see in a moment, only under 
certain conditions is it the sum of potential and kinetic energies). 

We can also deduce directly from Lagrange's equations, rather than 
from Hamilton's Principle, that $H$ is constant for a system that is 
both:\\
\indent \indent (i) monogenic and holonomic (so that Lagrange's 
equations take the familiar form \ref{eq;lag2}), and\\
\indent \indent (ii)
scleronomous, so that $L$ is not an explicit function of time.\\
The deduction simply applies Lagrange's equations to the expansion of 
$dL/dt$: 
\be
\frac{dL}{dt} = \Sigma_i \frac{\partial L}{\partial q_i}{\dot q_i} + 
\frac{\partial L}{\partial {\dot q_i}}{\ddot {q_i}}  .
\ee 
We get immediately that 
\be
H := \Sigma_i \dot{q_i}p_i - L 
\ee
is a constant of the motion.

Now let us add the assumption that the system is conservative in the 
sense that (the applied) forces are derived from potentials, and that 
the potentials are velocity-independent. Then, using again the 
assumption that the system is scleronomous, we can show that $H$ is 
the sum of potential and kinetic energies---as follows.\\
\indent These assumptions imply that $p_i := \frac{\partial 
L}{\partial {\dot q_i}} = \frac{\partial T}{\partial {\dot q_i}}$. Now 
recall from the discussion of equation \ref{equation;Tingeneral}, that  
the system being scleronomous implies that $T$ is a homogeneous 
quadratic function of the $\dot{q_i}$'s. This means that the first 
term of $H$, i.e. $\Sigma_i \dot{q_i}p_i = \Sigma_i 
\dot{q_i}\frac{\partial T}{\partial {\dot q_i}}$ must be $2T$ (Euler's 
theorem). So we have:-
\be
H = 2T - (T - V) = T + V.
\ee 

\subsection{Cyclic coordinates and their elimination}
\label{ssecCycl}
\subsubsection{The basic result}\label{basicCycl}
We say a coordinate $q_i$ is {\em cyclic} if $L$ does not depend on 
$q_i$. (The term comes from the example of an angular coordinate of a 
particle subject to a central force. Another term is: {\em 
ignorable}.) Then the Lagrange equation for a cyclic coordinate $q_n$ 
say, viz. ${\dot p}_n = \frac{\partial L }{\partial q_n}$, becomes 
${\dot p}_n = 0$, implying
\be
p_n = \mbox {constant, $c_n$ say}.
\label{pnconstant}
\ee
So: {\em the generalized momentum conjugate to a cyclic coordinate is 
a constant of the motion}.

 In other words, thinking of $p_n$ as a function of the $2n+1$ 
variables $q, {\dot q},t$, $p_n = p_n(q, {\dot q},t)$, and using the 
terminology of Paragraph 2.1.3.A (iii): the motion of the system is 
confined to a unique level set $p_n^{-1}(c_n)$.  And assuming 
scleronomous constraints, so that we work with the velocity phase 
space (tangent bundle) $TQ$: this level set is a $(2n-1)$-dimensional 
sub-manifold of $TQ$.\\
\indent  So finding a cyclic coordinate\footnote{Beware: The condition 
${\pl L}/{\pl  q_n} = 0$ is of course a property not just of the 
coordinate $q_n$, but of the entire coordinate system, since partial 
derivatives depend on what other variables are held constant. So one 
can have another coordinate system $q'$ such that $q'_n =q_n$ but 
${\pl L}/{\pl  q'_n} \neq {\pl L}/{\pl  q_n}$. Besides, one can have 
${\pl L}/{\pl  {\dot q'_n}} \neq {\pl L}/{\pl  {\dot q_n}}$, i.e. the 
momenta conjugate to $q'_n = q_n$ are distinct.} simplifies the 
dynamical system, i.e. the problem of integrating its equations of 
motion. The number of variables (degrees of freedom of the problem) is 
reduced by one---the merit (Reduce) of my moral (Scheme). 

This result is simple  but important. For first, it is straightforward 
to show that it encompasses the elementary theorems of the 
conservation of momentum, angular momentum and energy; (Goldstein et 
al (2002: 56-63)). Let us take as a simple example, the angular 
momentum of a free particle. The Lagrangian is, in spherical polar 
coordinates,
\be
L = \frac{1}{2}m({\dot r}^2 + r^2{\dot \theta}^2 + r^2{\dot 
\phi}^2\sin^2\theta) 
\label{LagSpherl}
\ee 
so that ${\pl L}/{\pl \phi} = 0$. So the conjugate momentum
\be
\frac{\pl L}{\pl {\dot \phi}} = mr^2{\dot \phi}\sin^2\theta \; ,
\label{AMfreeSpherl}
\ee
which is the angular momentum about the $z$-axis, is conserved.

Secondly, this result leads into important  theoretical developments 
about how to use symmetries, so as to simplify (reduce) problems. The 
rest of this Section is devoted to two such developments: in this 
Subsection, Routhian reduction; and later (Section \ref{NoetherLag}), 
Noether's theorem. Noether's theorem will be more general in the sense 
that Routhian reduction requires us to have identified cyclic 
coordinates, while Noether's theorem does not.

\subsubsection{Routhian reduction}\label{Routhian} 
My discussion of Routhian reduction has two main aims:\\
\indent (i): in this Subsection,  to illustrate using symmetries to 
secure (Reduce); and\\
\indent (ii) in Section \ref{ssecTime},  to vindicate Section 
\ref{ELsPLA}'s principle of least action from the perspective of 
Hamilton's Principle. ({\em Warning}: This use of Routhian reduction 
is a topic in the foundations of classical mechanics, not much related 
to my philosophical morals: it will not be used later, and can be 
skipped.)

So suppose $q_n$ is cyclic. To exploit this fact so as to reduce the 
number of variables in a mechanical problem, we can proceed in either 
of two ways: (a) after writing down Lagrange's equations; or (b) 
before doing so. I treat these in order; Routhian reduction is (b).

(a): Notice first that since $q_n$ does not occur in $L$, it does not 
occur in $\frac{\pl L}{\pl {\dot q}_n}$, so that we can solve $p_n = 
\frac{\partial L }{\partial {\dot q}_n} = c_n$ for ${\dot q}_n$ as a 
function of the other variables, i.e.
\be
{\dot q}_n = {\dot q}_n(q_1,\dots,q_{n-1},{\dot q}_1,\dots,{\dot 
q}_{n-1},c_n,t)\label{eq;solveforqn}
\ee  
and substitute the right-hand side of this into Lagrange's equations. 
This  reduces the problem of integrating Lagrange's equations to a 
problem in $n-1$ variables. Once solved we can find ${\dot q}_n$ by 
eq. \ref{eq;solveforqn}, and then find $q_n$ by quadrature. (Recall 
that `quadrature' is jargon for integration of a given function: if we 
cannot do the integral analytically, we do it numerically.) 

(b): But we can also instead reduce the problem {\em ab initio}, i.e. 
when it is formulated as a variational problem. This is Routhian 
reduction. It is important both historically and conceptually. As to 
history, it was Routh who first emphasised the importance of cyclic 
coordinates (and his work led to e.g. Hertz' programme in mechanics). 
And as to conceptual aspects, Routhian reduction yields a proper 
understanding of how the principle of least action is based on 
Hamilton's Principle. Indeed, since the principle of least action 
preceded Hamilton's principle, and thus also this understanding, the 
historical and conceptual roles are related---as we shall see in the 
next Subsection, Section \ref{ssecTime}.\footnote{Beware: many 
textbooks (including fine ones like Goldstein et al) only treat 
Routhian reduction as an aspect of Hamiltonian theory: in short as 
involving a Legendre transformation on only the cyclic 
coordinates---details in the companion paper. That lacuna is another 
reason for describing  Routhian reduction within the Lagrangian 
framework. My discussion will  follow Lanczos (1986: 125-140).}

Note first that we cannot just replace ${\dot q}_n$ in Hamilton's 
principle 
\be
\dd \int L(q_1,\dots,q_{n-1},{\dot q}_1,\dots,{\dot q}_{n-1},{\dot 
q}_n,t) = 0 
\ee
by the right-hand side  of eq. \ref{eq;solveforqn}. For eq. 
\ref{eq;solveforqn} makes ${\dot q}_n$ a function of the non-cyclic 
variables in the sense that $p_n \equiv \frac{\partial L }{\partial 
{\dot q}_n} = c_n$ is to hold not just for the actual motion but also 
for the varied motions. This restriction on the variations is not 
objectionable (since the integral's variation is to vanish for {\em 
arbitrary} variations). But the fact that $q_n$ is obtained by a 
quadrature means that the variation of $q_n$ does not vanish at the 
end-points. Rather we have (cf. eq \ref{eq;bdaryterminHamPrinc}):
\be
\dd \int L \; dt = [p_n \dd 
q_n]^{t_2}_{t_1}.\label{eq;cyclicbdaryter
m}
\ee
But $p_n$ is constant along the system's (actual or possible) 
trajectory in configuration space, so that
\be
[p_n \dd q_n]^{t_2}_{t_1} = p_n \; \dd \int {\dot q}_n \; dt = \dd 
\int p_n{\dot q}_n \; dt
\ee
so that eq. \ref{eq;cyclicbdaryterm} becomes
\be
\dd \int (L - p_n{\dot q}_n)\; dt = 0.
\ee
To sum up: our problem is reduced to extremizing a modified  integral 
in $n-1$ variables 
\be
\dd \int (L - c_n{\dot q}_n) \; dt = 0 ;
\ee
where $q_n$ does not occur in the modified Lagrangian ${\bar L}:= L - 
c_n{\dot q}_n$; {\em and nor does} ${\dot q}_n$ explicitly occur, 
since it is eliminated by using eq. \ref{eq;solveforqn}, i.e. the 
momentum first integral $\frac{\partial L}{\partial {\dot q}_n}= 
c_n$.\\
\indent Once this problem in $n - 1$ variables is solved, we  again 
(as at the end of (a) above) find ${\dot q}_n$ by using the momentum 
integral eq. \ref{eq;solveforqn}; and then we find $q_n$ by 
quadrature. 

We can easily adapt the argument of (b) to the case where the given 
problem has more than one cyclic coordinate. The modified Lagrangian 
$\bar L$ subtracts the corresponding sum over cyclic coordinates:
\be
{\bar L} := L - \Sigma_{k \mbox{ ignorable}} \; \; c_k{\dot q}_k \; .
\ee

Three final comments. (1): Anticipating the companion paper a 
little:--- This kind of subtraction of $p_n{\dot q}_n$ from the 
Lagrangian will be crucial in the discussion of the Legendre 
transformation which carries us back and forth between the Lagrangian 
and Hamiltonian frameworks. This is the reason why, as mentioned 
above, Routhian reduction is often treated in textbooks  just as an 
aspect of Hamiltonian theory: in short as involving a Legendre 
transformation on only the cyclic coordinates.

(2): This reduction has consequences for the form of the modified 
Lagrangian ${\bar L}:= L - c_n{\dot q}_n$; (details in Lanczos: 
128-132).\\
\indent First: there is a new velocity-independent  potential term. 
(An historical aside: This fostered Hertz's (1894) speculation that 
mechanics could be {\em forceless}, i.e. that all macroscopically 
observable forces are analysable in terms of monogenic forces whose 
potential energy   arises in just this way from the elimination of 
cyclic microscopic coordinates. Hertz's proposal reflects, and 
contributed to,  a long tradition of philosophical suspicion of 
forces. The root idea, present already in the seventeenth century, was 
that forces are unobservable while other quantities in mechanics, 
especially mass, length and time, are observable. For details, cf. 
Lutzen (1995).)\\
\indent Second: in general, the reduction also gives  new kinetic 
terms in ${\bar L}$ which are linear in the generalized velocities 
${\dot q}_i, i = 1,2,\dots, n-1$ (called `gyroscopic terms').

(3): Note that Routhian reduction assumes we have identified cyclic 
coordinates. In Section \ref{NoetherLag}, Noether's theorem will 
provide a perspective on symmetry and constants of the motion that 
does not assume this.

\subsection{Time as a cyclic coordinate; the principle of least 
action; Jacobi's principle}
\label{ssecTime}
{\em Warning:} This Section can be skipped: it is not used later on. 
But I include it on two grounds:---\\
\indent (i): It is worth seeing how to use Routhian reduction to 
understand the principle of least action (and another principle, 
Jacobi's), with which we began in  Section \ref{ELsPLA}.\\
\indent (ii): It illustrates the idea of a parameter-independent 
integral, which came up in Paragraph 3.3.2.B, (2), in connection with 
the fact that the Hessian condition can fail; though this paper will 
not pursue the topic. 

Any holonomic  system whose Lagrangian does not contain time 
explicitly provides an important case of the theory of Section 
\ref{ssecCycl}. For with such a system we can take the time itself as 
a cyclic coordinate.  If furthermore the system is conservative (i.e. 
forces are derived from potentials, and the potentials are 
velocity-independent) and also the system is scleronomous,
 so that $\Sigma p_j {\dot q}_j = 2T$ (cf. end of Section 
\ref{ssecGeneMom}), then the theory of Section \ref{ssecCycl} yields 
the principle of least action---i.e. the variational principle at the 
centre of Lagrange's and Euler's formulations of analytical mechanics.

But there are subtleties about this principle. As we shall see, the 
correct form of this principle is due to Jacobi. And on the other 
hand, we shall also be able to explain why earlier authors like 
Lagrange were able to get the right Euler-Lagrange equations for 
holonomic  conservative scleronomous systems---even though some of 
these authors lacked Hamilton's Principle.
 
If the Lagrangian $L$ of a holonomic system  does not contain time 
explicitly, we can  treat time $t$ `like the $q$s', to give a problem 
in $n+1$ variables. That is, we write the integral to be extremized 
(by Hamilton's principle) in terms of a differentiable  parameter 
$\tau = \tau(t)$, which is such that $d \tau / dt > 0$ but is 
otherwise arbitrary; (for more discussion of why this can be useful, 
cf. e.g. Butterfield (2004c: Sections 5,7)). So with a prime 
indicating differentiation with respect to  $\tau$, and $dt \equiv 
\frac{dt}{d\tau}{d\tau} \equiv t' d\tau$, Hamilton's Principle becomes
\be
\dd \int L 
\left(q_1,\dots,q_n;\frac{q_{1}'}{t'},\dots,\frac{q_{n}'}{t'}
\right) t' \;d \tau = 0 .
\ee 
So the generalized momentum conjugate to $t$ must be conserved. One 
immediately calculates that this momentum is the negative of the 
Hamiltonian as defined in general by eq. \ref{eq;defineH}.  That is:
\be
p_t := \frac{\partial (Lt')}{\partial t'} = L -\left(\Sigma_j \; 
\frac{\partial L}{\partial {\dot q}_j}\frac{q_{j}'}{t'^2}\right)t' = L 
- \Sigma_j \; p_j{\dot q}_j = \mbox{ constant.}
\footnote{As we also saw in Section \ref{ssecGeneMom}, if the system 
is conservative and scleronomous, the Hamiltonian is the sum of the 
kinetic and potential energies, so that we here get {\em yet another} 
derivation of the conservation of energy: (Reformulate) again!}
\label{eq;timemomentumconsved}
\ee

But now let us apply the theory of Section \ref{ssecCycl}. That is: 
let us eliminate $t$, to get a reduced variational problem---which 
will determine the path in configuration space without regard to the 
passage of time. The modified Lagrangian is:
\be
{\bar L}: = Lt' - p_{t}t' = \Sigma_j \; p_j{\dot q}_j \; t'
\ee
so that the reduced variational problem is 
\be
\dd \; \int \Sigma_j \; p_j{\dot q}_j \; t' \; d\tau = 0 .
\ee
If furthermore the system is conservative and scleronomous,
 so that $\Sigma p_j {\dot q}_j = 2T$, we can write this as:
\be
\dd \;\; 2 \int T t' \; d\tau = 0 .\label{eq;princleastaction}
\ee

Eq. \ref{eq;princleastaction} is the principle of least action of 
Section \ref{ELsPLA}. But as Jacobi emphasised, we should not write 
this 
\be
\dd \;\; 2 \int T \; dt = 0 ,
\label{PLAnaughty}
\ee  
since the cyclic coordinate $t$ obviously cannot be used as the 
independent variable of our problem: $t$ does not occur in eq. 
\ref{eq;princleastaction}.\\
\indent However, eq. \ref{PLAnaughty} {\em does} occur in the earlier  
authors using the principle of least action, e.g. Euler and Lagrange 
themselves.\footnote{Arnold (1989: 246) joyously quotes Jacobi who 
says (in his {\em Lectures on Dynamics}, 1842-1843): `In almost all 
textbooks, even the best, this principle is presented so that it is 
impossible to understand'. Arnold continues ironically: `I do not 
choose to break with tradition.  A very interesting ``proof'' of [the 
principle of least action] is in Section 44 of the mechanics textbook 
of Landau and Lifshitz.'  Arnold's own exposition (1989: 242-248) is 
of course admirably lucid and rigorous. But it is abstract and hard, 
not least because it is cast within Hamiltonian mechanics.} At the end 
of this Subsection, I will report how the practice of Euler and his 
contemporaries was in fact justified, by using Lagrange's method of 
multipliers (the $\lambda$-method). 

Let us now apply the theory of Section \ref{ssecCycl} to the principle 
of least action, i.e. to our modified Lagrangian problem eq. 
\ref{eq;princleastaction}. We need to undertake two stages:\\
\indent (i) to eliminate the corresponding velocity $t'$, by solving 
for this velocity the equation saying that the corresponding momentum 
is constant, i.e. by solving eq. \ref{eq;timemomentumconsved} for 
$t'$;\\
\indent (ii) to integrate the resulting equation for $t'$, to find 
$t$.

As to stage (i), we use eq. \ref{eq;Tdefinesline} and 
\ref{eq;Tonconfigmass1}. Recall that the latter is:
\be
 T = \frac{1}{2} \left(\frac{ds}{dt} \right)^2, \label{eq;Tonconfig}
\ee 
so that now using $\tau$ as independent variable, 
\be
 T = \frac{1}{2} \left(\frac{ds}{d \tau} \right)^2 / t'^2. 
\label{eq;Tonconfigusingtau}
\ee
This implies that solving the energy conservation equation, eq. 
\ref{eq;timemomentumconsved}, for $t'$ yields
\be
t' = \frac{1}{{\surd{(2(E - V))}}}\frac{ds}{d \tau}\label{eq;fort'}
\ee
Also eq. \ref{eq;Tonconfigusingtau} implies that the principle of 
least action, eq. \ref{eq;princleastaction}, can be written as:
\be
\dd  \; \int {\surd(2(E - V))}\frac{ds}{d \tau}d \tau = 
0\label{eq;jacobi}  
\ee 
This form is known as {\em Jacobi's principle}. It determines the path 
in configuration space without regard to the passage of time. This 
completes stage (i).

NB: Though one might write Jacobi's principle as 
\be 
\dd  \; \int {\surd(2(E - V))} ds = 0,\label{eq;jacobijustds}
\ee
{\em beware} that $ds$ is not an exact differential, and is not the 
differential of the independent variable in eq. \ref{eq;jacobi}. {\em 
Some} parameter $\tau$ must be chosen as the independent variable, 
e.g. one of the $q_j$, say $q_n$, so that all the other $q_j$ are 
functions of $q_n$: which clearly reduces the problem from $n$ to 
$n-1$ degrees of freedom. 

As to stage (ii), one integrates eq. \ref{eq;fort'} to get $t$ as a 
function of $\tau$. This determines how the motion through 
configuration space occurs in time.

I end this Subsection with four comments on the principle of least 
action. The first two are technical; but since they connect the 
principle of least action to the geometry of configuration space, they  
illustrate the moral (Reformulate), and the merit (Wider) of (Scheme).  
The third and fourth comments will return us to Section \ref{ELsPLA}'s 
treatment of the principle of least action.

 (1): {\em The analogy with optics}:---\\
  For a single particle, the line-element $ds$ is just ordinary 
spatial distance (in arbitrary curvilinear coordinates). Jacobi's 
principle is then very like Fermat's principle of least time in 
geometric  optics (mentioned in Paragraph 3.3.2.B, (2)), which 
determines the optical path by minimizing the integral
\be
\int n \; ds
\ee
where $n$ is the refractive index, which can change from point to 
point (cf. $\surd(2(E - V))$). This optico-mechanical analogy is deep 
and important: it plays a role both in Hamilton-Jacobi theory and in 
quantum theory. (Butterfield (2004c: Sections 7-9) gives details and 
references.)  But here I just note that the analogy concerns the path, 
not how the motion occurs in time---which is different in the two 
cases.

(2): {\em Geodesics}:---\\
 Eq. \ref{eq;jacobijustds} suggests that we think of Jacobi's 
principle as determining the path in configuration space as the 
shortest path (geodesic), according to a new line-element, $d 
\sigma^2$ say, defined by
\be
d \sigma^2 := (E - V)ds^2
\ee  
so that Jacobi's principle is the statement that the motion minimizes 
the integral $\int d \sigma$. Besides, if the system is free i.e. $V = 
0$, $d \sigma^2$ and $ds^2$ just differ by a multiplicative constant 
$E$, and Jacobi's principle now says that the system travels a 
geodesic of the original line-element $ds^2$ defined in terms of $T$. 
Furthermore, the conservation of energy and 
\be
E = T = \frac{1}{2}\left(\frac{ds}{dt}\right)^2
\ee   
implies that the representative point moves at constant velocity in 
configuration space.

(3): {\em Euler's and Lagrange's practice}:---\\
  We have deduced the principle of least action, eq. 
\ref{eq;princleastaction}, for holonomic  conservative scleronomous 
systems, as an example of Routhian reduction applied to Hamilton's 
Principle. But  now let us ask how Euler and Lagrange wrote down the 
principle of least action, and got the right Euler-Lagrange equations 
for such systems, without starting from Hamilton's Principle. The 
short answer is that Lagrange et al. regard the energy conservation 
equation eq. \ref{eq;fort'} as an auxiliary condition, and treat it by 
Lagrange's $\lambda$-method. 

Thus recall that in essence, Jacobi's principle involves two 
steps:---\\
\indent (A): In the kinetic energy, replace differentiation with 
respect to $t$ by differentiation with respect to the parameter 
$\tau$:
\be
T' = \frac{1}{2} \Sigma \; a_{ik}q'_{i}q'_{k} = Tt'^2 \; .
\ee
\indent (B):  Minimize the action integral (cf. eq. 
\ref{eq;princleastaction})
\be
2 \; \int \frac{T'}{t'} d \tau
\ee
after eliminating $t'$ by using the energy relation (cf. eq. 
\ref{eq;fort'})
\be
\frac{T'}{t'^2} + V = E .
\label{eq;enyforPLA}
\ee

Lagrange instead proposes to treat eq. \ref{eq;enyforPLA} by his 
$\lambda$-method. So his integral to be extremized is
\be
\int \left[ 2 \frac{T'}{t'} + \lambda \left(\frac{T'}{t'^2} + V\right) 
\right] d \tau .
\ee
Since $t'$ is one of our variables, we can find $\lambda$ by 
minimizing with respect to $t'$, getting
\be
- \frac{2T'}{t'^2} - \frac{2 \lambda T'}{t'^2} = 0 ; \;\;\; \mbox{  
giving  } \lambda = - t'.
\ee
Then the integral becomes 
\be
\int \left(\frac{T'}{t'^2} - V \right) t' \; d \tau = \int (T - V) t' 
\; d \tau.
\ee
But now that the variational problem is a free problem, i.e. has no 
auxiliary conditions, there is no reason {\em not} to use $t$ as the 
independent  variable. Doing so, we get
\be
\int (T - V) \; dt .
\ee
Thus we are led back to Hamilton's Principle; and thus Lagrange et al. 
could obtain the usual, correct, equations of motion for holonomic  
conservative scleronomous systems from their principle of least 
action. 

(4): {\em Two kinds of variation, revisited}:---\\
 In Section \ref{ELsPLA}, I introduced the principle of least action 
using a notion of variation different from the type I have considered 
throughout this Subsection. There I used a variation in which the 
energy $H$ is conserved, so that the transit time varies from one path 
to another. Indeed, the principle of least action is often discussed 
using this kind of variation. But I shall not go into details, nor 
relate this notion of variation to comment (3) above; (for details, 
cf.  Goldstein et al (2002: 356-362) or Arnold (1989: 242-248)).

\subsection{Noether's theorem}\label{NoetherLag}
\subsubsection{Preamble: a modest plan}\label{NoetherPreamble}
Any discussion of symmetry in Lagrangian mechanics must include a 
treatment of ``Noether's theorem''. The scare quotes are to indicate 
that there is more than one Noether's theorem. Quite apart from 
Noether's work in other branches of mathematics, her paper (1918) on 
symmetries and conservation principles (i.e. constants of the motion) 
in Lagrangian theories has several theorems. I will be concerned {\em 
only} with applying her first theorem to finite-dimensional systems. 
In short: it provides, for any symmetry of a system's Lagrangian (a 
notion I will define), a constant of the motion; the constant is 
called the `momentum  conjugate to the symmetry'.

\indent I stress at the outset that the great majority of subsequent 
applications and commentaries (also for her other theorems, besides 
her first) are concerned with versions of the theorems for infinite 
(i.e. continuous) systems. In fact, the context of Noether's 
investigation was contemporary debate about how to understand 
conservation principles and symmetries in the ``ultimate continuous 
system'', viz. gravitating matter as described by Einstein's general 
relativity. This theory can be given a Lagrangian formulation: that 
is, the equations of motion, i.e. Einstein's field equations, can be 
deduced from a Hamilton's Principle with an appropriate Lagrangian.  
The contemporary debate was especially about the conservation of 
energy and the principle of general covariance (aka: diffeomorphism 
invariance). General covariance prompts one to consider how a 
variational principle transforms under  spacetime coordinate 
transformations that are arbitrary, in particular varying from point 
to point. This leads to the idea of ``local'' symmetries, which since 
Noether's time has been immensely fruitful in both classical and 
quantum physics.\footnote{An excellent anthology of philosophical 
essays about symmetry is Brading and Castellani (2003): apart from its 
papers specifically about Noether's theorem, the papers by Wallace, 
Belot and Earman (2003) are closest to this paper's concerns.}\\
\indent So I agree that from the perspective of Noether's work, and 
its enormous later development, this Section's application of the 
first theorem to finite-dimensional systems is, as they say, trivial. 
Furthermore, this application is easily understood, {\em without} 
having to adopt that perspective, or even having to consider infinite 
systems.  In other words:  its statement and proof are natural, and 
simple, enough that no doubt several nineteenth century masters of 
mechanics, like Hamilton, Jacobi and Poincar\'{e}, could recognize it 
in their own work---allowing of course for adjustments to modern 
language. In fact, versions of it for the Galilei group of Newtonian 
mechanics and the Lorentz group of special relativity were published a 
few years before Noether's paper; (Brading and Brown (2003: 90); for 
details, cf. Kastrup (1987)).\footnote{Here again, `versions of it' 
needs scare-quotes. For in what follows, I shall be more limited than 
these proofs. I limit myself, as I did in Paragraph 3.3.2.E, both to 
time-independent Lagrangians and to time-independent transformations: 
so my discussion does not encompass boosts.}

\indent Nevertheless, for this paper's purposes, it is worth 
expounding the finite-system version of Noether's first  theorem. For 
it  generalizes Section \ref{ssecCycl}'s result about cyclic 
coordinates (and so the elementary theorems of the conservation of 
momentum, angular momentum and energy which that result encompasses).  
There is also a pedagogic reason for expounding it. Many books (e.g. 
Goldstein et al 2002: 589f.) concentrate on the versions of Noether's 
theorems  for infinite systems: for the reasons given above, that is 
understandable---but it can unwittingly give the impression that there 
is no version for finite systems. (Noether's theorem also has an 
important  analogue in Hamiltonian mechanics.)

I should also give a warning at the outset about the sense in which 
Noether's theorem generalizes Section \ref{ssecCycl}'s result about 
cyclic coordinates. I said at the very end of Section \ref{ssecCycl} 
that (unlike Routhian reduction) the theorem does not assume we have 
identified cyclic coordinates.\\
\indent Indeed so: but every symmetry in the Noether  sense will arise 
from a cyclic coordinate in some system of generalized coordinates. In 
fact, this will follow from the Basic Theorem (what Arnold dubs the 
``rectification theorem') of the theory of ordinary differential 
equations; cf. Paragraph 2.1.3.A (i).\\
\indent So the underlying point here will be the important one we have 
seen before. Namely: the Basic Theorem secures the existence of a 
coordinate system in which ``locally, the problem is completely 
solved'': i.e.,  $n$ first-order ordinary differential equations have, 
locally, $n-1$ functionally independent first integrals. But that does 
not mean it is easy to find the coordinate system!\\
\indent In other words:  as I emphasised already in Section 
\ref{schemes}: analytical mechanics provides no ``algorithmic'' 
methods for finding the best coordinate systems for solving problems. 
In particular,  Noether's theorem, for all its power,  will  not be a  
magic device for finding  cyclic coordinates!

My own exposition of the theorem is a leisurely pedagogic expansion of 
Arnold's concise geometric proof (1989: 88-89).\footnote{Other brief 
expositions of Noether's theorem for finite-dimensional systems  
include: Desloge (1982: 581-586), Lanczos (1986: 401-405: emphasizing 
the variational perspective) and Johns (2005: Chapter 13).} This will 
involve, following Arnold, two main limitations of scope:\\
\indent (i): I limit myself, as I did in Paragraph 3.3.2.E, both to 
time-independent Lagrangians and to time-independent transformations. 
Formally, this will mean $L$ is a scalar function on the 
$2n$-dimensional  velocity phase space (aka: tangent bundle) $TQ$ 
coordinatized by $q,{\dot q}$: $L: TQ \rightarrow \mathR$.\\
\indent (ii): I will take a symmetry of $L$ (or $L$'s being invariant) 
to require that $L$ be the {\em very same}. That is: a symmetry does 
not allow the addition to $L$ of the time-derivative of a function 
$G(q)$ of the coordinates $q$---even though, as discussed in Paragraph 
3.3.2.D (and Section \ref{ssecHami}), such a time-derivative  makes no   
difference to the Lagrange (Euler-Lagrange) equations.

\indent So my aims are modest. Apart from pedagogically expanding 
Arnold's proof, my only addition will be to contrast with (ii)'s 
notion of symmetry, {\em another} notion. Although  this notion is 
{\em not} needed for my statement and proof of Noether's theorem, it 
is so important to this paper's theme of general schemes for 
integrating differential equations (from Section \ref{Solv} onwards!)   
that I must mention it briefly.\\
\indent This is the notion of a symmetry of the set of solutions of a 
differential equation: (aka: a dynamical symmetry). This notion 
applies to all sorts of differential equations, and systems of them; 
not just to differential equations of this paper's sort---i.e. 
derived, or derivable, from an variational principle. In short, this 
sort of symmetry is a map that sends any solution of the given 
differential equation  (in effect: a dynamically possible history of 
the system---a curve in the state-space of the theory) to some other 
solution. Finding such symmetries, and groups of them, is a central 
part of the modern theory of integration of differential equations 
(both ordinary and partial). (This notion, and the theory based on it, 
were pioneered by Lie.)\\
\indent It will turn out that broadly speaking, this notion is more 
general than that of a symmetry of $L$ (the notion needed for 
Noether's theorem). Not only does it apply to many other sorts of 
differential equation than the Euler-Lagrange equations. Also, for the 
latter equations: a symmetry of $L$ is (with one {\em caveat}) a 
symmetry of the solutions, i.e. a dynamical symmetry---but the 
converse is false.\\
\indent An excellent account of this modern integration theory, 
covering both ordinary and partial differential equations, is given by 
Olver (2000). He also covers the Lagrangian case (Chapter 5 onwards), 
and gives many historical details about Lie's and others' 
contributions.

\indent The plan is as follows. Starting from cyclic coordinates, I 
first develop the idea of the Lagrangian being invariant under a 
transformation (Section \ref{InvarcefromCyclic}). This leads to 
defining:\\
\indent (i): a {\em  symmetry} as a vector field (on configuration 
space) that generates a family of transformations  under which the 
Lagrangian is invariant  (Section \ref{VecfieldsSymmies});\\
\indent (ii): the {\em momentum conjugate to a vector field}, as 
(roughly) the rate of change of the Lagrangian with respect to the 
${\dot q}$s in the direction of the vector field.\\ 
\indent Together, these definitions lead directly to Noether's theorem 
(Section \ref{Noetsubsubsec}): that the momentum conjugate to a 
symmetry is a constant of the motion. One might guess, in the light of 
the rough definitions just given in (i) and (ii), that proving this 
statement promises to be easy work. And so it is: after all the 
stage-setting, the proof in Section \ref{Noetsubsubsec} will be a 
one-liner application of Lagrange's equations.\footnote{If in (ii), 
the conjugate momentum had been defined simply as the rate of change 
of the Lagrangian in the direction of the vector field, then of course 
the theorem would be truly trivial: it would not require Lagrange's 
equations.}

\subsubsection{From cyclic coordinates  to the invariance  of the 
Lagrangian}\label{InvarcefromCyclic}
I begin by restating Section \ref{ssecCycl}'s result that the 
generalized momentum conjugate to a cyclic coordinate is constant, in 
terms of coordinate transformations. This leads to the correspondence 
between passive and active transformations, and between their 
associated definitions of `invariance'.\\
\indent The passive-active correspondence   is of course entirely 
general: it applies to any space on which invertible differentiable 
coordinate transformations are defined. But it is a notoriously 
muddling subject and so worth expounding {\em slowly}! And although my 
exposition will be in the context of cyclic coordinates in Lagrangian 
mechanics, it will be clear that the correspondence is general.

So let $q_n$ be cyclic, and consider a coordinate transformation $q 
\rightarrow q'$ that just shifts the cyclic coordinate $q_n$ by an 
amount $\epsilon$:
\be
q'_i := q_i + \epsilon \delta_{in} \;\;\;\; {\dot q}'_i = {\dot q}_i 
\;\;.
\label{shiftE}
\ee
So to write the Lagrangian in terms of the new coordinates, we 
substitute, using the reverse transformation, i.e.
\be
L_{\epsilon}(q',{\dot q}',t) = L(q(q',t),\; {\dot q}(q',t),\; t) \;\; 
.
\label{LagE}
\ee 
That is: the Lagrangian is a {\em scalar} function on the space with 
points labelled $(q,{\dot q},t)$. (In Paragraph 3.3.2.B, we called 
this the `extended velocity phase space', or without the time 
argument, just `velocity phase space'; in the geometric description of 
Paragraph 3.3.2.E, we called it the `tangent bundle'.) It is just that 
we use $L_{\epsilon}$ to label its functional form in the new 
coordinate system. 

Now we let $\epsilon$ vary, so that we have a one-parameter family of 
transformations. Differentiating $L_{\epsilon}$ with respect to 
$\epsilon$, we get using the chain rule 
\be
\frac{\pl L_{\epsilon}}{\pl \epsilon} = \Sigma_i \;
\frac{\pl L}{\pl q_i}\frac{\pl q_i}{\pl \epsilon} + \Sigma_i \;
\frac{\pl L}{\pl {\dot q}_i}\frac{\pl {\dot q}_i}{\pl \epsilon} = 
\Sigma_i \;
- \frac{\pl L}{\pl q_i} \delta_{in} = 
- \frac{\pl L}{\pl q_n} \;\; .
\label{diffteLagE}
\ee 
Now we use the fact $q_n$ is cyclic. This implies that eq. 
\ref{diffteLagE} is equal to zero: 
\be
\frac{\pl L_{\epsilon}}{\pl \epsilon} \equiv - \frac{\pl L}{\pl q_n} = 
0 \;\; ;
\label{difflinvarceL}
\ee
and thus that the Lagrangian has the same functional form in the new 
coordinates as in the old ones: which we can express as
\be
L_{\epsilon}(q,{\dot q},t) = L(q,{\dot q},t) \;\; ; \; \mbox{ or 
equivalently }\;\; L_{\epsilon}(q',{\dot q}',t) = L(q',{\dot q}',t) \; 
.
\label{LagInvart}
\ee      

So far we have viewed the coordinate transformation in the usual way, 
as a {\em passive} transformation: cf. the comment after eq. 
\ref{LagE}. But for Noether's theorem, we need to express these same 
ideas in terms of active transformations. So I will now describe the 
usual  correspondence between passive  and active transformations, in 
the context of the velocity phase space.\\
\indent (Again, I stress that my discussion assumes time-independent 
Lagrangians and transformations; so I will not work with extended 
velocity phase space. But some of the following discussion  could be 
straightforwardly generalized to that context. For example, the vector 
fields $X$ that in the next Subsection will represent symmetries would  
become time-dependent vector fields, so that their rates of change 
pick up a partial time-derivative $\frac{\pl X}{\pl t}$.) 

\indent I will temporarily label the velocity phase space  $S$ (for 
`space'). And let us write $s$ for a point (however coordinatized) in 
$S$, and indicate scalar functions on $S$ (however expressed in terms 
of coordinates) by a {\em bar}. So the Lagrangian is a scalar function 
${\bar L}: s \in S \mapsto {\bar L}(s) \in \mathR$.\\
\indent Furthermore, in this Subsection's discussion of the 
corresponding active transformations, the distinction between the $q$s 
and the ${\dot q}$s will play no role. As in the  discussion above, we 
will start with  a passive coordinate transformation on $S$ (both $q$s 
and ${\dot q}$s) that is induced by a passive coordinate 
transformation on just the configuration space (just $q$s); and then 
we will define a corresponding active transformation. But since the $q 
-{\dot q}$-distinction plays no role---the ${\dot q}$s just ``carry 
along'' throughout---it will be clearest to temporarily drop the 
${\dot q}$s from the notation. So in this Subsection, when I talk of a 
passive $q \rightarrow q'$ coordinate transformation inducing an 
active one, you can think indifferently of there being:\\
\indent either (i) $n$ $q$s on configuration space, with the 
discussion ``lifting'' to the $2n$-dimensional $S$, \\
\indent or (ii) $2n$ $q$s on $S$, so that the $q \rightarrow q'$ 
transformation need not be induced by a transformation on just the 
$n$-dimensional  configuration space.\\
\indent In the next Subsection, the $q -{\dot q}$-distinction will of 
course come back into play. 
   
\indent A passive coordinate transformation $q \rightarrow q'$ defines 
an active transformation, $\theta$ say, as follows. We will allow the 
coordinate transformation to be local, i.e. defined only on a patch 
(to be precise: an open subset) $U$  of $S$. Then $\theta: U 
\rightarrow U$ is defined by the rule that for any $s \in U$ the 
coordinates of $\theta(s)$ in the $q'$-system are to be the same 
numbers as the coordinates of $s$ in the $q$-system; that is
\be
q'_i(\theta(s)) = q_i(s) \;\; \mbox{ for all }i \; .
\label{definetheta}
\ee
This definition implies that $\theta$'s functional form in the 
$q'$-system is the transformation $q' \rightarrow q$, i.e. the inverse 
of our original coordinate transformation. That is: $\theta: s \mapsto 
\theta(s)$ is expressed in the $q'$-system by
\be
q'(s) \mapsto q'(\theta(s)) = q(s)\;\; .
\label{fnalexpressionoftheta}
\ee

\indent On the other hand, a scalar function ${\bar L}: U \rightarrow 
\mathR$ can be ``dragged along'' by composition with $\theta$. That 
is, we define ${\bar L} \circ \theta: s \in U \mapsto {\bar 
L}(\theta(s)) \in \mathR$.

\indent  Putting these points together, we deduce that the functional 
forms of ${\bar L}$ and ${\bar L} \circ \theta$, in the coordinate 
systems $q$ and $q'$ respectively, {\em match.}\\
\indent That is: let the functional form of  ${\bar L}$ in the 
$q$-system be $L$: so ${\bar L}(s)$ is calculated in the $q$-system as 
$L(q(s))$. But eq. \ref{fnalexpressionoftheta} implies that the 
functional form of  ${\bar L} \circ \theta$ in the $q'$-system is $q' 
\rightarrow q \rightarrow L(q)$: which is the same. That is, this 
functional form is also $L(q(s)) \equiv L(q(q'(s)))$. (The occurrence 
of $q(q')$ here corresponds to the use of the inverse coordinate 
transformation under the passive view; cf. eq. \ref{LagE}.)

\indent Furthermore, we could undertake this construction in reverse. 
That is, we could  instead start with differentiable invertible active 
maps $\theta$ on $S$, and thereby define coordinate transformations, 
with the property that a scalar function and its ``drag-along'' have 
the same functional expression  in the two coordinate systems. 
(Exercise! Fill in these details; and fill in the details of the above 
discussion, so as to respect the $q - {\dot q}$ distinction.)  

This passive-active correspondence obviously applies to any space on 
which invertible differentiable coordinate transformations are 
defined. For our construction of the corresponding active 
transformation made no appeal to the three special features of 
Lagrangian mechanics and cyclic coordinates that we used in our 
passive discussion, eq. \ref{shiftE} to \ref{LagInvart}. Namely, the 
three features:\\
\indent (i): the $q - {\dot q}$ distinction on $S$;\\
\indent (ii): the use of a translation in just one coordinate to give 
a one-parameter family of transformations, labelled by $\epsilon$; \\
\indent (iii): the idea of a cyclic coordinate and $L$ being 
invariant.

So we will now re-introduce these features. We begin with the most 
important one: (iii), invariance.\\
\indent (We will re-introduce feature (i), the $q - {\dot q}$ 
distinction, in the next Subsection. And it will be clear that the 
main role of (ii), in both passive and active views, is to provide a 
differential notion of invariance; cf. how eq. \ref{difflinvarceL} is 
the differential form of eq. \ref{LagInvart}.) 

\indent We saw that on the passive view, the invariance of the 
Lagrangian means that it has the same functional form in both 
coordinate systems; eq. \ref{LagInvart}.\\
\indent On the active view, it is natural to define: ${\bar L}$ is 
{\em invariant} under the map $\theta: U \rightarrow U$ iff    ${\bar 
L}$ and ${\bar L} \circ \theta$ are the {\em  same} scalar function $U 
\rightarrow \mathR$. 

But we also saw that on the active view, now replacing $\theta$ by 
$\epsilon$ in the obvious way: for {\em any} ${\bar L}$ and its 
drag-along ${\bar L}_{\epsilon}$, the functional forms in the 
$q$-system and $q'$-system respectively, match. This implies that 
${\bar L}$ being invariant means that ${\bar L}$ itself has the same 
functional form in the $q$-system and $q'$-system.

So to sum up: the  correspondence between passive and active (local) 
transformations implies an equivalence between two definitions of what 
it is for a scalar such as the Lagrangian to be invariant---where this 
invariance is understood as above, and not just as being a scalar on 
the space $S$! Namely: that $L$ have the same functional form in two 
coordinate systems, and that it be identical with its ``drag-along''. 

\subsubsection{Vector fields and  symmetries---variational and 
dynamical}\label{VecfieldsSymmies}
The last Subsection used, albeit briefly, the idea of a one-parameter 
family of transformations on the $n$-dimensional configuration space 
$Q$, and how it ``lifts'' to the $2n$-dimensional velocity phase space 
that I labelled $S$.\\
\indent I now need to state these ideas more carefully; and 
especially, to be explicit about the ``differential version'' of such 
a family of transformations. This is the idea of a {\em vector field} 
on $Q$. Here my discussion borders on the relatively abstract ideas of 
modern geometry: ideas which this paper has eschewed, apart from the 
geometric interlude Paragraph 3.3.2.E. But fortunately, in proving 
Noether's theorem  I will be able to make do with an elementary notion 
of a vector field. More specifically, I need to expound four topics:\\
\indent (1): the idea of a vector field on $Q$;\\
\indent (2): how such vector fields ``lift'' to velocity phase space; 
\\
\indent (3): the definition of a (variational) symmetry;  \\
\indent (4): the contrast between (3) and the idea of symmetry of the 
equations of motion (aka: a dynamical symmetry). {\em Warning}: the 
material in (4) will not be needed for Section \ref{Noetsubsubsec}.

(1): {\em Vector fields on $Q$} :----\\
The intuitive idea of a vector field on configuration space $Q$ is 
that it is an assignment to each point of the space, of a 
infinitesimal  displacement, pointing to a nearby point in the space. 
The assignment is to be continuous in the sense that close points get 
similar infinitesimal  displacements assigned to them, the 
displacements tending to each other as the points get closer. For 
present purposes, we will furthermore require that a vector field  be 
differentiable; (this is defined along similar lines to its being 
continuous).\\
\indent In this way, it is intuitively clear that a differentiable 
vector field on an $n$-dimensional configuration space is represented 
in a coordinate system $q = (q_1,\dots,q_n)$ by $n$ first-order 
ordinary differential equations
\be
\frac{d q_i}{d \epsilon} = f_i(q_1,\dots,q_n) \;\; .
\label{vecfieldintuitive}
\ee
At each point $(q_1,\dots,q_n)$, the assigned infinitesimal 
displacement has a component in the $q_i$ direction equal to 
$f_i(q_1,\dots,q_n) d \epsilon$.\\
\indent Thus we return to the basic ideas of solving ordinary 
differential equations, expounded in Paragraph 2.1.3.A, (1). In 
particular, a vector field has through every point a local integral 
curve (solution of the differential equations). In this way, a vector 
field generates a one-parameter family of active transformations: that 
is,  passage along the vector field's integral  curves, by a varying 
parameter-difference $\epsilon$, is such a family of transformations. 
The vector field is called the {\em infinitesimal generator} of the 
family. It is in this sense that a vector field is the ``differential 
version'' of such a family. (Using the last Subsection's 
passive-active correspondence, one could define what it is for a 
vector field to generate a one-parameter family of coordinate 
transformations. But I will not need this idea: exercise to write it 
down!)

I turn to a more precise definition of a vector field on $Q$: 
informal, by the standards of modern geometry, but adequate for 
present purposes. The idea of the definition is that a vector field 
$X$ on $Q$ is to assign the same small displacements irrespective of a 
choice of coordinate system on $Q$. So first we  recall that the 
expressions  in two coordinate systems $q$ and $q'$ of a small 
displacement are related by 
\be
\dd q'_i = \Sigma_j \; \frac{\pl q'_i}{\pl q_j} \dd q_j + 
\mbox{O}(\epsilon^2) \;\; .
\label{trsfmdisplmts}
\ee
We therefore define: a {\em vector field} $X$ on $Q$ is an assignment 
to each coordinate system $q$ on $Q$, of a set of $n$ real-valued 
functions $X_i(q)$, with the different sets meshing according to the 
transformation law (for the coordinate transformation $q \rightarrow 
q'$):
\be
X'_i = \Sigma_j \; \frac{\pl q'_i}{\pl q_j} X_j .
\label{trsfmvecfield}
\ee
The functions $X_i(q)$ are  called the {\em  components} of the vector 
field in the coordinate system $q$.

(2): {\em Vector fields on $TQ$; lifting fields from $Q$ to 
$TQ$}:----\\
Now consider the velocity phase space, the $2n$-dimensional space of 
configurations and generalized velocities taken together. This has a 
natural structure induced by $Q$, essentially because a coordinate 
system $q$ on $Q$,  defines a corresponding coordinate system $q,{\dot 
q}$ on the velocity phase space. Indeed, the velocity phase space is 
called the {\em tangent bundle} of $Q$', written $TQ$---where $T$ 
stands, not for `time', but for `tangent'. The reason for `tangent' 
lies in modern geometry; (cf. Paragraph 3.3.2.E for more detail). But 
in short, the reason is as follows.\\
\indent Consider any smooth curve in configuration space, $\phi: I 
\subset \mathR \rightarrow Q$, with coordinate expression in the 
$q$-system $t \in I \mapsto q(\phi(t)) \equiv q(t) = q_i(t)$. 
Mathematically, $t$ is just the parameter of the curve $\phi$: but the 
physical interpretation is of course  that $\phi$ is a possible 
motion, and $t$ is the time, so that if we differentiate the $q_i(t)$ 
the dot ${\dot {}}$ stands for time, and the $n$ functions ${\dot 
q}_i(t)$ together define the generalized velocity vector, for each 
time $t$.  Besides: for each $t$, the values ${\dot q}_i(t)$ of these 
$n$ functions together form the tangent vector to the curve $\phi$ 
where it passes through the point in $Q$ with coordinates $q(t) \equiv 
q(\phi(t))$. We think of this tangent vector as attached to the space 
$Q$, at the point.\\
\indent It will be helpful to have a notation for this point in $Q$, 
independent of its coordinate expression (here, $q(t)$). Let us write 
it as $x \in Q$. Then: considering all the various possible curves 
$\phi$ that pass through $x$  (with various different directions and 
speeds), we get all the various possible generalized velocity vectors. 
They naturally form a $n$-dimensional vector space, which we call the 
{\em tangent space} $T_x$ attached to $x \in Q$.\\
\indent Then the set (space) $TQ$ is defined by saying that an element 
(point) of $TQ$ is a pair, comprising a point $x \in Q$ together with 
a vector in $T_x$. So: adopting the coordinate system $q$ on (a patch 
of) $Q$, there is a corresponding coordinate system $q,{\dot q}$  on 
(a corresponding patch of) $TQ$.

$TQ$ is a ``smooth space''---formally, a differential manifold---so 
that we can define vector fields on it, on analogy with (1) above. In 
full generality, a vector field on $TQ$ will be an assignment to any 
coordinate system $\zeta^{\al} = (\zeta^1,\dots,\zeta^{2n})$ on $TQ$ 
of a set of $2n$ real-valued functions $X_{\al}(\zeta)$, with the 
different sets meshing according to the analogue of eq. 
\ref{trsfmvecfield}, i.e.  for a coordinate transformation $\zeta 
\rightarrow \zeta'$:
\be
X'_{\al} = \Sigma^{2n}_{\beta = 1} \; \frac{\pl {\zeta}'_{\al}}{\pl 
{\zeta}_{\beta}} X_{\beta} .
\label{trsfmvecfield2n}
\ee  
But we will be interested only in vector fields on $TQ$ that ``mesh'' 
with the structure of $TQ$ as a tangent bundle, i.e. with vector 
fields on $TQ$ that are induced by vector fields on $Q$---in the 
following natural way.

This induction has two ingredient ideas.\\
\indent  First, any curve in configuration space $Q$ defines a 
corresponding curve in $TQ$---intuitively, because the functions 
$q_i(t)$ define the functions ${\dot q}_i(t)$. More formally: given 
any curve in configuration space, $\phi: I \subset \mathR \rightarrow 
Q$, with coordinate expression in the $q$-system $t \in I \mapsto 
q(\phi(t)) \equiv q(t) = q_i(t)$, we define its {\em extension} to 
$TQ$ to be the curve $\Phi: I \subset \mathR \rightarrow TQ$ given in 
the corresponding coordinates by $q_i(t),{\dot q}_i(t)$.\\
\indent Second, any vector field $X$ on $Q$ generates displacements in 
any possible state of motion, represented by a curve in $Q$ with 
coordinate expression $q_i = q_i(t)$. (So here $t$ is the parameter of 
the state of motion, not of the integral curves of $X$.) Namely: for a 
given value of the parameter  $\epsilon$, the displaced state of 
motion is represented by the curve in Q
\be
q_i(t) + \epsilon X_i(q_i(t)) \;\; .
\label{displacedinQ}
\ee 
\indent Putting these ingredients together: we first displace a curve 
within $Q$, and then extend the result to $TQ$. Namely, the extension 
to $TQ$ of the (curve representing) the displaced state of motion is  
given by the $2n$ functions, in two groups each of $n$ functions, for 
the $(q,{\dot q})$ coordinate system 
\be
q_i(t) + \epsilon X_i(q_i(t)) \;\; \mbox{ and } \;\;
{\dot q}_i(t) + \epsilon Y_i(q_i(t), {\dot q}_i) \;\; ;
\ee
where $Y$ is defined to be the vector field on $TQ$ that is the 
derivative along the original state of motion of $X$. That is: 
\be
Y_i(q,{\dot q}) := \frac{d X_i}{dt} = \Sigma_j \; \frac{\pl X_i}{\pl 
q_j}{\dot q}_j.
\label{defineY}
\ee
In this sense, displacements by a vector field within $Q$  can be 
``lifted'' to $TQ$. The vector field $X$ on $Q$ lifts to $TQ$ as 
$(X,\frac{d X}{dt})$; i.e. it lifts to the vector field that sends a 
point $(q_i,{\dot q}_i) \in TQ$ to $(q_i + \epsilon X_i, {\dot q}_i + 
\epsilon \frac{d X_i}{dt})$.\footnote{I have discussed this in terms 
of some system $(q,{\dot q})$ of coordinates. But the definitions of 
extensions and displacements are in fact coordinate-independent. 
 Besides, one can show that
the operations of displacing a curve within $Q$, and extending it to 
$TQ$, commute to first order in $\epsilon$: the result is the same for 
either order of the operations.}

(3): {\em Definition of `symmetry'} :----\\
To define symmetry, I begin with the integral  notion and then give 
the differential notion. I will also simplify, as I have often done, 
by speaking ``globally, not locally'', i.e. by writing as if the 
relevant scalar functions, vector fields etc. are defined on all of 
$Q$ or $TQ$: of course, they need not be.

We return to the idea at the end of Section \ref{InvarcefromCyclic}: 
the idea of the Lagrangian $L$ being invariant under an active 
transformation $\theta$, i.e. equal to its drag-along $L \circ 
\theta$.  (So here $L$ is  the coordinate-independent scalar function 
on $TQ$, not a functional form. But we could use Section 
\ref{InvarcefromCyclic} to recast what follows in terms of a passive 
notion of symmetry as sameness of $L$'s functional form in different 
coordinate systems: exercise!)\\
\indent Now we consider an entire one-parameter family of (active) 
transformations $\theta_s : s \in I \subset \mathR$. We define the 
family to  be a {\em symmetry} of $L$ if the Lagrangian is invariant 
under the transformations, i.e. $L = L \circ \theta_s$. (But see (4) 
below for why `{\em variational symmetry}' is a better word for this 
notion.)

For the differential notion of symmetry, we use the idea of a vector 
field. We define a vector field $X$ to be a {\em symmetry} of $L$ if 
the Lagrangian is invariant, up to first-order in $\epsilon$, under 
the displacements generated by $X$. (But again, see (4) below for why 
`{\em variational symmetry}' is a better word.) More precisely, and 
now allowing for a time-dependent Lagrangian (so that for each time 
$t$, $L$ is a scalar function on $TQ$): we say $X$ is a symmetry iff
\be
L(q_i + \epsilon X_i, {\dot q}_i + \epsilon Y_i, t) \; = \; 
L(q_i,{\dot q}_i,t) + \mbox{O}(\epsilon^2)
\;\; 
\mbox{ with } \; Y_i = \Sigma_j \;\frac{\pl X_i}{\pl q_j}{\dot q_j} \; 
.
\label{definevarnlsymmy}
\ee    
An equivalent definition is got by explicitly setting the first 
derivative with respect to $\epsilon$ to zero. That is:  $X$ is a 
symmetry iff
\be
\Sigma_i \; X_i \frac{\pl L}{\pl q_i} \; + \; \Sigma_i \; Y_i 
\frac{\pl L}{\pl {\dot q_i}} \;\; = 0 \;\; 
\mbox{ with } \; Y_i = \Sigma_j \;\frac{\pl X_i}{\pl q_j}{\dot q_j}
\label{definevarnlsymy2}
\ee

(4): {\em A Contrast: symmetries of the equations of motion} :----\\
{\em Warning}:--- As I said at the end of Section 
\ref{NoetherPreamble}, the material here is not needed for Section 
\ref{Noetsubsubsec}'s presentation of Noether's theorem. But the 
notion of a symmetry of equations of motion (whether Euler-Lagrange or 
not) is so important that I must mention it,  though  only to contrast 
it with (3)'s notion.

The general definition is roughly as follows. Given any system of 
differential equations, ${\cal E}$ say, a {\em (dynamical) symmetry} 
of the system is an (active) transformation $\zeta$ on the system 
$\cal E$'s space of both independent variables, $x_j$ say, and 
dependent variables $y_i$ say, such that any solution of $\cal E$, 
$y_i = f_i(x_j)$ say, is carried to another solution. For a precise 
definition, cf. Olver (2000: Def. 2.23, p. 93), and his ensuing 
discussion of the induced action (called `prolongation') of the 
transformation $\zeta$ on the spaces of (in general, partial) 
derivatives of the $y$'s with respect to the $x$s (called `jet 
spaces').\\
\indent As I said in Section \ref{NoetherPreamble}, groups of 
symmetries in this sense play a central role in the modern theory of 
differential equations: not just in finding new solutions, once given 
a solution, but also in integrating the equations. For some main 
theorems stating criteria (in terms of prolongations) for groups of 
symmetries, cf. Olver (2000: Theorem 2.27, p. 100, Theorem 2.36, p. 
110, Theorem 2.71, p. 161).

But for present purposes, it is enough to state the rough idea of a 
one-parameter group of dynamical symmetries (without details about 
prolongations!) for  the Lagrangian equations of motion in the usual 
familiar form,  eq. \ref{eqn;lag} or \ref{eq;lag2}. In this simple 
case, there is just one independent variable $x := t$, so that we are 
considering ordinary, not partial, differential equations; and there 
are $n$ dependent variables $y_i := q_i(t)$. \\
\indent Furthermore, following the discussion in Section \ref{Anal}, 
these equations mean that the constraints are holonomic, scleronomous 
and ideal, and the system is monogenic with a velocity-independent and 
time-independent work-function. As I have often emphasised (especially  
Sections \ref{ssecGeneMom} and \ref{ssecTime}), this means that the 
system obeys the conservation of energy, and  time is a cyclic 
coordinate.  And {\em this} means that we can define dynamical 
symmetries $\zeta$ in terms of the familiar active transformations  on 
the configuration space, $\theta: Q \rightarrow Q$, discussed since 
Section \ref{InvarcefromCyclic}.  In effect, we define a $\zeta$ by 
just adjoining to any such $\theta: Q \rightarrow Q$ the identity map 
on the time variable $i: t \in \mathR \mapsto t$. (More formally: 
$\zeta: (q,t) \in Q \times \mathR \mapsto (\theta(q),t) \in Q \times 
\mathR$.)\\
\indent Then we define in the usual way what it is for a one-parameter 
family of such maps   $\zeta_s : s \in I \subset \mathR$ to be a 
one-parameter group of dynamical symmetries (for Lagrange's equations 
eq. \ref{eqn;lag}): namely, if any solution curve $q(t)$ (or 
equivalently: its extension $q(t),{\dot q}(t)$ to $TQ$) of the 
Lagrange equations is carried by each $\zeta_s$ to another solution 
curve, with the $\zeta_s$ for different $s$ composing in the obvious 
way.\\
\indent And finally: we also define (in a manner corresponding to 
(3)'s discussion) a differential, as against integral, notion of 
dynamical symmetry. Namely, we say a vector field $X$ is a dynamical 
symmetry if it is the infinitesimal generator of such a one-parameter 
family $\zeta_s$.

For us, the important point is that this notion of a dynamical 
symmetry is {\em different} from (3)'s notion of a variational 
symmetry. Hence it is best to use different words.\footnote{Since the 
Lagrangian $L$ is especially associated with variational principles, 
while the dynamics is given by equations of motion, calling (3)'s 
notion `variational symmetry', and the new notion `dynamical symmetry' 
is a good and widespread usage. But beware: it is not universal. Many 
treatments call (3)'s notion `dynamical symmetry'---understandably 
enough in so far as, for the systems being considered, $L$ determines 
the dynamics.}

\indent Many discussions of Noether's theorem  only mention (3)'s 
notion, variational symmetry. Arnold himself  just says that any 
variational symmetry is a dynamical symmetry, adding in a footnote 
that several textbooks mistakenly assert the converse 
implication---but he does not give a counterexample (1989: 88).  But 
in fact there is a subtlety also about  the first implication, from 
variational symmetry to dynamical symmetry. Fortunately, the same 
simple example will serve both as a counterexample to the converse 
implication, and to show the subtlety about the first implication. 
Besides, it is an example we have seen before: viz.,  the 
two-dimensional harmonic oscillator with a single frequency (Paragraph 
3.3.2.D).\footnote{All the material from here to the end of this 
Subsection is drawn from Brown and Holland (2004a); cf. also their 
(2004). Many thanks to Harvey Brown for explaining these matters. The 
present use of the harmonic oscillator example also occurs in Morandi 
et al (1990: 203-204).}

Recall from eq. \ref{Lag2DHO;eqns} and eq. \ref{OddLag2DHO} that  the 
usual and unfamiliar Lagrangians are respectively 
(with cartesian coordinates written as $q$s):
\be
L_1 =  \frac{1}{2}\left[{\dot q_1}^2 + {\dot q_2}^2 \; - \; \omega^2 
(q^2_1 + q^2_2) \right] \;\; ; \;\; L_2 = {\dot q_1}{\dot q_2} - 
\omega^2 q_1 q_2 \;\; .
\label{BothLags2DHO}
\ee
These inequivalent Lagrangians give the same Lagrange equations eq. 
\ref{eqn;lag}---or using Hamilton's Principle, the same Euler-Lagrange 
equations: viz.
\be
{\ddot q}_i + \omega^2 q_i = 0 \;\;, i = 1,2.
\label{eqnsfor1freq2DHO}
\ee
The rotations in the plane are of course a variational symmetry  of 
$L_1$, and a dynamical symmetry of eq. \ref{eqnsfor1freq2DHO}. But 
they are {\em not} a variational symmetry of $L_2$. So a dynamical 
symmetry need not be a variational one. Besides, eq.s 
\ref{BothLags2DHO} and \ref{eqnsfor1freq2DHO} contain another example 
to the same effect. Namely, the ``squeeze'' transformations 
\be
q'_1 := e^{\eta}q_1 \; , \; q'_2 := e^{-\eta}q_2
\label{squeeze}
\ee
are a dynamical symmetry of eq. \ref{eqnsfor1freq2DHO}, but not a 
variational symmetry of $L_1$. So again: a dynamical symmetry need not 
be a variational one.\footnote{In the light of this, you might ask 
about a more restricted implication: viz. must every dynamical 
symmetry of a set of equations of motion be a variational symmetry of 
{\em some or other Lagrangian} that yields the given equations as 
Euler-Lagrange equations? Again, the answer is No for the simple 
reason that there are many (sets of) equations of motion  that are not 
Euler-Lagrange equations of {\em any} Lagrangian, and yet have 
dynamical symmetries in the sense discussed, i.e. transformations that 
move solution curves to solution curves.

Wigner (1954) gives an example. The general question of under what 
conditions is a set of ordinary differential equations the 
Euler-Lagrange equations of some Hamilton's Principle is called the 
{\em inverse problem} of Lagrangian mechanics. It is a large subject, 
with a long history; cf. e.g. Santilli (1979), Lopuszanski (1999).}  

I turn to the first implication: that every variational symmetry is a 
dynamical symmetry. This is true: general and abstract proofs 
(applying also to continuous systems i.e. field theories) can be found 
in Olver (2000: theorem 4.14, p. 255; theorem 4.34, p. 278; theorem 
5.53, p. 332).\\
\indent  But beware of a condition of the theorem. It requires that 
all the variables $q$ (for continuous systems: all the fields $\phi$) 
be subject to Hamilton's Principle. The need for this condition  is 
shown by rotations in the plane, which are a variational symmetry  of 
the harmonic oscillator's familiar Lagrangian  $L_1$. But it is easy 
to show that such a rotation is a dynamical symmetry of one of 
Euler-Lagrange equations, say the equation for the variable $q_1$ 
\be
{\ddot q}_1 + \omega^2 q_1 = 0 \;\;, 
\label{eqnsforq1for1freq2DHO}
\ee
{\em only if} the corresponding Euler-Lagrange equation holds for 
$q_2$.

\subsubsection{The conjugate momentum of a vector 
field}\label{NoetherConjMomm}
Now we define {\em the momentum conjugate to a vector field} $X$ to be 
the scalar function on $TQ$:
\be
p_X: TQ \rightarrow \mathR \;\; ; \;\; p_X = \Sigma_i \; X_i \frac{\pl 
L}{\pl {\dot q}_i}
\label{defconjugmommX}
\ee
(For a time-dependent Lagrangian, $p_X$ would be a scalar function on 
$TQ \times \mathR$, with $\mathR$ representing time.)

\indent We shall see in examples below that this definition 
generalizes in an appropriate way our previous definition of the 
momentum conjugate to a coordinate $q$, in Section 
\ref{ssecGeneMom}.\\
\indent For the moment, I just note that it is an {\em improvement} in 
the sense that, as I said in Section \ref{basicCycl} (footnote 49), 
the momentum conjugate to a coordinate $q$ depends on the choice made 
for the other coordinates. But the momentum $p_X$ conjugate to a 
symmetry $X$ is independent of the coordinates chosen. I will (i) 
explain why, in intuitive terms (expanding Arnold 1989: 89); and then 
(ii) give a proof. {\em Warning}:--- (i) and (ii) are not needed for 
the statement and proof of Noether's theorem in Section 
\ref{Noetsubsubsec}.

\indent (i): This independence is suggested by the definition of 
$p_X$. For think of how in elementary calculus the rate of change 
(directional derivative) of a function $f: \mathR^3 \rightarrow 
\mathR$ along a line $\phi: t \mapsto \phi(t) \in \mathR^3$ is a 
coordinate-independent notion; and it is given by contracting the 
gradient of $f$ with the line's tangent vector, like eq. 
\ref{defconjugmommX}. That is: Taking cartesian coordinates, so that 
the tangent vector of the line is $(\frac{d x_1}{dt},\frac{d 
x_2}{dt},\frac{d x_3}{dt} )$, the directional derivative of $f$ is 
given by 
\be
\frac{d f}{dt} = \Sigma_i \; \frac{d x_i}{dt}\frac{\pl f}{\pl x_i} \; 
.
\ee
Then on analogy with the case in elementary calculus, we have: $p_X$ 
as defined by eq. \ref{defconjugmommX} is given by contracting the 
``gradient of $L$ with respect to the ${\dot q}$s'' with the vector 
$X$.\\
\indent Arnold (1989: 89) makes much the same point in terms of:\\
\indent \indent (i) the one-parameter family of transformations 
generated by $X$, call it $\theta_s$ with $s=0$ corresponding to the 
identity at a point $q \in Q$;\\
\indent \indent  (ii) the idea introduced in (2) of Section 
\ref{VecfieldsSymmies}, that the various possible velocity vectors 
${\dot q}$ form the tangent space $T_q$ at $q \in Q$.\\
\indent Thus Arnold says $p_X$ is
   the rate of change of $L(q,{\dot q})$ when the vector ${\dot q}$ 
`varies inside the tangent space $T_q$ with velocity 
$\frac{d}{ds}\mid_{s=0} \theta_s(q)$'.

(ii) To prove independence, we first apply the chain-rule to $L = 
L(q'(q), {\dot q}'(q,{\dot q}))$ and ``cancellation of the dots'' 
(i.e. eq. \ref{canceldots}, but now between arbitrary coordinate 
systems), to get:
\be
\frac{\pl L}{\pl {\dot q}_i} = \Sigma_j \; \frac{\pl L}{\pl {\dot 
q}'_j}\frac{\pl {\dot q}'_j}{\pl {\dot q}_i} 
= \Sigma_j \; \frac{\pl L}{\pl {\dot q}'_j}\frac{\pl q'_j}{\pl q_i}
\ee
Then  using eq. \ref{trsfmvecfield}, and relabelling $i$ and $j$, we 
deduce:
\be
p'_X = \Sigma_i \; X'_i \frac{\pl L}{\pl {\dot q}'_i} = \Sigma_{ij} \;
 X_j \frac{\pl q'_i}{\pl q_j}\frac{\pl L}{\pl {\dot q}'_i}
 = \Sigma_{ij} \;
 X_i \frac{\pl q'_j}{\pl q_i}\frac{\pl L}{\pl {\dot q}'_j}
 = \Sigma_i \; X_i \frac{\pl L}{\pl {\dot q}_i} \equiv p_X \; .
\ee 
Finally,  I remark incidentally that in a geometric formulation of 
Lagrangian mechanics, the coordinate-independence of $p_X$ becomes, 
unsurprisingly, a triviality. Namely: $p_X$ is obviously the 
contraction of $X$ with the canonical  one-form
\be
\theta_L := \frac{\pl L}{\pl {\dot q}^i}dq^i \;\;.
\label{defpdq2nd}
\ee
that we defined in eq. \ref{defpdq} of Paragraph 3.3.2.E (3).

\subsubsection{Noether's theorem; and examples}\label{Noetsubsubsec}
Given just the definition of conjugate momentum, eq. 
\ref{defconjugmommX}, the proof of Noether's theorem is  immediate. 
(The interpretation and properties of this momentum, discussed in the 
last Subsection, are not needed.)
The theorem says:
\begin{quote}
If $X$ is a (variational) symmetry of a system with Lagrangian 
$L(q,v,t)$, then $X$'s conjugate momentum is a constant of the motion.
\end{quote}
Proof: We just calculate the derivative of the momentum eq. 
\ref{defconjugmommX} along the solution curves in $TQ$, and apply the 
definitions of $Y_i$ eq. \ref{defineY}, and of symmetry eq. 
\ref{definevarnlsymy2}:
\begin{eqnarray}
\frac{d p}{dt} = \Sigma_i \; \frac{d X_i}{dt}\frac{\pl L}{\pl {\dot 
q}_i} \; + \;
\Sigma_i \;  X_i \frac{d }{dt}\left(\frac{\pl L}{\pl {\dot 
q}_i}\right) \\ \nonumber
= \; \Sigma_i \; Y_i\frac{\pl L}{\pl {\dot q}_i} \; + \; \Sigma_i \; 
X_i \frac{\pl L}{\pl q_i} \;\; = 0 \;\;.
\label{provenoet}
\end{eqnarray}

All of which, though neat, is a bit abstract! So here are two 
examples, both of which return us to examples we have already seen.

\indent The first example is a shift in a cyclic coordinate $q_n$: 
i.e. the case with which our discussion of Noether's theorem began, in 
Section \ref{InvarcefromCyclic}. So suppose $q_n$ is cyclic, and 
define a vector field $X$ by
\be
X_1 = 0, \dots, X_{n-1} = 0, \; X_n = 1 \; .
\ee
So the displacements generated by $X$ are translations by an amount 
$\epsilon$ in the $q_n$-direction. Then $Y_i := \frac{d X_i}{dt}$ 
vanishes, and the definition of (variational) symmetry eq. 
\ref{definevarnlsymy2} reduces to
\be
\frac{\pl L}{\pl q_n} = 0 \;
\ee
So since $q_n$ is assumed to be cyclic, $X$ is a symmetry. And the 
momentum conjugate to $X$, which Noether's theorem tells us is a 
constant of the motion, is the familiar one  
\be
p_X := \Sigma_i \; X_i \frac{\pl L}{\pl {\dot q}_i} =  \frac{\pl 
L}{\pl {\dot q}_n} \;\; .
\ee
Furthermore, as mentioned in this Section's Preamble, Paragraph 4.7.0, 
this example is {\em universal}, in that it follows from the Basic 
Theorem about solutions of ordinary differential equations (the 
`rectification theorem': Paragraph 2.3.1.A (i)) that every symmetry 
$X$ arises from a cyclic coordinate in some system of coordinates.

But for good measure, let us nevertheless look at our previous 
example, the angular momentum of a free particle (Section 
\ref{basicCycl}), in the {\em cartesian} coordinate system, i.e. a 
coordinate system without cyclic coordinates. So let $q_1 := x, q_2 := 
y, q_3 := z$. Then a small rotation about the $x$-axis
\be
\dd x = 0,  \;\; \dd y = - \epsilon z, \;\; \dd z = \epsilon y
\ee
corresponds to a vector field $X$ with components 
\be
X_1 = 0, \;\; X_2 = - q_3, \;\; X_3 = q_2
\ee
so that the $Y_i$ are
\be
Y_1 = 0, \;\; Y_2 = - {\dot q}_3, \;\; X_3 = {\dot q}_2 \;\; .
\ee
For the Lagrangian
\be
L = \frac{1}{2}m({\dot q}^2_1 + {\dot q}^2_2 + {\dot q}^2_3)
\ee
$X$ is a (variational) symmetry since the definition  of symmetry eq. 
\ref{definevarnlsymy2} now reduces to
\be
\Sigma_i \; X_i \frac{\pl L}{\pl q_i} \; + \; \Sigma_i \; Y_i 
\frac{\pl L}{\pl {\dot q_i}}  = - {\dot q}_3 \frac{\pl L}{\pl {\dot 
q}_2} + {\dot q}_2 \frac{\pl L}{\pl {\dot q}_3} = 0 \; .
\ee
So Noether's theorem them tells us that $X$'s conjugate momentum
\be
p_X := \Sigma_i \; X_i \frac{\pl L}{\pl {\dot q}_i} = X_2\frac{\pl 
L}{\pl {\dot q}_2} + X_3\frac{\pl L}{\pl {\dot q}_3} 
= -mz{\dot y} + my{\dot z}  
\ee 
which is indeed the $x$-component of angular momentum.

\newpage

\section{Envoi}\label{EnvoiLag}
Two of this paper's themes have been: praise of eighteenth century 
mechanics; and criticism of conceiving physical theorizing as 
``modelling''. So let me end by quoting two  passages concordant with 
those themes. My praise is summed up by Lanczos in the Preface to his 
wise book:
\begin{quote}
[T]he author ... again and again ... experienced the extraordinary 
elation of mind which accompanies a preoccupation with the basic 
principles and methods of analytical mechanics.
\indent \indent \indent  (Lanczos 1986: vii)
\end{quote}
And as an antidote to elation! ... My criticism is illustrated by  
Truesdell, famous not only as a distinguished mathematician and 
historian of mechanics, but also as an acerbic polemicist against all 
manner of shallow and fashionable ideas:
\begin{quote}
Nowadays people who for their equations and other statements about 
nature claim exact and eternal verity are usually dismissed as cranks 
or lunatics. Nevetheless, we lose something in this surrender to 
lawless uncertainty: Now we must tolerate the youth who blurts out the 
first, untutored, and uncritical thoughts that come into his head, 
calls them ``my model'' of something, and supports them by five or ten 
pounds of paper he calls ``my results'', gotten by applying his model 
to some numerical instances which he has elaborated by use of the 
largest machine he could get hold of, and if you say to him, ``Your 
model violates NEWTON's laws'', he replies ``Oh, I don't care about 
that, I tackle the physics directly, by computer.''\\
  (Truesdell 1987: 74; quoted by Papastavridis 2002: 817) 
\end{quote}

{\em Acknowledgements}:--- I am grateful to various audiences and 
friends  for comments on talks; and to Alexander Afriat, Harvey Brown, 
Jordi Cat, Tim Palmer, Graeme Segal and David Wallace for 
conversations and correspondence. I am especially grateful for 
comments on previous versions, to: Harvey Brown, Anjan Chakravartty, 
Robert Bishop, Larry Gould, Oliver Johns, Susan Sterrett, Michael 
St\"{o}ltzner and Paul Teller. I am also grateful to Oliver Johns for 
letting me read Chapters of his forthcoming (2005). 

\section{References}
V. Arnold (1973), {\em Ordinary Differential Equations}, MIT Press.

V. Arnold (1989), {\em Mathematical Methods of Classical Mechanics}, 
Springer, (second edition).

J. Bell (1987), {\em Speakable and Unspeakable in Quantum Mechanics}, 
Cambridge University Press.

G. Belot (2000), Geometry and Motion, {\em British Journal for the 
Philosophy of Science}, vol {\bf 51}, pp. 561-596.

G. Belot (2003), `Notes on symmetries', in Brading and Castellani 
(ed.s) (2003), pp. 393-412.

E. Borowski and J. Borwein (2002), {\em The Collins Dictionary of 
Mathematics}, Harper Collins; second edition.

U. Bottazini (1986), {\em The Higher Calculus: A History of Real and 
Complex Analysis from Euler to Weierstrass}, Springer-Verlag.

C. Boyer (1959), {\em The History of the Calculus and its Conceptual 
Development}, New York: Dover.

K. Brading and H. Brown (2003), `Symmetries and Noether's theorems', 
in Brading and Castellani (ed.s) (2003), p. 89-109.

K. Brading and E. Castellani (ed.s) (2003), {\em Symmetry in Physics}, 
Cambridge University Press.

H. Brown and P. Holland (2004), `Simple applications of Noether's 
first theorem in quantum mechanics and electromagnetism`, {\em 
American Journal of Physics} {\bf 72} p. 34-39. Available at: 
http://xxx.lanl.gov/abs/quant-ph/0302062 and 
http://philsci-archive.pitt.edu/archive/00000995/

H. Brown and P. Holland (2005), `Dynamical vs. variational symmetries: 
Understanding Noether's first theorem', {\em Molecular Physics}, 
forthcoming.
 
J. Butterfield (2004), `On the persistence of homogeneous matter',\\
Available at Los Alamos arXive: 
http://xxx.soton.ac.uk/abs/physics/0406021; and Pittsburgh archive: 
http://philsci-archive.pitt.edu/archive/00001760.

J. Butterfield (2004a), `Classical mechanics is not {\em 
pointilliste}, and can be perdurantist'. In preparation.

J. Butterfield (2004b), `Some philosophical morals of catastrophe 
theory'. In preparation.

J. Butterfield (2004c), `On Hamilton-Jacobi Theory as a Classical Root 
of Quantum Theory', in {\em Quo Vadis Quantum Mechanics?}, ed. A. 
Elitzur et al., Proceedings of a Temple University conference; 
Springer.\\
Available at Pittsburgh archive: 
http://philsci-archive.pitt.edu/archive/00001193, and at Los Alamos 
arXive: http://xxx.lanl.gov/abs/quant-ph/0210140 or\\    
http://xxx.soton.ac.uk/abs/quant-ph/0210140

J. Butterfield (2004d), `David Lewis Meets Hamilton and Jacobi', 
forthcoming in {\em Philosophy of Science}. Available at Pittsburgh 
archive:\\ http://philsci-archive.pitt.edu/archive/00001191

J. Butterfield (2004e), `Some Aspects of Modality in Analytical 
mechanics',  in {\em Formal Teleology and Causality}, ed. M. 
St\"{o}ltzner, P. Weingartner, Paderborn: Mentis. \\
Available at Los Alamos arXive: 
http://xxx.lanl.gov/abs/physics/0210081 or \\   
http://xxx.soton.ac.uk/abs/physics/0210081;
 and at Pittsburgh archive: 
http://philsci-archive.pitt.edu/archive/00001192.

J. Butterfield and C.Isham (1999), `The Emergence of time in quantum 
gravity', in J. Butterfield ed., {\em The Arguments of Time}, British 
Academy and Oxford University Press.

N. Cartwright (1989), {\em Nature's Capacities and their Measurement}, 
Oxford University Press. 

N. Cartwright (1999), {\em The Dappled World}, Cambridge  University 
Press. 

J. Casey (1991), The principle of rigidification, {\em Archive for the 
History of the Exact Sciences} {\bf 43}, p. 329-383.

R. Courant and D. Hilbert (1953), {\em Methods of Mathematical 
Physics}, volume I, Wiley-Interscience (Wiley Classics 1989).

R. Courant and D. Hilbert (1962), {\em Methods of Mathematical 
Physics}, volume II, Wiley-Interscience (Wiley Classics 1989).

M. Crowe (1985), {\em A History of Vector Analysis: the evolution of 
the idea of a vectorial system}, Dover.

E. Desloge (1982), {\em Classical Mechanics}, John Wiley.

F. Diacu and P. Holmes (1996), {\em Celestial Encounters: the Origins 
of Chaos and Stability}, Princeton University Press. 

J. Earman (2003), `Tracking down gauge: an ordinary differential 
equation to the constrained Hamiltonian formalism', in Brading and 
Castellani (ed.s) (2003), pp. 140-162.

J. Earman and J. Roberts (1999), `{\em Ceteris paribus}, there is no 
problem of provisos', {\em Synthese} {\bf 118}, p. 439-478.

G. Emch (2002), `On Wigner's different usages of models', in {\em 
Proceedings of the Wigner Centennial Conference, Pecs 2002};paper No. 
59; to appear also in {\em Heavy Ion Physics}.

G. Emch and C. Liu (2002), {\em The Logic of Thermo-statistical 
Physics}, Springer.

C. Fox (1987), {\em An Introduction to the Calculus of Variations}, 
Dover.

C. Fraser (1983), `J.L. Lagrange's early contributions to the 
principles  and methods of dynamics', {\em Archive for the History of 
the Exact Sciences} {\bf 28}, p. 197-241. Reprinted in C. Fraser 
(1997). 

C. Fraser (1985), `J.L. Lagrange's changing approach to the 
foundations of the calculus of variations', {\em Archive for the 
History of the Exact Sciences} {\bf 32}, p. 151-191. Reprinted in C. 
Fraser (1997).  

C. Fraser (1985a), `D'Alembert's Principle: the original formulation 
and application in Jean d'Alembert's {\em Trait\'{e} de Dynamique} 
(1743), {\em Centaurus} {\bf 28}, p. 31-61, 145-159. Reprinted in C. 
Fraser (1997).  

C. Fraser (1994), `The origins of Euler's variational calculus', {\em 
Archive for the History of the Exact Sciences} {\bf 47}, p. 103-141. 
Reprinted in C. Fraser (1997).  

C. Fraser (1997), {\em Calculus and Analytical Mechanics in the Age of 
the Enlightenment}, Ashgate: Variorum Collected Studies Series.

R. Giere (1988), {\em Explaining Science: a Cognitive Approach}, 
University of Chicago Press.

R. Giere (1999), {\em  Science without Laws}, University of Chicago 
Press.

H. Goldstein (1966), {\em Classical Mechanics}, Addison-Wesley. (third 
printing of a 1950 first edition)

H. Goldstein et al. (2002), {\em Classical Mechanics}, Addison-Wesley, 
(third edition) 

J. Gray (2000), {\em Linear Differential Equations and Group Theory 
from Riemann to Poincar\'{e}}, Boston: Birkhauser.

H. Hertz  (1894), {\em Die Principien der Mechanik}, Leipzig: 
J.A.Barth; trans. by D. Jones and J. Whalley as {\em The Principles of 
Mechanics}, London: Macmillan 1899; reprinted Dover, 1956.

O. Johns (2005), {\em Analytical Mechanics for Relativity and Quantum 
Mechanics}, Oxford University Press, forthcoming.

J. Jos\'{e} and E. Saletan (1998), {\em Classical Dynamics: a 
Contemporary Approach}, Cambridge University Press.

H. Kastrup (1987), `The contributions of Emmy Noether, Felix Klein and 
Sophus Lie to the modern concept of symmetries  in physical  systems', 
in {\em Symmetries in Physics (1600-1980)}, Barcelona: Bellaterra, 
Universitat Autonoma de Barcelona, p. 113-163.

M. Kline (1972), {\em Mathematical Thought from Ancient to Modern 
Times}, Oxford: University Press.

T. Kuhn (1962), {\em The Structure of Scientific Revolutions}, second 
edition with a Postscript (1970); University of Chicago Press.

C. Lanczos (1986), {\em The Variational Principles of Mechanics}, 
Dover; (reprint of the 4th edition of 1970).
 
D. Lewis (1986), {\em Philosophical Papers, volume II}, Oxford 
University Press.

J. Lopuszanski (1999), {\em The Inverse Variational Problem in 
Classical Mechanics}, World Scientific.

D. Lovelock and H. Rund (1975), {\em Tensors, Differential Forms and 
Variational Principles}, John Wiley.

J. Lutzen (1995), {\em Denouncing Forces; Geometrizing Mechanics: 
Hertz's Principles of Mechanics}, Copenhagen University Mathematical 
Institute Preprint Series No 22. 

J. Lutzen (2003), `Between rigor and applications: developments in the 
concept of function in mathematical analysis', in {\em Cambridge 
History of Science, vol. 5: The modern physical and mathematical 
sciences},  ed. M.J. Nye, p. 468-487.

J. Marsden and T. Hughes (1983), {\em Mathematical Foundations of 
Elasticity}, Prentice-Hall; Dover 1994.

E. McMullin (1985), `Galilean idealization',  {\em Studies in the 
History and Philosophy of Science} {\bf 16}, p. 247-273. 

G. Morandi et al (1990), `The inverse problem of the calculus of 
variations and the geometry of the tangent bundle', {\em Physics 
Reports} {\bf 188}, p. 147-284.

M. Morgan and M. Morrison (1999), {\em Models as Mediators}, Cambridge 
University Press. 

E. Noether (1918), `Invariante Variationprobleme', {\em G\"{o}ttinger 
Nachrichten, Math-physics. Kl.}, p. 235-257.

P. Olver (2000), {\em Applications of Lie Groups to Differential 
Equations}, second edition: Springer-Verlag.

J. Papastavridis (2002), {\em Analytical Mechanics}, Oxford University 
Press.

R.M. Santilli (1979), {\em Foundations of Theoretical Mechanics, vol. 
I}, Springer-Verlag

V. Scheffer (1993) `An inviscid flow with compact support in 
spacetime', {\em Journal  Geom. Analysis} {\bf 3(4)}, pp. 343-401

J H Schmidt (1997), 'Classical Universes are perfectly Predictable!', 
{\em Studies in the History and Philosophy of Modern Physics}, volume 
28B, 1997, p. 433-460.

J H Schmidt (1998), 'Predicting the Motion of Particles in Newtonian 
Mechanics and Special Relativity', {\em Studies in the History and 
Philosophy of Modern Physics}, volume 29B, 1998, p. 81-122.

A. Shnirelman (1997), `On the non-uniqueness of weak solutions of the 
Euler equations', {\em Communications on Pure and Applied mathematics} 
{\bf 50}, pp. 1260-1286.

St\"{o}ltzner, M.  (2003), `The Principle of Least Action as the 
Logical Empiricist's Shibboleth',  {\em Studies in History and 
Philosophy of Modern Physics} {\bf 34B}, p. 285-318.

St\"{o}ltzner, M.  (2004), `On Optimism and Opportunism in Applied 
Mathematics (Mark Wilson Meets John von Neumann on Mathematical 
Ontology), {\em Erkenntnis} {\bf 60}, pp. 121-145. Available at 
http://philsci-archive.pitt.edu/archive/00001225

F. Suppe (1988), {\em The Semantic Conception of Scientific Theories 
and Scientific Realism}, University of Illinois Press.

P. Teller (1979), `Quantum mechanics and the nature of continuous 
physical quantities', {\em Journal of Philosophy} {\bf 76}, p. 
345-361. 

R. Torretti (1999), {\em The Philosophy of Physics}, Cambridge 
University Press.

C. Truesdell (1956), `Introduction', in L. Euler {\em Opera Omnia} 
(four series from 1911: Leipzig/Berlin/Zurich/Basel): series 2, {\bf 
13}, pp. ix-cv.

C. Truesdell (1960), `The rational mechanics of flexible or elastic 
bodies 1638-1788', in L. Euler {\em Opera Omnia} (four series from 
1911: Leipzig/Berlin/Zurich/Basel): series 2, {\bf 11}, Section 2.

C. Truesdell (1987), {\em Great Scientists of Old as Heretics in ``The 
Scientific Method''}, University Press of Virginia. 

C. Truesdell (1991), {\em A First Course in Rational Continuum 
Mechanics}, volume 1; second edition; Academic Press. 

D. Wallace (2003), `Time-dependent symmetries: the link between gauge 
symmetries and indeterminism', in Brading and Castellani (ed.s) 
(2003), pp. 163-173.

E. Whittaker (1959), {\em A Treatise on the Analytical Dynamics of 
Particles and Rigid Bodies}, Cambridge University Press (4th edition).

E. Wigner (1954), `Conservation laws in classical and quantum 
physics', {\em Progress of Theoretical Physics} {\bf 11}, p. 437-440. 

M. Wilson, (1985), `What is this Thing called Pain?---the Philosophy 
of Science behind the Contemporary Debate', {\em Pacific Philosophical 
Quarterly} {\bf 66}, p. 227-267.

M. Wilson, (1993), `Honorable Intensions', in {\em Naturalism}, ed. S. 
Wagner and R.Warner, South Bend: University of Notre Dame Press, p. 
53-94.

M. Wilson (1997), `Reflections on Strings',  in {\em Thought 
Experiments in Science and Philosophy}, ed. T. Horowitz and G. Massey, 
University of Pittsburgh Press p.?? 

M. Wilson, (2000), `The Unreasonable Uncooperativeness of Mathematics 
in the Natural Sciences', {\em The Monist} {\bf 83}, 296-314.

N. Woodhouse (1987), {\em Introduction to Analyical Dynamics}, Oxford 
University Press.

W. Yourgrau and S. Mandelstam (1979), {\em Variational Principles in 
Dynamics and Quantum Theory}, Dover.  

A Youschkevitch (1976), 'The Concept of Function up to the Middle of 
the nineteenth Century', {\em Archive for the History of the Exact 
Sciences} {\bf 16}, p. 37-85. 

\end{document}